\documentclass[prx,aps,amsart,amssymb,epsfig,showpacs,eqsecnum,superscriptaddress,twocolumn]{revtex4-1}
\usepackage{longtable}
\usepackage{lipsum}
\usepackage[hidelinks]{hyperref}
\usepackage{graphicx}
\usepackage[table,dvipsnames]{xcolor}
\usepackage{amsmath}
\usepackage{amsfonts}
\usepackage{amssymb}
\usepackage{dsfont}
\usepackage[retainorgcmds]{IEEEtrantools}
\usepackage{physics}
\usepackage{mhchem}
\usepackage[normalem]{ulem}
\usepackage[export]{adjustbox}

\usepackage{etoolbox}
\makeatletter
\patchcmd{\@IEEEeqnarray}{\relax}{\relax\intertext@}{}{}
\makeatother

\usepackage{floatrow}
\usepackage[caption=false]{subfig}
\usepackage{placeins}

\definecolor{spinless_green2}{rgb}{0.95, 1.0, 0.9}
\definecolor{spinless_green1}{rgb}{0.85, 1.0, 0.80}
\definecolor{spinfull_blue1}{rgb}{0.90, 0.93, 1.0}
\definecolor{spinfull_blue2}{rgb}{0.94, 0.97, 1.0}

\newcommand{\mytilde}{\raise.17ex\hbox{$\scriptstyle\mathtt{\sim}$}}

\begin{document}
\floatsetup[figure]{style=plain,subcapbesideposition=top}
\captionsetup[subfigure]{position=top}

\title{Symmetry-enforced topological band crossings in orthorhombic crystals: \\
Classification and materials discovery}

\author{Andreas Leonhardt}	
\email{a.leonhardt@fkf.mpg.de}
\affiliation{Max-Planck-Institut f\"ur Festk\"orperforschung, Heisenbergstrasse 1, D-70569 Stuttgart, Germany} 

\author{Moritz~M.~Hirschmann}
\affiliation{Max-Planck-Institut f\"ur Festk\"orperforschung, Heisenbergstrasse 1, D-70569 Stuttgart, Germany} 

\author{Niclas Heinsdorf}
\affiliation{Max-Planck-Institut f\"ur Festk\"orperforschung, Heisenbergstrasse 1, D-70569 Stuttgart, Germany} 

\author{Xianxin Wu}
\affiliation{Max-Planck-Institut f\"ur Festk\"orperforschung, Heisenbergstrasse 1, D-70569 Stuttgart, Germany} 

\author{Douglas H. Fabini}
\affiliation{Max-Planck-Institut f\"ur Festk\"orperforschung, Heisenbergstrasse 1, D-70569 Stuttgart, Germany} 

\author{Andreas P. Schnyder} 
\email{a.schnyder@fkf.mpg.de}
\affiliation{Max-Planck-Institut f\"ur Festk\"orperforschung, Heisenbergstrasse 1, D-70569 Stuttgart, Germany} 

\date{\today}

\begin{abstract}
We identify all symmetry-enforced band crossings
in nonmagnetic orthorhombic crystals with and without
spin-orbit coupling and discuss their topological properties.
We find that orthorhombic crystals can host a large number of different
band degeneracies, including movable Weyl and Dirac points with hourglass dispersions,
fourfold double Weyl points, Weyl and Dirac nodal lines, almost movable nodal lines,
nodal chains, and topological nodal planes. 
Interestingly, spin-orbit coupled materials in the space groups 18, 36, 44, 45, and 46 can
have band pairs with only two Weyl points in the entire Brillouin zone.
This results in a simpler connectivity of the Fermi arcs and more pronounced topological responses
 than in materials with four or more Weyl points.
In addition, we show that the symmetries of the space groups 56, 61, and 62 
enforce nontrivial weak $\mathbb{Z}_2$ topology in materials with strong spin-orbit coupling,
leading to helical surface states.
With these classification results in hand,
we perform extensive database searches 
for orthorhombic materials crystallizing in the relevant space groups. 
We find that Sr$_2$Bi$_3$ and Ir$_2$Si have bands crossing the Fermi energy
with a symmetry-enforced nontrivial $\mathbb{Z}_2$ invariant, CuIrB possesses nodal 
chains near the Fermi energy, Pd$_7$Se$_4$ and Ag$_2$Se
exhibit fourfold double Weyl points, the latter one even in the absence of spin-orbit coupling,
whereas the fourfold degeneracies in AuTlSb are made up from intersecting nodal lines.
For each of these examples we compute the \textit{ab-initio} band structures,
discuss their topologies, and for some cases also calculate the surface states.
\end{abstract}

\maketitle

\section{Introduction}

The discovery of topological insulators more than ten years ago~\cite{hsieh_BiSb_nature_08,hsieh_science_2009,kane_hasan_review_2010} has kicked off a classification program of topological band structures~\cite{schnyder_classification_2008,kitaev_classification_2009,chiu_RMP_16}, 
which is still ongoing today.
Insulators as well as Dirac and Weyl-type semimetals have been classified, both in terms of nonspatial symmetries~\cite{schnyder_classification_2008,kitaev_classification_2009} and crystalline symmetries~\cite{ando_fu_review,chiu_ryu_PRB_13,chiu_schnyder_PRB_14,ando_fu_review,binghai_review_weyl_17}.
The methods used to establish these classifications are wide ranging, including Clifford algebra extensions~\cite{kitaev_classification_2009,morimoto_PRB_13,zhao_schnyder_PRL_16,song_schnyder_PRB_17,Stone_2010}, minimal Dirac models~\cite{chiu_ryu_PRB_13,chiu_schnyder_PRB_14}, 
K theory~\cite{kitaev_classification_2009}, compatibility relations between irreps~\cite{Elcoro_double_cryst,hexagonals_PhysRevMaterials.2.074201,trigonals_PhysRevMaterials.3.124204}, 
symmetry eigenvalue analyses~\cite{hexagonals_PhysRevMaterials.2.074201,trigonals_PhysRevMaterials.3.124204,hirschmann2021symmetryenforced,bzdusek_soluyanov_nature_16,furusaki_2017,ryo_murakami_PRB_17,yang_furusaki_PRB_2017,tsirkin_vanderbilt_PRB_17},
symmetry-based indicators~\cite{tang_science_advances_2019}, as well as the study of elementary band representations~\cite{bradlyn_bernevig_Nature_2017,vergniory_PRE_17}. 
These classification works lay down the basis for the study of topological systems in general and can be used
as guiding principles for the design and discovery of new topological systems.
Indeed, when combined with materials databases, such as the
ICSD from FIZ Karlsruhe~\cite{ICSD_link} and the Materials Project database~\cite{materials_project_2013,Ong2015}, these classifications have lead to the prediction of
new topological materials~\cite{hexagonals_PhysRevMaterials.2.074201,trigonals_PhysRevMaterials.3.124204,hirschmann2021symmetryenforced,tang_science_advances_2019,zhang_chen_Nature_2019,vergniory_wang_Nature_19}.

Topological materials are not only of fundamental interest, but are also sought after for
their functional properties that can be harnessed for applications~\cite{topology_applications_nature_reviews_2019}.
For example, the spin-momentum locking of the surface states could be used for low-dissipation transport
in future post-silicon devices~\cite{tian_scientific_reports_2015,tian_spin_pol_TI_sci_adv_17}. Moreover, 
the high mobility and large magnetoresistance of many Weyl semimetals could be
useful for future high-speed electronics and spintronics~\cite{parikin_IEEE_03,binghai_review_weyl_17}.
Hence, there is a need for new topological materials with these properties.
One strategy to discover new topological semimetals is to focus on symmetry-enforced
topologies, i.e., topological features that are enforced to exist by symmetry alone, independent
of the band dispersion, orbital content, and chemical composition of the material. 
Once the space group (SG) symmetries that enforce the desired band topologies are identified,
a suitable material can be found by browsing databases of known materials by SG number.

Previously, we have applied this strategy to hexagonal, trigonal, and tetragonal crystals~\cite{hexagonals_PhysRevMaterials.2.074201,trigonals_PhysRevMaterials.3.124204,hirschmann2021symmetryenforced}. 
These investigations have uncovered, among other things, the accordion states in trigonal Te~\cite{crepaldi_accordion_Te_PRL20}, and identified ZrIrSn and NaSn$_5$ as nodal line materials with twofold and fourfold Weyl lines, respectively~\cite{hexagonals_PhysRevMaterials.2.074201,hirschmann2021symmetryenforced}. 
In this paper, we continue this classification program by investigating symmetry-enforced band crossings in orthorhombic crystals.
Our results are summarized in Tables~\ref{SGs_noinv_nosoc}, \ref{SGs_noinv_soc}, \ref{SGs_inv_nosoc}, and~\ref{SGs_inv_soc},
which classify all symmetry-enforced topological features in band structures of orthorhombic crystals with time-reversal symmetry.
 
Tables~\ref{SGs_noinv_nosoc} and~\ref{SGs_inv_nosoc} apply to band structures with time-reversal symmetry that squares to $+1$, e.g., to electronic bands without spin-orbit coupling (SOC). 
This type of time-reversal symmetry is present in materials with light elements, where SOC can be neglected. It also occurs in the excitation spectra of bosonic quasiparticles, e.g., phonon or magnon bands~\cite{topological_phononics_review_20,topological_magnons_tutorial_21}, and in synthetic materials, such as photonic crystals~\cite{top_photonics_RMP_19} and electric circuit networks~\cite{lee_comm_phys_18,rui_zhao_NSR_2020}.
We find that without SOC the band structures can exhibit movable Weyl points, due to screw rotations, movable and almost movable nodal lines, due to mirror symmetries,
and nodal planes, due to the combination of screw rotations with time-reversal symmetry.
By the bulk-boundary correspondence, the Weyl points and nodal lines lead to Fermi arc and drumhead surface states, respectively.

Tables~\ref{SGs_noinv_soc} and~\ref{SGs_inv_soc} list all symmetry-enforced topological features in band structures with time-reversal symmetry that squares to $-1$, e.g., in electronic bands with strong SOC. We find that with SOC the bands can possess various types of different point degeneracies, namely,
Kramers-Weyl points, fourfold points with zero Chern number ($\mathcal{C}=0$), 
movable Weyl and Dirac points with hourglass dispersions, and
fourfold double Weyl points with $|\mathcal{C}|=2$.
Line degeneracies in the presence of SOC also exist in different varieties:
(almost) movable Weyl lines with hourglass dispersion,
Dirac lines protected by off-centered symmetries,
and (interlinked) nodal chains.
Finally, there are nine SGs with nodal planes at the BZ boundary,
of which the chiral SGs 17--19, and 20 can have a nonzero topological charge.
Remarkably, in SG~19 the nodal planes must always be topological by symmetry,
irrespective of the orbital content and dispersions of the bands (Sec.~\ref{sec_nodal_planes}).

Furthermore, we identify two SGs, namely 56 and 62, which in the presence of SOC exhibit symmetry-enforced nontrivial weak $\mathbb{Z}_2$ invariants (Sec.~\ref{sec_enforced_Z2}).
The same holds for SG 61, assuming that elementary band representations with different inversion eigenvalues do not mix.
Similarly, SGs 52 and 60 have nontrivial weak $\mathbb{Z}_2$ invariants, albeit only with weak SOC that does not induce band inversions.
By the bulk-boundary correspondence, spin-orbit coupled materials in all these SGs exhibit helical surface states, that could potentially be used for low-dissipation (spin) transport.

Lastly, we determine those SGs where the number of Weyl points formed by a spin-orbit coupled band pair 
can be as low as four (or even two), i.e., where only four (or two) of the eight time-reversal invariant 
momenta (TRIMs) host Weyl points, while the other TRIMs are part of nodal lines or nodal planes.
Assuming that there are no other accidental Weyl points, band structures in these 
SGs have large Fermi arc surface states with a simple connectivity. Moreover,
the topological responses, such as the anomalous (spin) Hall effect, are expected
to be enhanced, since Weyl points with opposite chiralities are separated
by a large distance in reciprocal space~\cite{minimumNumberArcs_MoTe_2016_prl,minimumWeyl_TaIrTe_2016_prb,belopolski2017signatures}.

Using these classification results as an input, we perform an extensive database search for
materials crystallizing in the relevant SGs (Sec.~\ref{sec_III} and Fig.~\ref{fig_heatmap}). This search yields six candidate materials, 
whose band topology we study using DFT calculations and by determining the irreducible representations at high-symmetry points.
From this, we find that Ag$_2$Se and Pd$_7$Se$_4$ possess fourfold double Weyl points and nodal planes (Sec.~\ref{mat_Pd7Se4}).
AuTlSb exhibits hourglass nodal lines and a fourfold point with $\mathcal{C} =0$, while
CuIrB has nodal chains near the Fermi level.
Both Sr$_2$Bi$_3$ and Ir$_2$Si have bands crossing the Fermi level
with nontrivial $\mathbb{Z}_2$ invariants.

The remainder of the paper is organized as follows. We start in Sec.~\ref{sec_II} by defining our naming conventions and by describing
the six different orthorhombic BZs and their TRIMs and high-symmetry lines.
Our selection criteria for the database search are described in Sec.~\ref{sec_III}, which yields six candidate materials.
The band topologies and dispersions of these materials are presented in those sections where the corresponding
SGs are discussed.
In Sec.~\ref{sec_IV} we study the band topologies for the rhombic disphenoidal SGs (SGs 16--24), which
are chiral due to the absence of both mirror and inversion symmetries.
Sections~\ref{sec_V} and \ref{sec_VI} contain the analyses of the rhombic pyramidal SGs (SGs 25--46)
and the rhombic dipyramidal SGs (SGs 47--74), respectively. 
Some concluding remarks are given in
Sec.~\ref{sec_VII}. Additional \textit{ab-initio} band structures of the example materials are presented in Appendix~\ref{app_A}.
In Appendix~\ref{app_B} we present low-energy models of the
fourfold degenerate points discussed in the main text.
Appendix~\ref{app_SG19} contains a minimal tight-binding model for
SG 19, which highlights the enforced features of chiral
orthorhombic SGs both with and without SOC. 

\section{Conventions}
\label{sec_II}

\subsection{Symmetries}

The orthorhombic SGs consist of all those SGs
with only twofold symmetries on the three mutual orthogonal
symmetry axes. Those symmetries are
twofold (screw) rotations and (glide) mirror symmetries.
In case of three orthogonal mirror symmetries there is also inversion symmetry.
Additionally, we always assume time-reversal symmetry to
be present. Since we will make frequent use of these symmetries,
we introduce the notation with some examples by 
explicitly writing out the transformed coordinates
of a point $(x,y,z)$ in Cartesian coordinates aligned with the conventional
unit cell. Additionally, we include the action in spin space in terms of 
the Pauli matrices $\sigma_i$, $i\in\{x,y,z\}$, and
$\sigma_0=\mathds{1}_{2\times2}$.
A translation by a
lattice vector $(a,b,c)$ in real space is denoted by
\begin{IEEEeqnarray}{lCl}
\IEEEeqnarraymulticol{3}{l}{ t(a,b,c):} \nonumber \\
(x,y,z) \mapsto  ( x+a, y+b, z+c) &\otimes& \sigma_0.          \label{def_translation}\\
\intertext{A twofold rotation about the $z$-axis with (fractional) translation vector $(a,b,c)$ is written as}
\IEEEeqnarraymulticol{3}{l}{ 2_{001}(a,b,c):} \nonumber \\
(x,y,z) \mapsto   (-x+a,-y+b, z+c) &\otimes& \mathrm{i}\sigma_z. \label{def_rot} \\ 
\intertext{Similarly, a mirror symmetry with $xz$ mirror planes and (fractional) translation vector $(a,b,c)$ is written as}
\IEEEeqnarraymulticol{3}{l}{ M_{010}(a,b,c):} \nonumber \\
(x,y,z) \mapsto  ( x+a,-y+b, z+c) &\otimes& \mathrm{i}\sigma_y. \label{def_mirror} 
\end{IEEEeqnarray}
For SGs with inversion symmetry $\mathcal{P}$, the origin choice
is always at an inversion center (origin choice 2
in~\cite{Hahn2002}),
\begin{IEEEeqnarray}{rcCl}
\mathcal{P}: &  (x,y,z)\mapsto(-x,  -y,  -z  ) &\otimes& \sigma_0.        \label{def_inversion}   \\
\intertext{
Finally, time-reversal symmetry is denoted by
}
\mathcal{T}: & (x,y,z)\mapsto( x,   y,   z  ) &\otimes& \mathrm{i}\sigma_y\mathcal{K}. \label{def_trs_last}
\end{IEEEeqnarray}
Our analysis covers spinful and spinless band structures of the
orthorhombic SGs. 
The spinful symmetry
groups are the double SGs with a distinguished element
for a $2\pi$-rotation, $\mathds{1}\otimes(-\sigma_0)$.

Spinless representations are relevant for bosonic band structures,
including in metamaterials, and materials with negligible spin-orbit coupling.
In the latter case, we make use of
the unbroken SU(2) symmetry. This allows us to use spinless
representations by removing the spin part of the
symmetries
in Eqs.\eqref{def_translation}-\eqref{def_trs_last}
using a properly chosen SU(2) rotation. 
This way each spin degree of freedom can be treated independently
without the need of using double SGs.
For our results in that case, we will not include the resulting
spin degeneracy, unless it is mentioned explicitly.
We use the parameter $\zeta=0,1$ to distinguish between the
spinless and the spinful case, for example the representation of a
$2\pi$ rotation is always $(-1)^\zeta$. 
% TODO: remove if the color in the tables is removed
For easier readability, the tables with the results for spinless
band structures, Tab.~\ref{SGs_noinv_nosoc} and
Tab.~\ref{SGs_inv_nosoc}, have a green background, whereas the
results for spinful band structures in Tables~\ref{SGs_noinv_soc}
and~\ref{SGs_inv_soc} are colored in blue.

\subsection{Orthorhombic lattices and Brillouin zones}

% Figure: Brillouin zones %%%%%%%%%%%%%%%%%%%%%%%%%%%%%%%%%%%%%%%%
\begin{figure*}[h!t]
\subfloat[ P: primitive ]{
\includegraphics[width=.24\linewidth,valign=t]{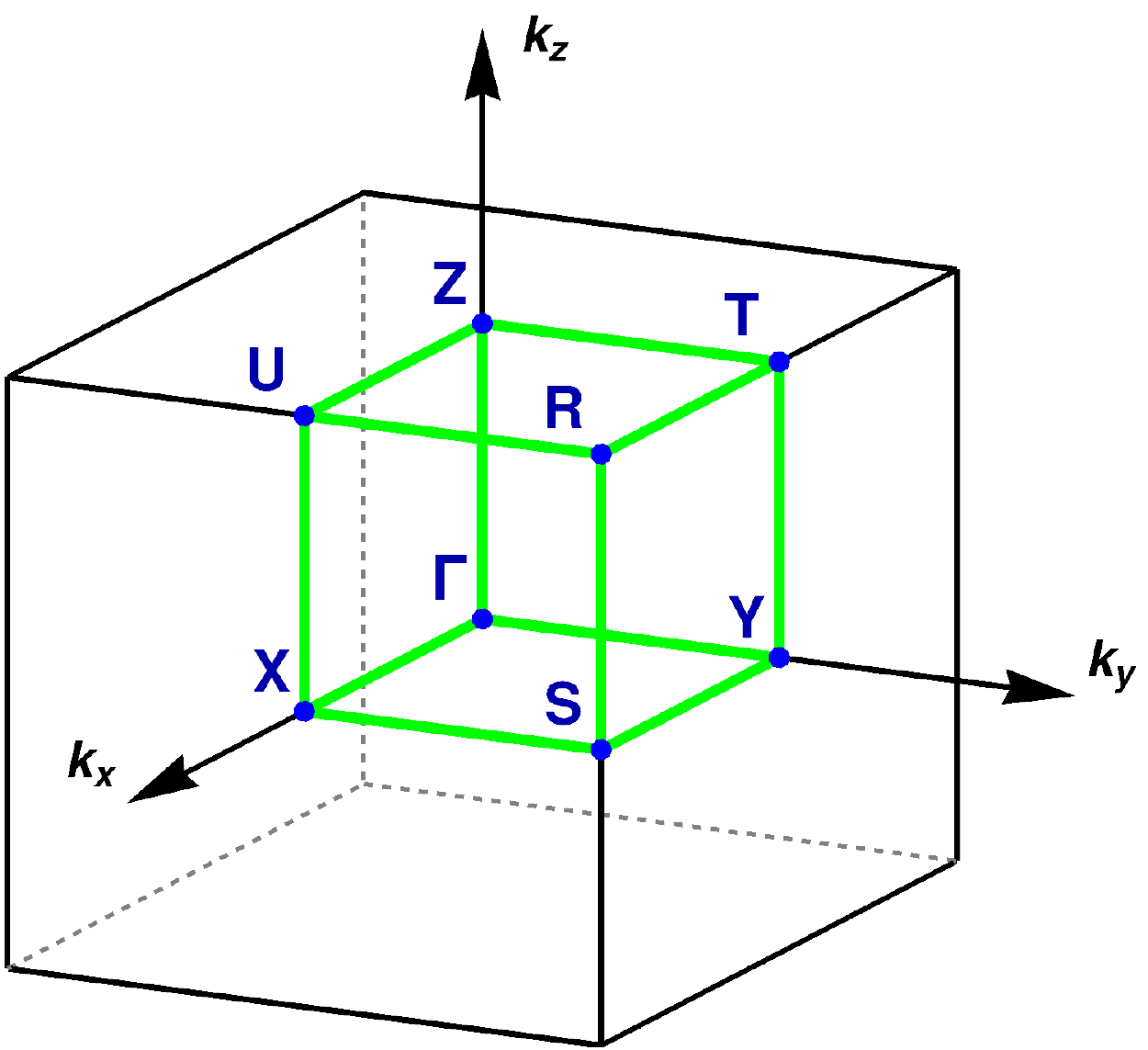}
\label{BZ_prim}
}
\subfloat[ C: base-centered ]{
\begin{minipage}[c][4cm]{.24\linewidth}
\includegraphics[width=1.0\linewidth,valign=t]{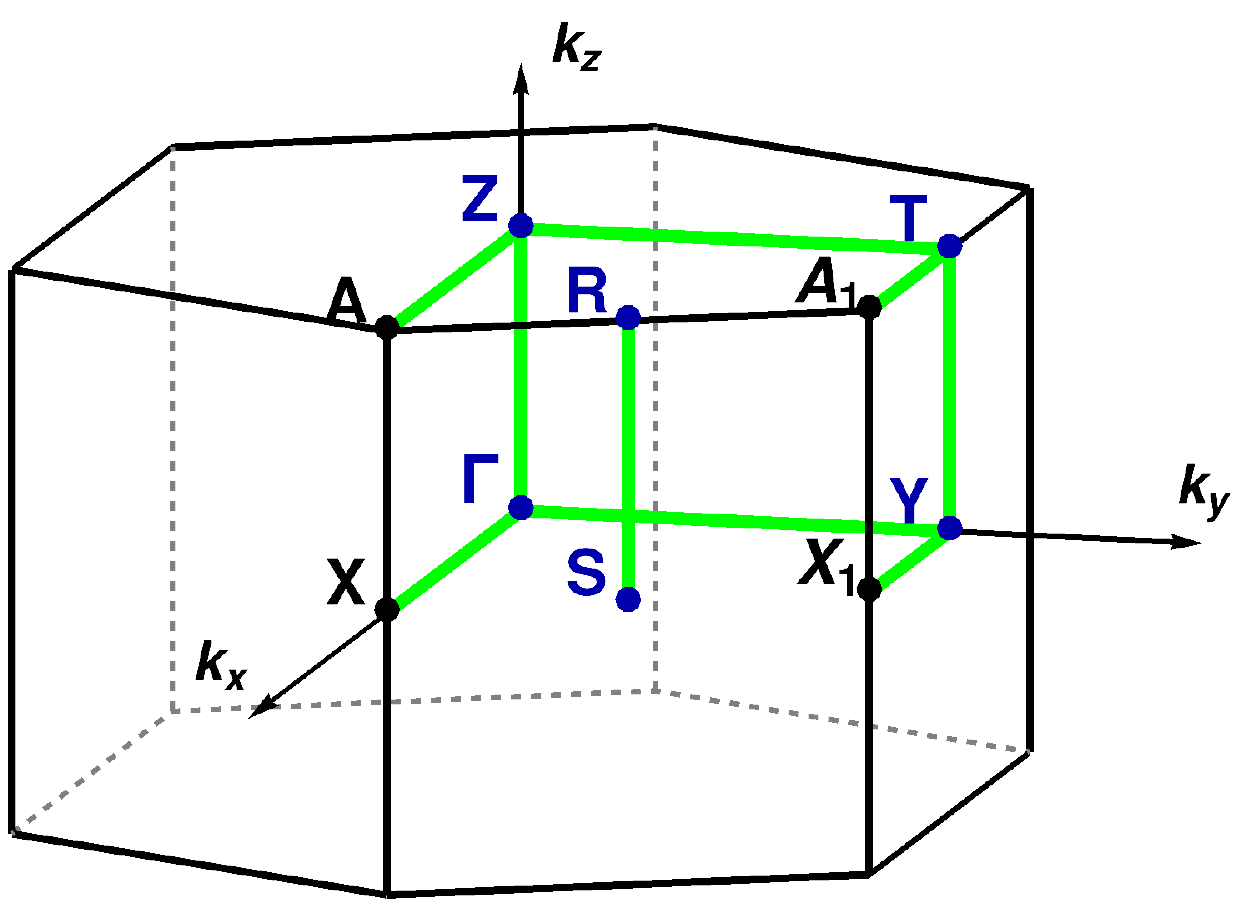}
\end{minipage}
\label{BZ_base}
}
\subfloat[ F: face-centered ]{
\includegraphics[width=.24\linewidth,valign=t]{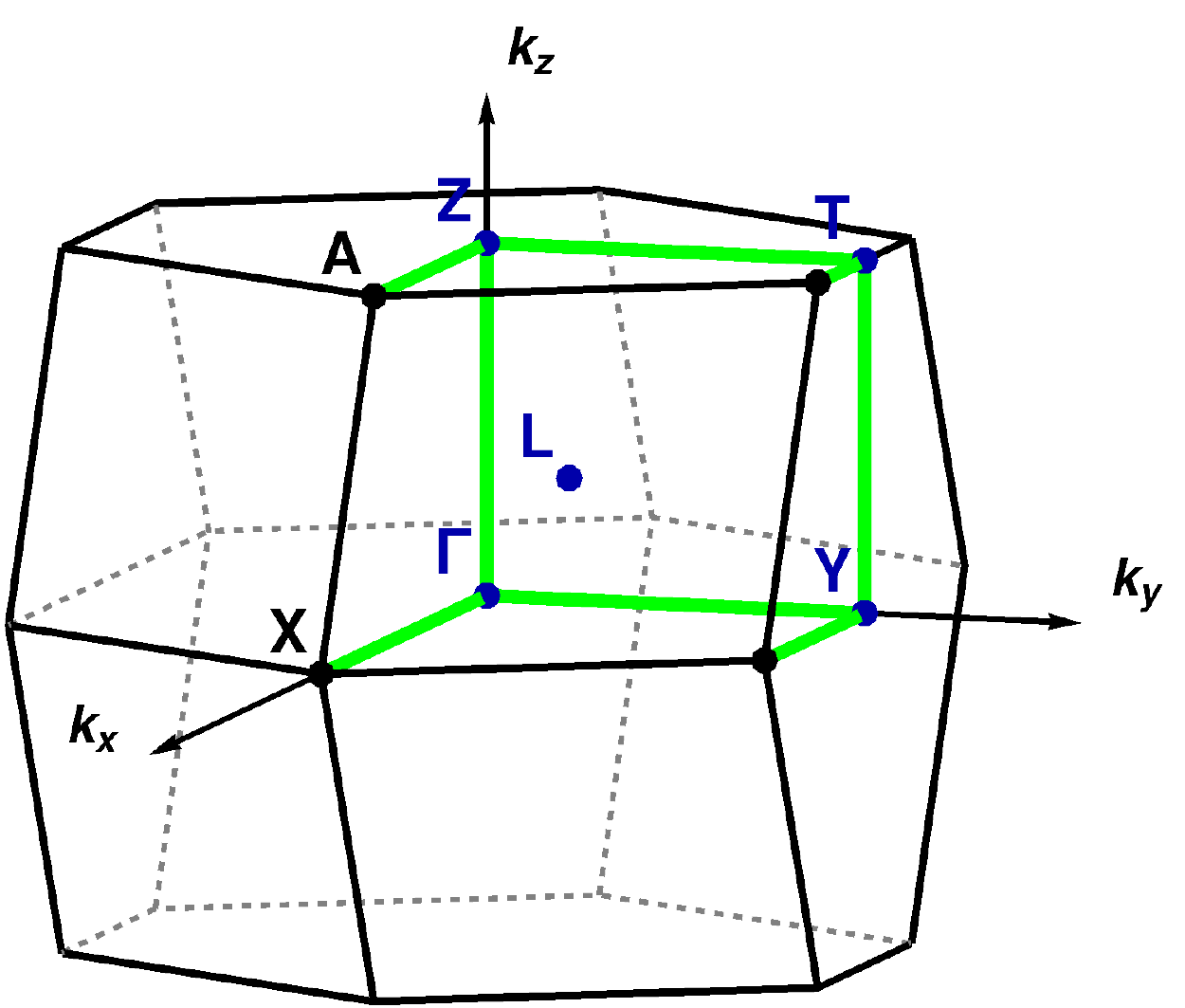}
\label{BZ_face}
}
\subfloat[ I: body-centered ]{
\includegraphics[width=.24\linewidth,valign=t]{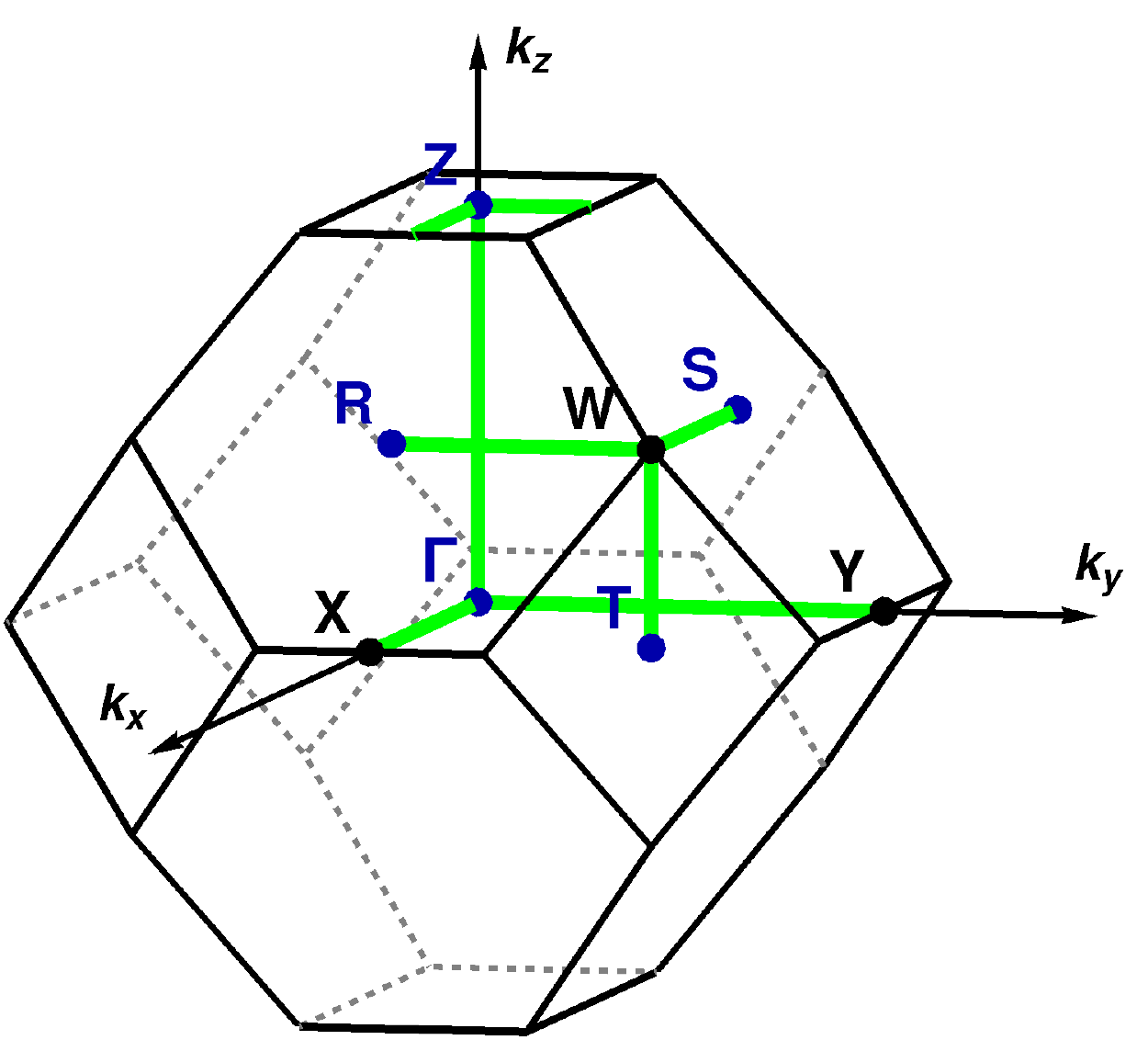}
\label{BZ_body}
}
\caption{
Brillouin zones (BZs) for lattices in the orthorhombic
SGs.
TRIMs are labeled in blue, rotation
axes are shown in green and labeled by the points they are
connecting.
}
\label{BZ_fig}
\end{figure*}

There are four Bravais lattices compatible with the above
combinations of symmetries, primitive, side- or base- centered,
face-centered and body-centered lattices. The conventional cells
of all orthorhombic lattices have right angles and three independent lengths,
$a,b$ and $c$. In the following, we will omit lengths by expressing
points in terms of the reciprocal lattice vectors of the
conventional lattice in the crystallographic definition, i.e.
excluding the factor $2\pi$.

Note that in the orthorhombic crystal system there is no unique primary symmetry
axis and there are various conventions in choosing the primary,
secondary and tertiary symmetry directions. Since our analysis is
based on symmetries alone, we align the symmetries according to
their position in the Hermann-Mauguin symbols in the conventional
setting~\cite{Hahn2002} with the directions a, b and c. See the
Hermann-Mauguin symbols in the first column of
Tables~\ref{SGs_noinv_nosoc} to \ref{SGs_inv_soc} for the
assignment of symmetries to the crystal axes. This convention is
also used by the Bilbao Crystallographic
server~\cite{Elcoro_double_cryst}. In high-throughput
calculations, the axes are often ordered such that $a<b<c$
regardless of symmetry~\cite{setyawan2010high} and labels need to
be interchanged accordingly to compare the results.

%%%%%%%%%%%%%%%%%%%%%%%%%%%%%%%%%%%%%%%%%%%%%%%%%%%%%%%%%%%%%%%%%%
There are always 8 TRIMs in the Brillouin zone (BZ). 
% primitive
In the BZ of primitive lattices these are also the points of
maximal symmetry, i.e., their little group is identical to the SG
and they are invariant under all symmetry operations.
Figure~\ref{BZ_fig}\subref{BZ_prim} shows the BZ of primitive
lattices, where all possible axes invariant under rotations are shown
in green. An axis will be denoted by the two labeled points on it, e.g.,
$\Gamma$-X for an axis parametrized by $(u,0,0)$. Contractions are
possible, i.e., the two axes $\Gamma$-Z and Z-R are compactly
written as $\Gamma$-Z-R. 

% base and side centered
In base- [side-] centered lattices, indicated by the lattice symbol C
[A], there is an additional lattice site at the center of the (001)
[(100)] face. Only four TRIMs have maximal symmetry,
$\Gamma$, Z, Y and T. For the remaining TRIMs S and R there are two
non-equivalent copies only invariant under symmetry of the third
(first) symmetry axis and the remaining symmetries map the two
copies onto each other, see Fig.~\ref{BZ_fig}\subref{BZ_base}. There are now
two rotation axes connecting $\Gamma$ and Y, and we introduce the
additional labels X and A to distinguish the axis
$\Gamma$-Y=$(0,v,0)$ from $\Gamma$-X=$(u,0,0)$ and equivalently
Z-T=$(0,v,\frac{\pi}{2})$ and Z-A=$(u,0,\frac{\pi}{2})$.
In the case $a>b$, the first BZ extends to Y in $[100]$ direction and 
the additional points appear along $[010]$. Our results, however, hold
always in relation to the orientation of symmetries.
Therefore, the points X and Y, as well as A and T should always be
read as shown in Fig.~\ref{BZ_fig}\subref{BZ_base} and parametrized above.
The BZ for side-centered lattices is only relevant for SGs with
crystallographic point group $mm2$. The rotation axis is always
chosen to be along the [001] direction, which requires the use of
the site-centered lattice for SGs 38-41. As convention for this
paper, we rotate the BZ in Fig.~\ref{BZ_fig}\subref{BZ_base}
around the $k_y$ axis. This way, $\Gamma-X$ and $Z-A$ always
correspond to the rotation axis ($0,0,k_z)$ and
$(\tfrac{1}{2},0,k_z)$, whereas the lines $\Gamma-Y$ and $Z-T$ are
to be understood as the lines perpendicular to it.

% face centered
In face-centered lattices we also find four TRIMs with maximal
symmetry, again $\Gamma$, Z, Y and T. The remaining four TRIMs are
symmetry related copies of L and have only the identity in their
site symmetry group. Depending on the relative length of the
lattice constants $a,b$ and $c$, there are several distinct shapes
the BZ can take, of which we show one example in
Fig.~\ref{BZ_fig}\subref{BZ_face}.
Again we introduce additional labels to allow for a unique
declaration of directions, $\Gamma$-X=$(u,0,0)$ connecting
$\Gamma$ and T, Y-T=$(0,2\pi,w)\sim(2\pi,0,w)$ and Z-A=$(u,0,2\pi)$
connecting Z with Y. The results presented in this paper do not
depend on the actual shape of the BZ, as long as the correct
orientation is chosen and high-symmetry lines should always be
understood as defined above and visualized in the exemplary
BZ.

%body centered
The last lattice type found for the orthorhombic SGs has a
body-centered conventional cell. Only $\Gamma$ and
Z=$(0,0,2\pi)\sim(2\pi,0,0)\sim(0,2\pi,0)$ have maximal symmetry. The
other TRIMs S, R and T have the symmetry of the first, second and
third symmetry axis, respectively, and exist in two symmetry
related copies. Additionally, there is the high-symmetry point W,
which is invariant under all SG symmetries but not under $\mathcal{T}$, see
Fig.~\ref{BZ_fig}\subref{BZ_body}.

In all cases we specify high-symmetry planes through
parametrization in Cartesian coordinates, e.g., $k_z=\pi$.

%%%%%%%%%%%%%%%%%%%%%%%%%%%%%%%%%%%%%%%%%%%%%%%%%%%%%%%%%%%%%%%%%%

\section{Screening for candidate materials}
\label{sec_III}

\begin{figure}
\includegraphics[width=.98\linewidth]{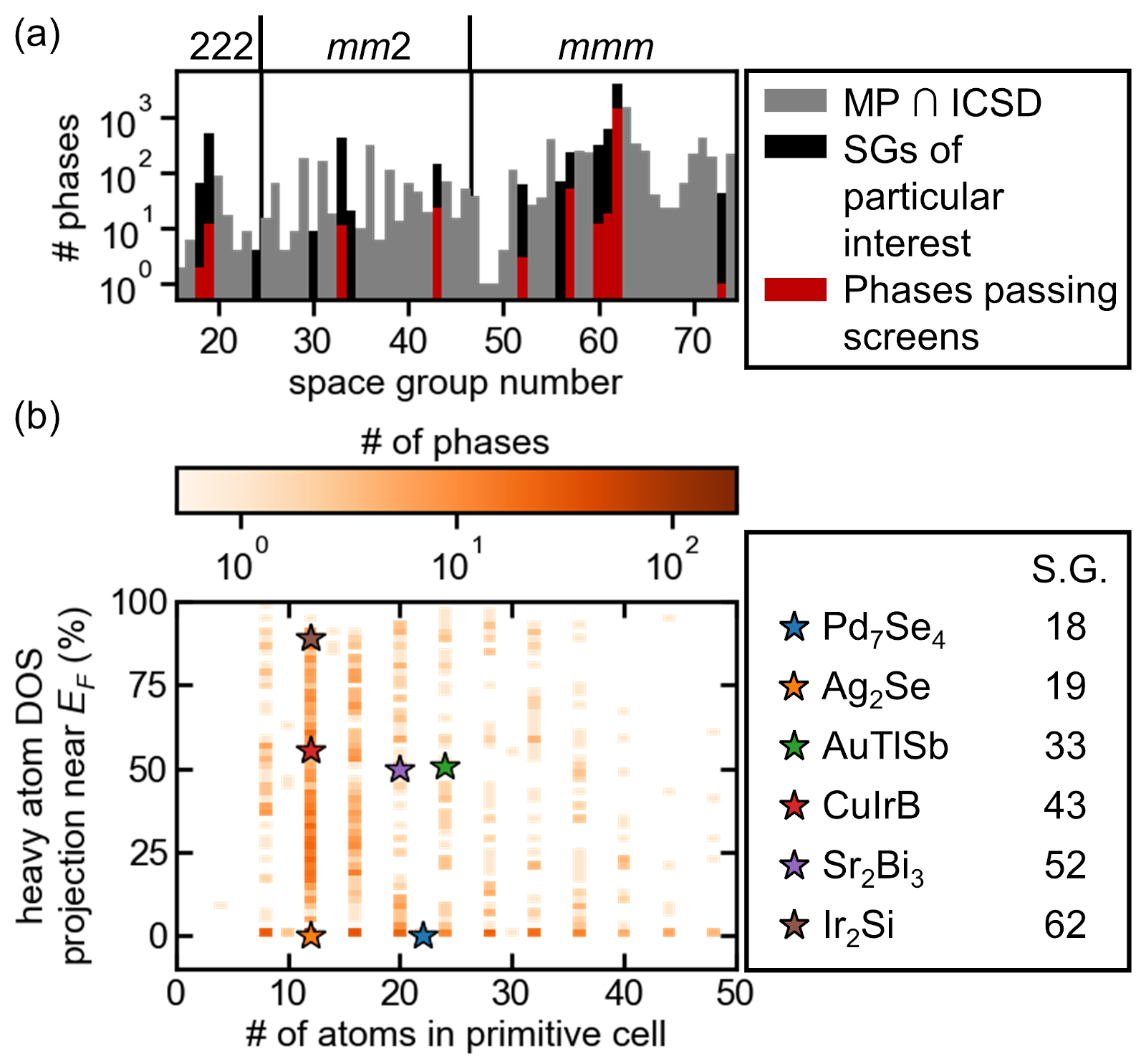}
\caption{
(a) Distribution of ordered, orthorhombic phases in the Materials
Project and ICSD by SG number (MP $\cap$ ICSD, $N$ =
12,292). 14 SGs of particular interest are indicated, as
well as the number of phases in these SGs passing the
automated screening criteria described in the text. 
The crystallographic point groups are indicated
above the plot.
(b) Distribution of phases passing the automated screening
criteria by fraction of the projected DOS near the Fermi energy
contributed by heavy atoms (see text and
supplement~\cite{screening_results}) 
and number of atoms in the
primitive unit cell ($N$ = 1259). The specific example materials
discussed in the text are highlighted.
}
\label{fig_heatmap}
\end{figure}

For 14 of the orthorhombic SGs (SGs 18, 19, 24,
30, 33, 34, 43, 52, 56, 57, 60, 61 and 62) that exhibit some of
the most notable symmetry-enforced features listed in
Tables~\ref{SGs_noinv_nosoc}, \ref{SGs_noinv_soc},
\ref{SGs_inv_nosoc}, and \ref{SGs_inv_soc}, we perform a
computationally-assisted screening of the Materials Project (MP)
database~\cite{materials_project_2013, Ong2015} for real material examples.
We restrict our search to phases which have corresponding entries
in the Inorganic Crystal Structure Database (ICSD, FIZ Karlsruhe)
-- essentially, the space of known, ordered, inorganic crystals.
The distribution of these phases by SG is depicted in
Figure~\ref{fig_heatmap}(a), with incidence varying by 3 orders of
magnitude.

We apply several preliminary screens on thermodynamic stability
(formation energy $\leq$50 meV atom$^{-1}$ off the convex hull),
structural complexity ($\leq$50 atoms in the primitive unit cell,
as larger unit cells tend to have more crowded and flatter bands),
and electronic band gap ($\leq$0.1 eV by the generalized gradient
approximation). We note that many insulating crystals have the
desired SG symmetries, but their more localized bands
tend to diminish the importance of the symmetry-enforced
crossings, for instance by making crossings poorly isolated in
momentum and energy from other states. In addition, many
insulating crystals may be difficult to degenerately dope to
position the Fermi energy near crossings of interest. 
As a result, we focus here on metals, semimetals, and nearly
gapless semiconductors. Together, these preliminary screens narrow
the search to \mytilde 1540 phases across these 14 SGs (ranging
from zero phases for SGs 24, 30, 34, and 56 to \mytilde 1400 phases for SG
62). To isolate cases of both strong and weak SOC near the Fermi
level, we use the electronic densities of states (DOS) computed
without SOC in the MP database (present for \mytilde 1260 of the remaining
phases, or 82\%) to assess what fraction of the orbital character
is due to heavy atoms. Specifically, we define a heavy atom DOS
fraction near $E_F$ as the integral of the DOS projection onto
orbitals from atoms indium or heavier ($Z\geq$49) divided by the
integral of the DOS projection onto all orbitals, evaluated on the
energy window [$E_F-1$ eV, $E_F+1$ eV]. Typical of plane-wave
codes, the total projections onto these atom-centered spherical
harmonics aren't quantitative due to the arbitrary definition of
the atomic radii, but this ratio is nonetheless indicative of the
contribution from heavy atoms. The phases passing the screening are given in the
supplement~\cite{screening_results}, 
the distribution of these
phases by heavy atom DOS projection near $E_F$ and number of atoms
in the primitive cell is given in Figure~\ref{fig_heatmap}(b), and
several example materials are annotated which will be discussed
subsequently. As a last step to identify high-quality material
examples, we examine the experimental literature to assess
synthesizability, evaluate whether the computed GGA band structure
is consistent with experimental observations (e.g. color and
luster, resistivity, magnetization) and confirm that the crystal
structure model in the desired SG has not been superceded
by one of a different symmetry in subsequent investigations.

For these example materials, we performed electronic structure
calculations using the VASP code~\cite{Kresse1996-PRB, Kresse1996-CMS}, 
which implements the Kohn$-$Sham
formulation of density functional theory (DFT) using a plane wave
basis set and the projector augmented wave formalism~\cite{Bloechl1994, Kresse1999}.
All calculations were performed with the PBE exchange$-$correlation
functional~\cite{Perdew1996}, and optimized
structures from the MP database were used directly.
The band structures calculated along high-symmetry paths are presented
in Appendix~\ref{app_A}.
Surface states were calculated via a Wannier model using
Wannier90~\cite{Mostofi2014} and WannierTools~\cite{WU2017}.

\rowcolors{1}{spinless_green1}{spinless_green2}
\setlength{\arrayrulewidth}{1.0pt}
\setlength{\LTcapwidth}{.9\textwidth}
\begin{longtable*}{ | r@{ }l || >{\raggedright}p{3.0cm} | >{\raggedright}p{3.5cm} | c | p{6.0cm} | }
\hline
\multicolumn{2}{|c||}{SG} &  \multicolumn{1}{c|}{points} &  \multicolumn{1}{c|}{lines} & nodal planes & \multicolumn{1}{c|}{notable features} 
\\
\hline
\endhead
%%% SPINLESS %%%%%%%%%%%%%%%%%%%%%%%%%%%%%%%%%%%%%%%%%%%%%%%%%%%%%
16 & ( P$222$ )         &   &   &               &   \\
17 & ( P$222_1$ )       &   &   & $k_z=\pi$     &   \\
18 & ( P$2_1 2_1 2$ )   &   &   & $k_{x/y}=\pi$ &   \\
19 & ( P$2_1 2_1 2_1$ ) & $\Gamma$-[X$\veebar$Y$\veebar$Z](4), R(4) & & $k_{x/y/z}=\pi$ & fourfold double Weyl point  \\
20 & ( C$222_1$ )       &   &   & $k_z=\pi$     &   \\
21 & ( C$222$ )         &   &   &               &   \\
22 & ( F$222$ )         &   &   &               &   \\
23 & ( I$222$ )         &   &   &               &   \\
24 & ( I$2_1 2_1 2_1$ ) & $\Gamma$-[X$\veebar$Y$\veebar$Z](2), W(2) &  &  & only 4 Weyl points \\
\hline
\hline
%%% SPINLESS %%%%%%%%%%%%%%%%%%%%%%%%%%%%%%%%%%%%%%%%%%%%%%%%%%%%%
25 & ( P$mm2$ )     &   &                       &           & \\
26 & ( P$mc2_1$ )   &   &                       & $k_z=\pi$ & \\
27 & ( P$cc2$ )     &   & Z-U-R-T-Z             &           & \\
28 & ( P$ma2$ )     &   & X-U-R-S-X             &           & \\
29 & ( P$ca2_1$ )   &   & ($U$-X-S-$R$;$U$-$R$) & $k_z=\pi$ & \\
30 & ( P$nc2$ )     &   & Z-U-R-S-Y-T-Z         &           & \\
31 & ( P$mn2_1$ )   &   & U-X-S-R               & $k_z=\pi$ & \\
32 & ( P$ba2$ )     &   & X-U-R-T-Y-S-X         &           & \\
33 & ( P$na2_1$ )   &   & (S-X-$U$;$U$-R), S-Y-T& $k_z=\pi$ & \\
34 & ( P$nn2$ )     &   & X-S-Y-T-Z-U-X         &           & \\
35 & ( C$mm2$ )     &   &                       &           & \\
36 & ( C$mc2_1$ )   &   &                       & $k_z=\pi$ & \\
37 & ( C$cc2$ )     &   & U-Z-T                 &           & \\
38 & ( A$mm2$ )     &   &                       &           & \\
39 & ( A$em2$ )     &   & R-S                   &           & \\
40 & ( A$ma2$ )     &   & A-Z-T                 &           & \\
41 & ( A$ea2$ )     &   & R-S, A-Z-T            &           & \\
42 & ( F$mm2$ )     &   &                       &           & \\
43 & ( F$dd2$ )     &   & A-Z-T-Y               &           & \\
44 & ( I$mm2$ )     &   &                       &           & \\
45 & ( I$ba2$ )     &   & R-W-S                 &           & \\
46 & ( I$ma2$ )     &   & R-W                   &           & \\
\hline

\caption{
\cellcolor{white}
Symmetry-enforced band crossings in \emph{spinless} band
structures of \emph{non-centrosymmetric} orthorhombic SGs. The
label of high-symmetry points and axes corresponds to the BZs
shown in Fig.~\ref{BZ_fig}.
The second column shows all point-like degeneracies. They can be
on a high-symmetry point (e.g. $\Gamma$), somewhere on an axis
connecting two high-symmetry points (e.g. $\Gamma$-Z) or on one of
several axes (e.g. $\Gamma$-[X$\veebar$Y$\veebar$Z]) with
$\veebar$ indicating the exclusive OR. 
Numbers in brackets indicate the number of bands involved. The
topological type of the degeneracies are listed under notable
features if it is different from a Weyl point.
The third column lists all symmetry-enforced (unpinned) nodal
lines. The notation $(A;B)$ indicates an hourglass nodal line
between two point- or line-like degeneracies $A$ with eigenvalue
pairing $(+,-)$ and $B$ with $(+,+)$ or $(-,-)$.
The fourth column lists all twofold degenerate nodal planes defined
through Cartesian coordinates as shown in Fig.~\ref{BZ_fig}.
}
\label{SGs_noinv_nosoc}
\end{longtable*}

\rowcolors{1}{spinfull_blue1}{spinfull_blue2}
\begin{longtable*}{ | r@{ }l || >{\raggedright}p{3.0cm} | >{\raggedright}p{3.5cm} | c | p{6.0cm} | }
\hline
\multicolumn{2}{|c||}{SG} &  \multicolumn{1}{c|}{points} &  \multicolumn{1}{c|}{lines} & nodal planes & \multicolumn{1}{c|}{notable features} 
\\
\hline
\endhead
%%% SPINFULL %%%%%%%%%%%%%%%%%%%%%%%%%%%%%%%%%%%%%%%%%%%%%%%%%%%%%
16 & ( P$222$ )         & all TRIMs                                                 &   &               &                       \\
17 & ( P$222_1$ )       & $\Gamma$, X, Y, S, $\Gamma$-Z(4)                          &   & $k_z=\pi$     &  only 4 Weyl points   \\
18 & ( P$2_1 2_1 2$ )   & $\Gamma$, Z, $\Gamma$-X(4), $\Gamma$-Y(4), S(4), R(4)     &   & $k_{x/y}=\pi$ & fourfold double Weyl points ($n$ even), only 2 Weyl points ($n$ odd)  \\
19 & ( P$2_1 2_1 2_1$ ) & $\Gamma$, $\Gamma$-Z(4), \mbox{$\Gamma$-X(4)}, \mbox{$\Gamma$-Y(4)}, S(4), T(4), U(4), \mbox{R-[S$\veebar$T$\veebar$U](8)} &   &$k_{x/y/z}=\pi$& top. nodal plane trio, fourfold double Weyl points  \\
20 & ( C$222_1$ )       & $\Gamma$, S, Y, $\Gamma$-Z(4)                             &   & $k_z=\pi$     & only 4 Weyl points    \\
21 & ( C$222$ )         & all TRIMs                                                 &   &               &                       \\
22 & ( F$222$ )         & all TRIMs                                                 &   &               &                       \\
23 & ( I$222$ )         & all TRIMs, W                                              &   &               &                       \\
24 & ( I$2_1 2_1 2_1$ ) & all TRIMs, \mbox{W-[R$\veebar$S$\veebar$T]}(4)        &   &               &                       \\
\hline
\hline
%%% SPINFULL %%%%%%%%%%%%%%%%%%%%%%%%%%%%%%%%%%%%%%%%%%%%%%%%%%%%%
25 & ( P$mm2$ )         &                       & $\Gamma$-Z, X-U, Y-T, S-R                                                             &           &                                                   \\
26 & ( P$mc2_1$ )       &                       & ($\Gamma$-$Z$,Y-$T$;$Z$-$T$), \mbox{(X-$U$,$R$-S;$U$-$R$)}                            & $k_z=\pi$ &                                                   \\
27 & ( P$cc2$ )         & Z(4), T(4), U(4), R(4)&  $\Gamma$-Z-U-R-T-Z, X-U, S-R, Y-T                                                    &           & fourfold points ($\mathcal{C}{=}0$)               \\
28 & ( P$ma2$ )         &                       & ($\Gamma$-Z;X,U), \mbox{(Y-T;S,R)}, X-S, U-R                                          &           &                                                   \\
29 & ( P$ca2_1$ )       &                       & ($\Gamma$-$Z$;$Z$-U,X), \mbox{(Y-$T$;S,R-$T$),} \mbox{(X-S;U-R)}                      & $k_z=\pi$ &                                                   \\
30 & ( P$nc2$ )         & Z(4), U(4)            & ($\Gamma$-$Z$-T;Y), \mbox{(X-$U$-R;S)}, \mbox{(Y-S;T,R)}, Z-U                         &           & nodal chain, fourfold points ($\mathcal{C}{=}0$)  \\
31 & ( P$mn2_1$ )       &                       &\mbox{($\Gamma$-$Z$,$T$-Y;$Z$-$T$)}, \mbox{($\Gamma$-$Z$-U;X)}, \mbox{(Y-$T$-R;S)}, X-S& $k_z=\pi$ &                                                   \\
32 & ( P$ba2$ )         & R(4), S(4)            & ($\Gamma$-Z;Y,T), \mbox{($\Gamma$-Z;X,T)}, X-S-Y, U-R-T, S-R                          &           & fourfold points ($\mathcal{C}{=}0$)               \\
33 & ( P$na2_1$ )       & S(4)                  & ($R$-S-X;$R$-U) \mbox{($\Gamma$-$Z$;$Z$-U,X)}, \mbox{($\Gamma$-$Z$-T;Y)}, Y-S         & $k_z=\pi$ & fourfold point ($\mathcal{C}{=}0$)                \\
34 & ( P$nn2$ )         & Z(4), S(4)            & ($\Gamma$-$Z$-T;Y), \mbox{(Y-$S$-R;T)}, \mbox{($\Gamma$-$Z$-U;X)}, \mbox{(X-S-R;U)}   &           & nodal chain, fourfold points ($\mathcal{C}{=}0$)  \\
35 & ( C$mm2$ )         & R, S                  &  $\Gamma$-Z, Y-T                                                                      &           & only 4 Weyl points                                \\
36 & ( C$mc2_1$ )       & S, R-S(4),            & ($\Gamma$-$Z$,Y-$T$;$Z$-$T$)                                                          & $k_z=\pi$ & only 2 (4) Weyl points for $n$ odd (even)                  \\
37 & ( C$cc2$ )         & R, S, Z(4), T(4)      &  $\Gamma$-Z-T-Y, Z-U                                                                  &           & fourfold points ($\mathcal{C}{=}0$), only 4 Weyl points   \\
38 & ( A$mm2$ )         &                       &  (R;$-$), (S;$-$), $\Gamma$-X, Z-A                                                    &           &                                                   \\
39 & ( A$em2$ )         &                       & ($\Gamma$-X;S), \mbox{(Z-A;R)}, R-S                                                   &           &                                                   \\ 
40 & ( A$ma2$ )         &                       & ($\Gamma$-X;Z,T), (R;$-$), (S;$-$), Z-T                                               &           &                                                   \\
41 & ( A$ea2$ )         &                       & ($\Gamma$-X;Z,T), \mbox{($\Gamma$-X;S)}, \mbox{(Z-T;R)}, R-S                          &           &                                                   \\
42 & ( F$mm2$ )         & L                     & $\Gamma$-Z, Y-T                                                                       &           & only 4 Weyl points                                \\
43 & ( F$dd2$ )         & L, Z(4)               & ($\Gamma$-$Z$-T;Y)$_{k_x=0}$, \mbox{($\Gamma$-$Z$-A;T)$_{k_z=0}$}                     &           & nodal chain, fourfold point, only 4 Weyl points   \\
44 & ( I$mm2$ )         & T                     & (R;$-$), (S;$-$), $\Gamma$-Z                                                          &           & only 2 Weyl points    \\
45 & ( I$ba2$ )         & T, T-W(4)             & ($\Gamma$-Z;S,R), R-W-S                                                               &           & only 2 (4) Weyl points for $n$ odd (even)         \\
46 & ( I$ma2$ )         & T                     & ($\Gamma$-Z;R), (S;$-$), R-W                                                          &           & only 2 Weyl points                                \\
\hline
\caption{
\cellcolor{white}
Symmetry-enforced band crossings in band structures \emph{with SOC} of
\emph{non-centrosymmetric} orthorhombic SGs.
The notation is identical to the one in Tab.~\ref{SGs_noinv_nosoc}. 
Additionally, almost movable lines are indicated by ($A$;$-$)
with $A$ being the high-symmetry point they run through as
described in Sec.~\ref{Sec_almost_movable_lines}.
}
\label{SGs_noinv_soc}
\end{longtable*}
\rowcolors{1}{white}{white}

\section{Rhombic disphenoidal: SG 16 - SG 24}
\label{sec_IV}

The chiral SGs in the orthorhombic crystal system, SGs 16 to 24, have
the crystallographic point group $D_2$, or $222$ in the
Hermann-Mauguin notation, consisting of three perpendicular
twofold rotations. Since neither mirror planes nor inversion are
present, these SGs are chiral.
 
Symmetry-enforced features are movable Weyl points, nodal planes
and fourfold degenerate points with a topological charge of
$|\mathcal{C}|=2$, called fourfold double Weyl points~\cite{hirschmann2021symmetryenforced}.

\subsection{Weyl points at high-symmetry points}
\label{Sec_Kramers}

Time-reversal symmetry $\mathcal{T}$ squares to a $2\pi$-rotation
and therefore $-1$ in spinful representations. Kramers theorem
implies twofold degeneracies at all TRIMs in spinful
band structures. Without further restrictions from additional
symmetries, these degeneracies are Weyl points with a chirality
$\mathcal{C}=\pm1$~\cite{chang2018topological,tsirkin_vanderbilt_PRB_17}.
In the chiral orthorhombic SGs 18-24 with strong SOC, every
Kramers degeneracy at a TRIM that is not part of a nodal plane
(see Sec.~\ref{sec_nodal_planes} below) is such a Kramers-Weyl point.
They are listed in the column ``points'' of
Tab.~\ref{SGs_noinv_soc}.

%%%%%%%%%%%%%%%%%%%%%%%%%%%%%%%%%%%%%%%%%%%%%%%%%%%%%%%%%%%%%%%%%%
% spinless in SG 24 at W
% is not a Kramers Weyl, but generated by off-centered rotations
In spinless representations, $\mathcal{T}$ squares to the identity
and Kramers theorem does not apply. Consequently, the bands at TRIMs are in
general not degenerate. There is however one case of a symmetry
enforced Weyl point at a high-symmetry point in SG 24. The
symmetries in this body-centered SG are three orthogonal,
off-centered rotations. While twofold rotations without
translational components always commute, the resulting translation
in the combination of two off-centered rotations depends on their
order. For the generators of SG 24 we find for example
\begin{multline}
2_{100}\!\left(0,0,\tfrac{1}{2}\right)2_{001}\!\left(0,\tfrac{1}{2},0\right) \\
    = (-1)^\zeta t\!\left(0,-1,0\right) 2_{001}\!\left(0,\tfrac{1}{2},0\right)2_{100}\!\left(0,0,\tfrac{1}{2}\right).
\end{multline}
There are three points in the Brillouin zone with maximal
symmetry, of which only W has translation eigenvalue $-1$ for 
the above translation by one lattice constant
in the conventional cell. Therefore, the spinless representations% ($\zeta=0$) 
of the rotations $U_{2_{001}}$ and $U_{2_{100}}$ anticommute and
must be at least two-dimensional. For an eigenstate $\ket{\pm}$ of
$U_{2_{001}}$ with eigenvalue $\pm1$ we find
\begin{IEEEeqnarray}{rCr}
U_{2_{001}} U_{2_{100}}\ket{\pm} &=& -U_{2_{100}}U_{2_{001}}\ket{\pm} \hphantom{,}
\nonumber \\
    &=& \mp U_{2_{100}}\ket{\pm},
\label{comm_rel_24}
\end{IEEEeqnarray}
which implies that for example $\ket{+}$ and $U_{2_{100}}\ket{+} \propto \ket{-}$ are two distinct states.
Because the Hamiltonian commutes with $U_{2_{001}}$ and $U_{2_{100}}$, they will be degenerate.
In comparison, SG 23 has no translational parts in its rotations
and the translation eigenvalue of $-1$ is missing in the
commutation relation. Instead, Eq.~\eqref{comm_rel_24} holds for
spinful representations.
In both cases states with opposite rotation eigenvalues are paired
at W and the point degeneracy carries a topological charge of
$\mathcal{C}=\pm1$~\cite{tsirkin_vanderbilt_PRB_17}. Notably, this Weyl
point is enforced even in the absence of time-reversal symmetry.

If the spin degrees of freedom in SU(2)-symmetric band structures
in SG 24 are considered, the topological charges of the Weyl
points at W in each spin space are identical and the total
topological charge is $\mathcal{C}_{\uparrow} +
\mathcal{C}_{\downarrow} =\pm2$.  Including SOC requires the use
of spinful representations, where the above relation does not hold,
due to an additional sign change from the spin part of the
symmetries in the commutation relation. As we will show in the
next chapter, there is a weaker requirement in that case for two
separate Weyl points on a rotation axis connecting to W. As long
as SOC is small, these two Weyl points remain close to W,
and the Chern number calculated on a surface enclosing the two
Weyl points is not changed by SOC.

\subsection{Movable Weyl points on rotation axes}
\label{Sec_movable_Weyl}

A movable Weyl point is not fixed to a high-symmetry point. Such Weyl points 
are enforced when compatibility relations between representations
at high-symmetry points and the axis connecting them requires an
exchange of bands with different symmetry eigenvalues. The
resulting band crossing is therefore symmetry protected, but its
position on the axis is not fixed, hence it is called movable.

\begin{figure}
\sidesubfloat[]{
\includegraphics[width=.90\linewidth]{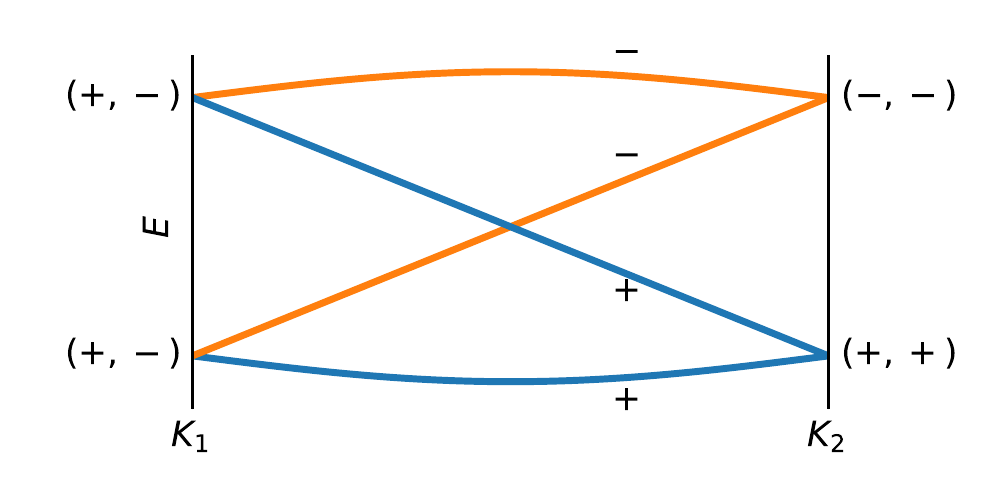}
\label{hourglass_sketch}
}

\sidesubfloat[]{
\includegraphics[width=.90\linewidth]{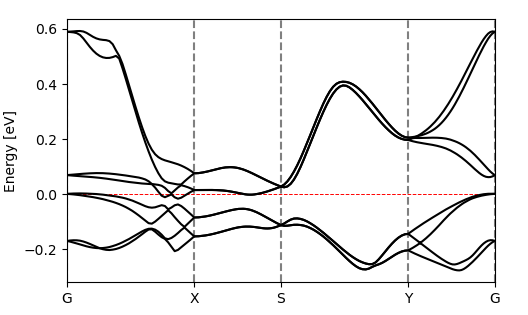}
\label{pd7se4_hourglass}
}
\caption{
(a)
Scheme of an hourglass dispersion along a path between two degenerate points $K_1$ and $K_2$
along a twofold rotation axis or in a mirror plane. Colors
indicate the sign of the symmetry eigenvalue of the twofold symmetry.
(b)
Band structure for \ce{Pd7Se4} in SG 18 with SOC
showing hourglass states on $\Gamma$-X (less clearly at
$\Gamma$-Y) and fourfold double Weyl points at S.
}
\label{fig_hourglass}
\end{figure}

Twofold screw rotations square to lattice translations in
the direction of the rotation axis. Hence, their eigenvalues are
the square roots of translation eigenvalues and therefore
$k$-dependent. For example, $2_{001}(a,b,c)$
has eigenvalues
\begin{IEEEeqnarray}{rCl}
U_{2_{001}}\ket{\pm} &=&
    \pm \mathrm{i}^\zeta \exp(\mathrm{i}k_z c)\ket{\pm},
\label{rot_eigvals}
\end{IEEEeqnarray}
where $c=\tfrac{1}{2}$ for a screw rotation.
Along the rotation axis the eigenvalue of each
band evolves smoothly with $k_z$, and 
a state can be labeled uniquely by 
the sign $\pm$ according to this definition.
Throughout the paper, we will use blue and orange in plots
for states with positive and negative signs, respectively.
Kramers partners at TRIMs have complex conjugate eigenvalues. For
a TRIM on the axis invariant under the rotation with $k_z=0$, the
eigenvalues $\pm\mathrm{i}$ are paired. We label such a pair in
terms of the signs $(+,-)$. At TRIMs with $k_z=\pi$ the
eigenvalues are $\pm(-1)$ and identical eigenvalues are paired,
$(+,+)$ or $(-,-)$. Connecting these pairs smoothly along the
rotation axis requires four states in total and at least one band
crossing in between the two TRIMs, leading to a so-called hourglass
dispersion~\cite{wang2016hourglass}, see Fig.~\ref{fig_hourglass}\subref{hourglass_sketch}
for an illustration. Without further restrictions from other
symmetries in the SG, such a band crossing is a Weyl point.  The
rotation axes on which these movable Weyl points occur are listed
in the column ``points'' in Tab.~\ref{SGs_noinv_soc} by specifying
the rotation axis with the number of bands involved given in
brackets. Movable Weyl points enforced by
this mechanism are found in spinful band structures of the chiral
SGs 17-20 on all screw axes.

% extended compatibility relations %%%%%%%%%%%%%%%%%%%%%%%%%%%%%%%
In spinless systems $\mathcal{T}$ squares to the identity and
there are in general no Kramers degeneracies at TRIMs and no
symmetry-enforced hourglass states.
In the case of three screw rotations in SG 19 and 24, however, a
movable crossing is still enforced by an extended compatibility
relation, meaning the combined compatibility relations on all
three rotation axes together with the group structure at their
intersections~\cite{furusaki_2017}. 

\begin{figure}
\sidesubfloat[]{
\includegraphics[width=.40\linewidth,valign=t]{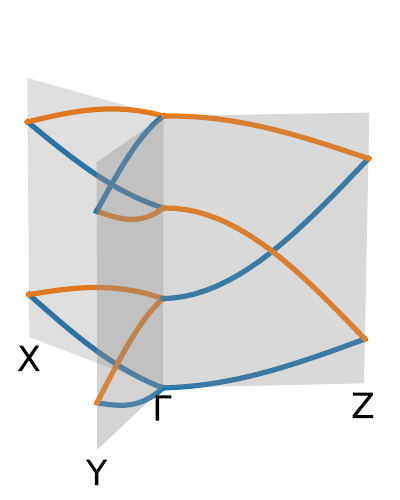}
\label{comprel_19}
}
\sidesubfloat[]{
\label{comprel_24}
\includegraphics[width=.40\linewidth,valign=t]{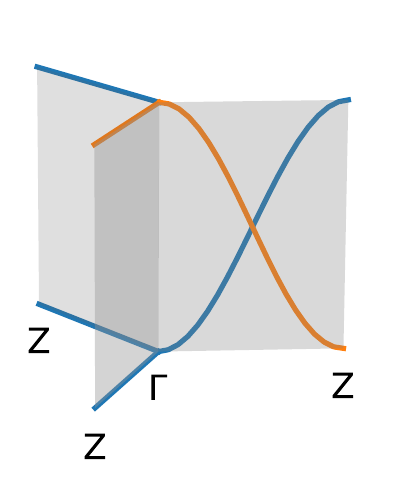}
}
\caption{
Band crossings enforced by compatibility relations comprising three
rotation axes.
(a)
Movable Weyl point in spinless band structures of SG 19 on one
of the three rotation axes connecting to $\Gamma$. Blue (orange)
colored bands correspond to bands labeled $+$ ($-$) according to
the definition in Eq.~\eqref{rot_eigvals}. 
(b)
Possible arrangement of spinless bands in SG 24 with rotation
eigenvalue $+1$ in blue and $-1$ in orange of the corresponding
axis. The product of eigenvalues is fixed to $+1$ at $\Gamma$ and
$-1$ at Z, enforcing a movable Weyl point on one of the axes.
}
\label{Fig_comprel_SG19_SG24_noSOC}
\end{figure}

Starting with SG 19, we note that spinless representations have
eigenvalues $\pm\mathrm{i}$ at the TRIMs X, Y and Z according to
Eq.~\eqref{rot_eigvals} for $2_{100}$, $2_{010}$ and $2_ {001}$,
respectively. There they are paired as $(+,-)$ by time-reversal
symmetry. At $\Gamma$, all three rotation eigenvalues are $\pm1$ and 
simultaneously good quantum numbers. 
We label a state $\ket{s_{100},s_{010},s_{001}}$ using the signs $s_i=\pm$
of all three rotations.
The product of all three rotations 
result in a translation with eigenvalue 1, and therefore the
product of rotation eigenvalues $\pm1$ must also be +1.
Consequently, only the combinations $\ket{+,+,+}$, $\ket{+,-,-}$
and permutations thereof are valid. Connecting these states
smoothly to the pairs $(+,-)$ at the other TRIM on each axis,
i.e., X, Y, or Z, requires all four possible combinations to be
present at $\Gamma$ and at least one crossing on one of the
rotation axes, see Fig.~\ref{Fig_comprel_SG19_SG24_noSOC}\subref{comprel_19}. 
Since crossings of bands with different eigenvalues are symmetry
protected, exchanging the bands can only move the crossing to
another axis, but not eliminate it, and additional crossings can
only be introduced pairwise. These movable Weyl points on one of
the three rotation axes are indicated in the column ``points'' by
\mbox{$\Gamma$-[X$\veebar$Y$\veebar$Z].} Here, the symbol $\veebar$
stands for ``exclusive or'' between the different high-symmetry
points, as only one crossing on one of the three rotation axes is
symmetry enforced. In the absence of additional accidental
crossings, which might appear on any of the rotation axes, SG~19
has only three point-like degeneracies for $4n+2$ filled bands and
the topological charge from the two symmetry related copies of the
movable point is compensated by the fourfold double Weyl point at
R. 

Next, we consider SG 24, which has a body-centered BZ where three
rotation axes connect $\Gamma$ to Z, see Fig.~\ref{BZ_fig}\subref{BZ_body}.
Both TRIMs $\Gamma$ and Z are non-degenerate in spinless
representations of SG 24. Because of the translational parts of
the rotations, the product of all three rotations results in a
lattice translation,
\begin{IEEEeqnarray}{rCl}
2_{100}(0,0,\tfrac{1}{2}) \,
2_{010}(\tfrac{1}{2},0,0) \,
2_{100}(0,\tfrac{1}{2},0) 
    &=&
        t(\tfrac{1}{2},-\tfrac{1}{2},\tfrac{1}{2}). \IEEEeqnarraynumspace
\end{IEEEeqnarray}
At $\Gamma$, the translation eigenvalue is 1 and the product of
all three rotation eigenvalues $\pm1$ has to be positive, as before. 
Note that the rotation eigenvalues of these off-centered rotations are not
$k$-dependent, cf. Eq.~\eqref{rot_eigvals} with $c=0$. At Z on the
other hand, eigenvalue of the translation is $-1$, therefore the only
valid combinations are $\ket{-,-,-}$ and $\ket{+,+,-}$ and
permutations thereof.
The two different requirements can only be fulfilled
simultaneously by two bands exchanging on at least one rotation
axis. An example for states $\ket{+,+,+}$ and $\ket{+,-,-}$ at
$\Gamma$ and $\ket{+,-,+}$ and $\ket{-,+,+}$ at Z is shown in
Fig.~\ref{Fig_comprel_SG19_SG24_noSOC}\subref{comprel_24}. While
the band representation at $\Gamma$ and Z in a real material is
induced from the orbital present at a certain Wyckoff
position~\cite{Elcoro_double_cryst} and the ordering of states in
energy is dependent on material parameters, the restriction on the
dispersion is always the same as in the example.  The charge of
this movable Weyl point is, in the absence of accidental
crossings, compensated by the pinned charge at W, which also
appears in two symmetry related copies.

The above relations leading to the movable crossing hold for spinless
representations only. However, treating weak SOC as a perturbation
to a spin-degenerate band structure will not remove the Weyl points.
As seen from a closed surface surrounding the band crossing, the
Chern number evaluated on this surface is not changed by SOC, as long as the
gap on the surface remains open. The two copies of the Weyl point
might split and leave the rotation axis, but only a large
perturbation can remove them from the band structure altogether.

% SG 24 with SOC: W-[R,T or S]
In addition, we find in spinful bands of SG 24 a movable Weyl
point on one of the three axes W-R, W-S or W-T. The compatibility
relations require the same connectivity as shown in 
Fig.~\ref{Fig_comprel_SG19_SG24_noSOC}\subref{comprel_19},
just with inverted
eigenvalues, because all three rotation eigenvalues at W
multiplied equal to $-1$.
If SOC vanishes exactly, the Weyl point in each spin subspace is
pinned to W, as has been shown above.  With increasing SOC, they
can split up and move onto an axis, where they are still related by
symmetry.

\subsubsection{Material example \ce{Pd7Se4}}
\label{mat_Pd7Se4}
An example of a movable Weyl point with hourglass dispersion is
found in \ce{Pd7Se4}, which crystallizes in SG 18~\cite{Matkovi1978}, exhibiting
metallic conductivity~\cite{Kukunuri2015}. Single crystals large
enough for diffraction were obtained by annealing the product
obtained from the melt~\cite{Matkovi1978}, but larger crystals may
be challenging as it is incongruently melting with a very small
exposed liquidus (e.g. for flux growth)~\cite{Olsen1979}.
The bands of \ce{Pd7Se4} are shown in
Fig.~\ref{fig_hourglass}\subref{pd7se4_hourglass} and exhibit, as
expected, an hourglass structure on the axes $\Gamma$-X and
$\Gamma$-Y. The full band structure along high-symmetry paths is
shown in Fig.~\ref{band_structures}. As we will discuss in
Sec.~\ref{sec_fourfold_double_weyl} the compound also hosts
fourfold double Weyl points.

\subsection{Topological nodal planes}
\label{sec_nodal_planes}

The combination of time-reversal symmetry and a twofold screw
rotation leads to twofold degeneracies on a plane.  These so
called nodal planes can contribute to the band
topology~\cite{wilde2021symmetry}. After discussing the mechanism
forming these nodal planes, we identify cases where the nodal
planes are enforced to be topologically nontrivial.

The combined symmetry of a screw rotation and time-reversal
symmetry
is antiunitary and squares to a full lattice translation
 along the rotation. The action on a point $\vb{k}$ in the BZ
is similar to a mirror symmetry with its invariant plane
orthogonal to the
rotation axis. 
For example, $2_{001}(0,0,\frac{1}{2})\mathcal{T}$ squares to 
$t(0,0,1)$ and transforms a point $\vb{k}$ according to
$(k_x,k_y,k_z)\rightarrow(k_x,k_y,-k_z)$. In the BZs of primitive
and base-centered lattices, the plane $k_z=\pi$ is equivalent to
$k_z=-\pi$ up to a reciprocal lattice translation and the
eigenvalue of the translation is
$\mathrm{e}^{\mathrm{i}k_z}=-1$.
Any antiunitary symmetry squaring to $-1$ enforces degeneracies on
invariant momenta through an extended Kramers
theorem~\cite{hirschmann2021symmetryenforced}.
Consequently, the whole plane $k_z=\pi$ is twofold degenerate.
Note that this holds equally for spinless and spinful
representations, since both time reversal and the rotation square
to a $2\pi$-rotation such that the eigenvalue of the latter
contributes no overall sign.
We list nodal planes 
in the 4th column of Tables~\ref{SGs_noinv_nosoc},
\ref{SGs_noinv_soc} and \ref{SGs_inv_nosoc}.
Notably, in SG 18 two such nodal planes exist on perpendicular
planes, called a nodal plane duo. In SG 19 there are three screw
rotations and the resulting nodal plane trio encloses the
whole BZ.

Like a point-like degeneracy, a nodal plane can act as a source or
sink of Berry curvature and carry a topological charge. In that
case we refer to it as a topological nodal plane. All nodal planes
in chiral SGs can carry a topological charge. 

For band structures with strong SOC in SG 19, this is not only
a possibility, but required by symmetry for the nodal plane trio.
In order to find the nodal charge, it
is sufficient to analyze all Weyl points that close the same gap
as the nodal plane trio. Because of the
Nielsen-Ninomiya theorem~\cite{nielsen_no_go}, their chiralities
sum up to the negative topological charge of the nodal plane trio.
In each elementary band representation there is a Kramers-Weyl
point at $\Gamma$ with a chirality of
$\pm1$~\cite{bradlyn_bernevig_Nature_2017,vergniory_PRE_17}.
Every additional Weyl point can only be at a $k$-point with a star
of order 2 or 4, because all other points of maximal symmetry
reside within the nodal planes. Because the different points of a
star are related by time-reversal symmetry or rotations, the Weyl
points have the same chirality. In total, we find 
\begin{IEEEeqnarray}{rCl}
\mathcal{C}_\text{total} &=& \pm 1 + 2n \ne 0 \quad n \in \mathbb{Z}.
\end{IEEEeqnarray}
Therefore, the topological charge of the nodal plane trio has to
be odd and $|\mathcal{C}_\text{NP}|\ge1$~\cite{CicumventingNoGo_yuxin}.

\subsection{Fourfold double Weyl points}
\label{sec_fourfold_double_weyl}

\begin{figure}
\includegraphics[width=.8\linewidth]{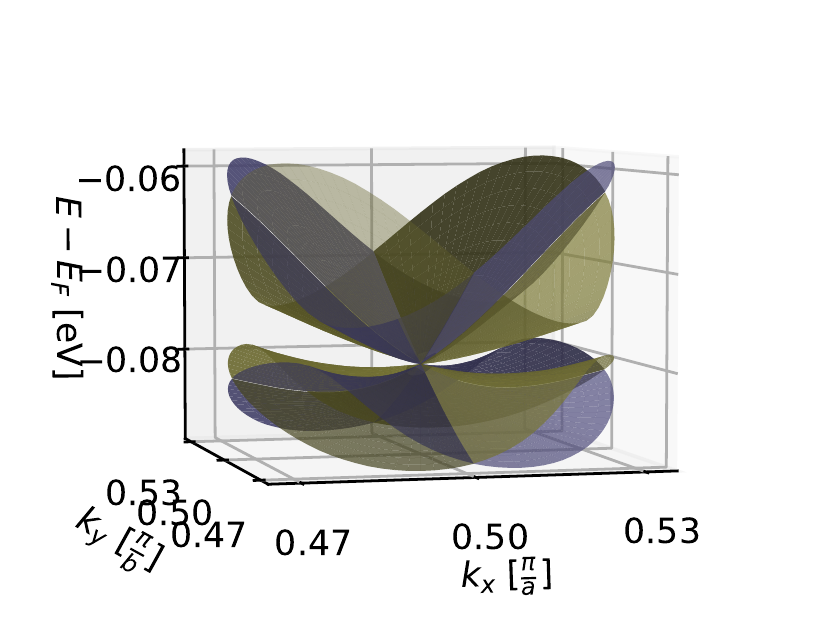}
\caption{
Dispersion of the fourfold double Weyl point in \ce{Pd7Se4} within the $k_z=0$ plane around
\mbox{S$=\pi(\tfrac{1}{2},\tfrac{1}{2},0)$}.
The colors indicate the two Weyl cones, related by the rotation
$2_{100}$ or $2_{010}$. The twofold degenerate lines along 
\mbox{$(k_x,\tfrac{\pi}{2},0)$} and \mbox{$(\tfrac{\pi}{2},k_y,0)$} are part of
nodal planes.
}
\label{4fold_weyl_fig}
\end{figure}

Fourfold double Weyl points are fourfold degenerate points with a
Chern number of
$\mathcal{C}=\pm2$~\cite{hirschmann2021symmetryenforced,PhysRevB.98.155145}.
Their topology is equivalent to two superimposed Weyl points with
equal Chern number. An example of the dispersion in the
proximity of a fourfold double Weyl point in \ce{Pd7Se4} is shown in
Fig.~\ref{4fold_weyl_fig}.

Two different mechanisms need to be at play simultaneously to
enforce such a feature. First, the representations of the three
twofold rotations at a TRIM need to anticommute and secondly,
their eigenvalues must be real, such that time-reversal symmetry
$\mathcal{T}$ relates identical eigenvalues. In orthorhombic SGs
they are found in band structures with strong SOC in SG 18 at the
TRIMs S and R, and in SG 19 at the TRIMs S, T and U.

We illustrate the mechanism for S in SG 18, where the symmetries
$\mathcal{T}$, $2_{100}(\frac{1}{2},\frac{1}{2},0)$ and
$2_{010}(\frac{1}{2},\frac{1}{2},0)$ are present at all TRIMs.
A direct calculation shows the relation
\begin{IEEEeqnarray}{l}
2_{100}\!\left(\tfrac{1}{2},\tfrac{1}{2},0\right) 
2_{010}\!\left(\tfrac{1}{2},\tfrac{1}{2},0\right)
\nonumber \\ \quad
    =
        (-1)^\zeta t\left(1,-1,0\right)
        2_{010}\!\left(\tfrac{1}{2},\tfrac{1}{2},0\right)
        2_{100}\!\left(\tfrac{1}{2},\tfrac{1}{2},0\right)
\label{comprel_rot}
\end{IEEEeqnarray}
and at the TRIMs S and R we find therefore $U_{2_{100}}U_{2_{010}}
=-U_{2_{010}}U_{2_{100}}$ in spinful representations, since there the
translation eigenvalue is $1$ for $t\left(1,-1,0\right)$. Applying $U_{2_{010}}$ to a state
with a positive $2_{100}$ eigenvalue results in a new state with
negative eigenvalue and vice versa. This requirement alone leads
to a Weyl point with chirality
$|\mathcal{C}|=1$~\cite{tsirkin_vanderbilt_PRB_17}. Because
$2_{100}$ is a screw rotation, the eigenvalues at R and S are
$\pm\mathrm{i}\mathrm{e}^{\mathrm{i}k_x/2} = \mp 1$. Kramers
partners therefore have the same symmetry eigenvalue. In terms of
$2_{100}$-eigenvalues, we therefore find the quadruple degeneracy
$(+,+,-,-)$. The chiralities of both time-reversal related copies
are identical and add up to $\pm2$.

The dispersion around a fourfold double Weyl point is linear to first
order. Because of the necessary presence of time-reversal symmetry
and screw rotation, all fourfold double Weyl points reside at
intersections of nodal planes. Within these planes, the states
remain twofold degenerate, otherwise they will split into four
non-degenerate bands, see Fig.~\ref{4fold_weyl_fig} for the
dispersion in the plane $k_z=0$ of a fourfold double Weyl point in
\ce{Pd7Se4}.

\begin{figure}
\sidesubfloat[]{
\hspace{1cm}
\includegraphics[width=.40\linewidth,valign=t]{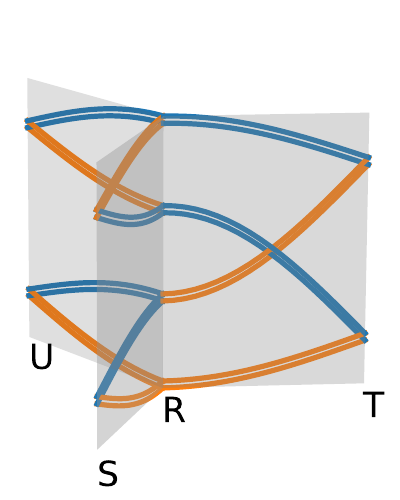}
\hspace{1cm}
\label{comprel_SG19_RSTU}
}

\sidesubfloat[]{
\includegraphics[width=.65\linewidth,valign=t]{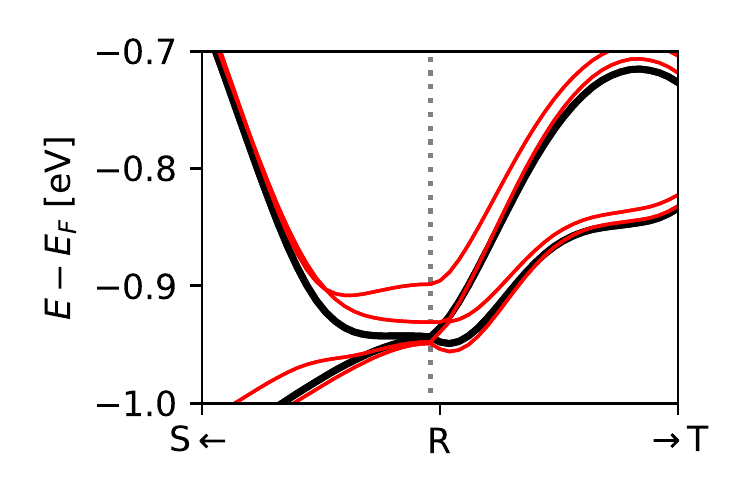}
\label{bands_ag2se_SRT}
}
\caption{
(a)
Compatibility relations at R in SG 19 for spinful bands. Colors
indicate the sign of the rotation eigenvalue according to
Eq.~\eqref{rot_eigvals}. Note that on the paths shown all bands are
twofold degenerate with identical rotation eigenvalues paired.
Figure adapted from Ref.~\cite{furusaki_2017}.
(b)
Fourfold double Weyl point in \ce{Ag2Se} without SOC at R (black).
Including SOC (red) moves the fourfold double Weyl point to the axis
R-T. In both cases bands are twofold degenerate.
}
\label{Fig_comprel_SG19_SOC}
\end{figure}

In spinless band structures there is one case of a symmetry
enforced fourfold double Weyl point at the TRIM R in SG 19. 
The principle remains the same, with the difference that the
commutation relation for the screw rotations reads
\begin{IEEEeqnarray}{l}
2_{100}\!\left(\tfrac{1}{2},\tfrac{1}{2},0\right) 
2_{010}\!\left(0,\tfrac{1}{2},\tfrac{1}{2}\right)
\nonumber \\ \quad
    =
        (-1)^\zeta t\left(1,-1,-1\right)
        2_{010}\!\left(0,\tfrac{1}{2},\tfrac{1}{2}\right)
        2_{100}\!\left(\tfrac{1}{2},\tfrac{1}{2},0\right).
\IEEEeqnarraynumspace
\label{comprel_rot_19}
\end{IEEEeqnarray}
In the case $\zeta=0$, the negative sign is provided by the
translation eigenvalue of $t(1,-1,-1)$ at the TRIM R.
Additionally, Kramers pairs are created by the combination of a
screw rotation with time-reversal symmetry, e.g.,
$2_{001}(\frac{1}{2},0,\frac{1}{2})\mathcal{T}$. Even though the
eigenvalues of $U_{2_{100}}$ and $U_{2_{010}}$ at R are purely
complex, identical eigenvalues are paired, since $U_{2_{001}}$
anticommutes with both representations, $U_{2_{100}}$ and
$U_{2_{010}}$, and this cancels the sign change from complex
conjugation.
%%%%%%%%%%%%%%%%%%%%%%%%%%%%%%%%%%%%%%%%%%%%%%%%%%%%%%%%%%%%%%%%%%
In the spinful case with nonzero SOC, there are fourfold double
Weyl points at the TRIMs S, U and T, and the fourfold double Weyl
point at R moves to one of the three axis S-R, U-R and
T-R~\cite{furusaki_2017}. This movable point is required to exist
due to the compatibility relations of the band structures that
must be satisfied on the three rotation axes simultaneously.
Figure~\ref{Fig_comprel_SG19_SOC}\subref{comprel_SG19_RSTU} shows
an example of an arrangement of bands fulfilling the compatibility
relations on the three rotation axes through R. All three axes are
part of nodal planes, therefore all bands are twofold degenerate.
Each pair is formed by two bands with the same eigenvalue, as
indicated by the double lines in the same color.

\subsubsection{Material example: \ce{Ag2Se}}
We have identified \ce{Ag2Se} in SG 19 as an example of a material with
fourfold double Weyl points and nodal plane trios.
\ce{Ag2Se} is a narrow gap semiconductor whose room temperature
polymorph crystallizes in SG 19~\cite{Wiegers1971}.  As
with the corresponding sulfide and telluride, it has been examined
for its thermoelectric properties and fast ion conduction at high
temperatures.
In the vicinity of R, the splitting of bands in \ce{Ag2Se} is
small with respect to the bandwidth. By comparison of DFT band
structures without and with SOC, the connection between the pinned
fourfold double Weyl point in the spinless case and how it splits
into two movable ones upon including SOC is demonstrated in
Fig.~\ref{Fig_comprel_SG19_SOC}\subref{bands_ag2se_SRT}. The black
lines show bands excluding contributions from SOC in which case
there is a fourfold degeneracy pinned at R. Note that all bands
are twofold degenerate because they are part of nodal planes.
Taking SOC into account removes the spin degeneracy, but the
twofold degeneracy in the nodal planes remains. There are now two
fourfold double Weyl points at the axis R-T, related by
time-reversal symmetry. Thus, the total topological charge is
conserved.

%%%%%%%%%%%%%%%%%%%%%%%%%%%%%%%%%%%%%%%%%%%%%%%%%%%%%%%%%%%%%%%%%%
%%%%%%%%%%%%%%%%%%%%%%%%%%%%%%%%%%%%%%%%%%%%%%%%%%%%%%%%%%%%%%%%%%
\section{Rhombic pyramidal: SG 25 - SG 46}
\label{sec_V}

SGs 25 to 46 have the crystallographic point group
$C_{2v}$ ($mm2$), consisting of a twofold rotation in the $[001]$-direction
and two mirror planes with normal along $[100]$ and $[010]$. Because
of the presence of mirror symmetries, these SGs are not
chiral, but they are polar in $[001]$-direction.

\subsection{Weyl points}
\label{mm2_Weyls}
There cannot exist any point degeneracies with nonzero Chern number in
mirror planes, since the Berry curvature transforms as a
pseudo-vector under reflections and the contributions to a closed
surface integral on either side of a mirror plane cancel.
In base-, side- and body-centered lattices however, not all TRIMs
reside in mirror planes. In band structures with strong SOC, these
TRIMs host Kramers-Weyl points whenever they are not in a nodal
plane. The difference to chiral SGs lies in the relative
sign of Chern numbers for symmetry related Weyl points. If two
Weyl points are mapped onto each other by a mirror symmetry, they
have opposite chirality. Consequently, the nodal planes $k_z=\pi$
in SGs 26, 29, 31, 33 and 37 cannot be topological, because the
chiralities of all Weyl points in the interior of the BZ will
always add up to zero.
 
Band structures in SGs 36 and 45 with SOC have additionally movable Weyl points. In the base
centered lattice of SG 36, the axis R-S has only the screw
rotation and translations in its little group and an
hourglass dispersion along this axis is enforced by the same
mechanism as in the chiral SGs, see~\ref{Sec_movable_Weyl}.
Because this axis is the only place for Weyl points, SG 36 has
only two symmetry-enforced Weyl points 
at S for an odd number of filled bands $n$ within a group of bands, while R is part
of a nodal plane. 
For $n$ even, there are four Weyl points in total on the two axes S-R.
Note that while there can be addition accidental Weyl points,
it is in principle possible to have only two Weyl points in
materials with SG 36. The existence of additional accidental Weyl
points depends on the particular band dispersions and needs to be
checked for a given material by inspection of the DFT bands.

The rotation axis T-W in SG 45 gets an hourglass dispersion from a
slightly different mechanism. The twofold rotation is not a screw
rotation, therefore the eigenvalues are $\pm\mathrm{i}$
on the whole axis. But W is not a TRIM, instead it is invariant under the
combined symmetry
$M_{010}(\tfrac{1}{2},\tfrac{1}{2},0)\mathcal{T}$. This symmetry
is antiunitary as well and enforces Kramers pairs on the axis R-W,
equivalently $M_{100}(\tfrac{1}{2},\tfrac{1}{2},0)\mathcal{T}$
with the invariant axis S-W. Because the representation for the
mirror symmetry anticommutes with the one for the rotation at W,
the sign change from complex conjugation is compensated and
identical eigenvalues are paired, $(+,+)$ and $(-,-)$, as opposed
to $(+,-)$ at T. Connecting these pairs leads again to the
hourglass dispersion shown in Fig.~\ref{fig_hourglass}\subref{hourglass_sketch}.

% minimal number of Weyl points
In the spinful case, SGs 36 and 45-46 have only two Weyl points
connecting pairs of bands. This is the lowest possible number of Weyl
points in any band structure and makes these SGs interesting
candidates for investigating the effects from such topological
point degeneracies. There are however additional degeneracies
closing the gap along lines and, in the case of SG 36, a nodal plane. 
Additionally it can be noted, that SGs 36 and 45 have four
Weyl points connecting pairs of bands.
Note that there are two non-equivalent copies of
S-R in the BZ, related by a mirror symmetry. Therefore the
topological charges of the two Weyl points will cancel in all
cases.
Four Weyl points at TRIM can also be found in SGs 35 and 37 at R and S,
and in the face-centered lattices of this group, namely in SG 42
and 43 at the TRIM L.
A material example for the movable Weyl points in SG~36 is
AsPb$_2$Pd$_3$, where the enforced Weyl points occur within about
300~meV above and below the Fermi energy
\cite{ExhaustiveHourGlass_2020}.

\subsection{Pinned nodal lines}
\label{Sec_pinned_lines}

As shown above, there are no Kramers-Weyl points in mirror planes.
However, Kramers theorem still holds and spinful band structures at TRIMs
must be at least twofold degenerate. These degeneracies are always
part of nodal lines and in most cases, these nodal lines are pinned
to high-symmetry lines.
Pinned nodal lines can be readily identified from tables of irreducible
representations for the little groups of the high-symmetry line,
e.g., using the Bilbao Crystallographic
Server~\cite{Elcoro_double_cryst}. 
For completeness and quick reference, we included them in
Tables~\ref{SGs_noinv_nosoc} to \ref{SGs_inv_soc} in the column
''lines´´.

For example, the rotation axis $\Gamma$-Z with little group $mm2$
is always twofold degenerate in spinful band structures, even in
the absence of $\mathcal{T}$. Because the spin parts of the
symmetries anticommute, all irreducible representations must be
at least two-dimensional and eigenvalues of all symmetries are
paired as $(+,-)$.

Another mechanism for pinned nodal lines results from the
combination of glide mirror symmetries with time-reversal
symmetry. A glide mirror symmetry has a translational part
$\frac{1}{2}\vb{t}_\bot$ within the mirror plane.  In double
groups, mirror symmetries square to a $2\pi$ rotation and in the
case of a glide mirror symmetry an additional lattice translation
$\vb{t}_\bot$ within the mirror plane.  The eigenvalues are
therefore of the form 
\begin{IEEEeqnarray}{rCl} U_M\ket{\pm} &=&
\pm\mathrm{i}^\zeta
\exp(\tfrac{\mathrm{i}}{2}\vb{k}\cdot\vb{t}_\bot)\ket{\pm}.
\label{reflection_eigvals} 
\end{IEEEeqnarray} 
The combination of the glide mirror with
$\mathcal{T}$ acts like a rotation in $k$-space with its invariant
axis parallel to the mirror normal and running through TRIMs in
the mirror plane, see the green lines in Fig.~\ref{BZ_fig}. On an
axis through a TRIM $\vb{K}$ with
$\vb{K}\cdot\vb{t}_\bot=\pm\pi$, the combined
symmetry squares to $-1$ and enforces Kramers pairs by the
extended Kramers theorem.
For example, the glide mirror symmetry $M_{010}(\tfrac{1}{2},0,0)$
in SG 28 combined with $\mathcal{T}$ enforces the lines X-S and
T-R to be twofold degenerate.

\subsection{Movable nodal lines}
\label{Sec_movable_lines}

The $k$-dependent eigenvalues of glide mirror symmetries defined
in Eq.~\eqref{reflection_eigvals} lead to the possibility of
pairing identical and opposite eigenvalues at different points in
the same mirror plane. This leads to an hourglass dispersion as
shown in Fig.~\ref{fig_hourglass}\subref{hourglass_sketch} for any path in the mirror plane
between these points. The movable crossings on all these paths
form a nodal line, which might either form a closed loop around one
of the points or extend through the BZ. 
For example, in SG 28 with strong SOC, the glide mirror symmetry
$M_{010}(\frac{1}{2},0,0)$ has eigenvalues $\pm\mathrm{i}$ at
$\Gamma$-Z and Y-T, paired on the whole line, and
eigenvalues $\pm1$ for $k_x=\pi$, forming Kramers pairs $(+,+)$
and $(-,-)$ at X, U, S and R.
The movable nodal line in the $k_y=0$
plane therefore runs between $\Gamma$-Z and X or U. A movable line
is given in Tables~\ref{SGs_noinv_nosoc} and \ref{SGs_noinv_soc}
in terms of a tuple, e.g., ($\Gamma$-Z;R,U), where the entries left
of the semicolon are the degeneracies  with
opposite eigenvalue pairing and the entries on the right the
points where identical eigenvalues are paired. The mirror plane
can be inferred from the high-symmetry points and lines in the
bracket, otherwise it is specified explicitly.  If a point appears
left and right of the semicolon, there is a fourfold degeneracy
with eigenvalues $(+,+,-,-)$ and the point is written in italic
font. See Sec.~\ref{mm2_fourfold_points} below for a discussion of
fourfold degenerate points.

Monolayer GaTeI can be considered as a subset of the 3D SG~31 with
SOC, which exhibits hourglass nodal lines that correspond to
($\Gamma$-Z-U;X) in our notation~\cite{GaTeI_2019}. While this
feature is at about 1.2~eV above the Fermi energy, it has been
suggested that due to the two-dimensional geometry 
it may be possible to introduce a significant number of carrier
by electrostatic gating.

\subsection{Nodal chain}
\label{Sec_nodal_chain}

When two mirror planes with movable nodal lines intersect, it can
be required that the nodal lines touch on the intersection and
form a nodal chain~\cite{bzdusek_soluyanov_nature_16}.
TRIMs at the intersection of two glide mirror symmetries
might have different eigenvalue pairing, 
i.e., $(+,-)$ for one mirror symmetry and $(+,+)$ for the other.
If the pairing is different for both mirror symmetries at another TRIM
on the same intersection, the requirements of an hourglass
dispersion need to be fulfilled for both mirror symmetries simultaneously. 
Because the total number of crossings must be odd, the two movable
nodal lines have to meet at one point on the axis. 
In case the TRIMs on the high-symmetry axis have opposite eigenvalue
pairing as mentioned above, 
the two loops form a so called nodal chain, found in 
SGs 30, 34 and 43 with SOC~\cite{bzdusek_soluyanov_nature_16}.
In Tab.~\ref{SGs_noinv_soc} they can be identified by two entries
for movable nodal lines with the two TRIMs changing sides with respect
to the semicolon. 
For example, in the hourglass relations for SG 43, T and Y appear
on each side once. In the $k_x=0$ plane the relation reads
($\Gamma$-$Z$-T;Y) and Y has identical $M_{100}$ eigenvalues
paired. On the other hand we find Y as part of the line Z-A on the
left in the relation for the $k_y=0$ plane, ($\Gamma$-$Z$-A;T),
and T has identical $M_{010}$ eigenvalues paired.
On the line Y-T, both eigenvalues are good quantum numbers and
both have to exchange as part of nodal lines. Consequently, the two
nodal lines have to intersect and form a chain.

\subsubsection{Material example: CuIrB}

\begin{figure}
\sidesubfloat[]{
\includegraphics[width=.60\linewidth,valign=t]{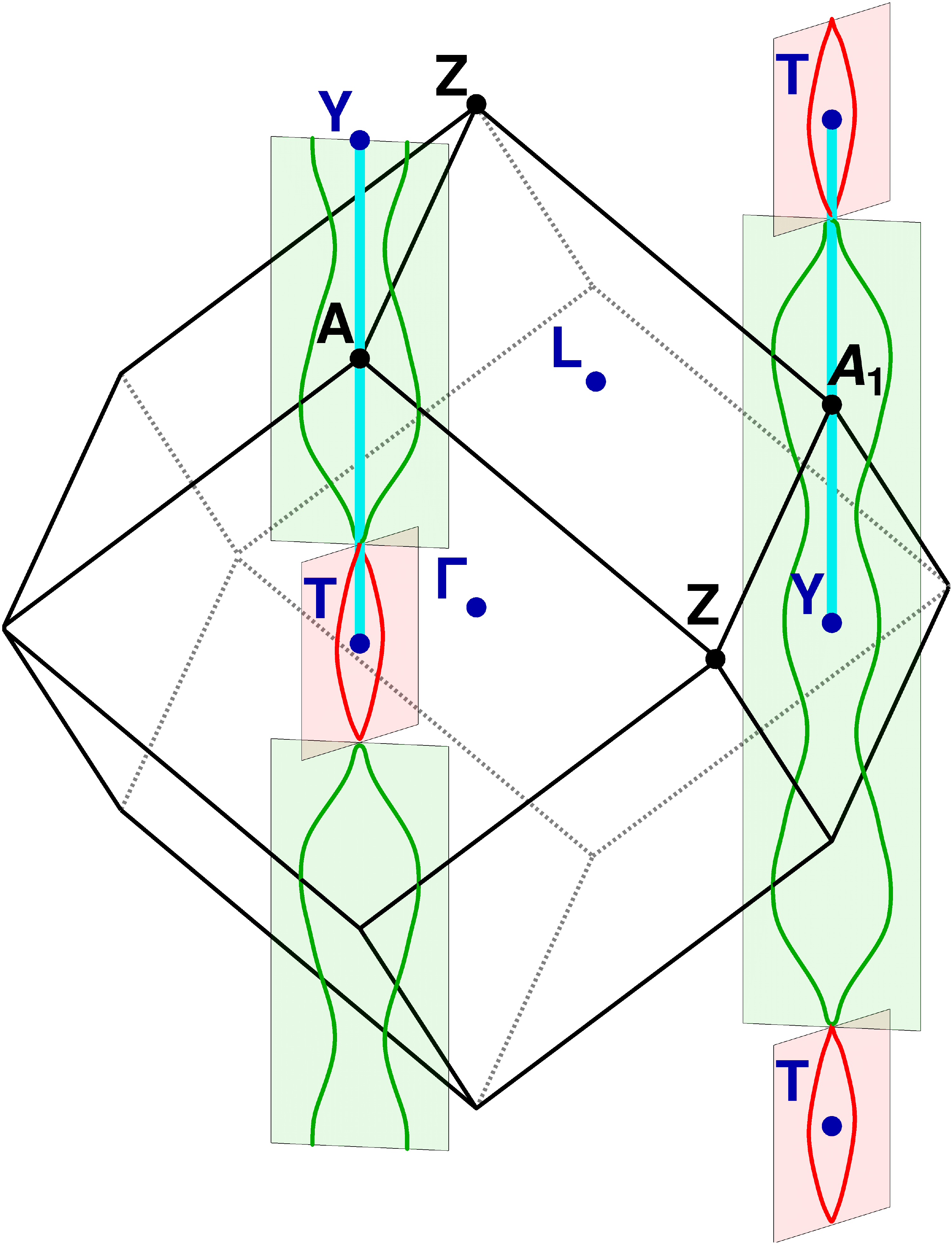}
\label{cuirb_chain}
}

\sidesubfloat[]{
\includegraphics[width=.60\linewidth,valign=t]{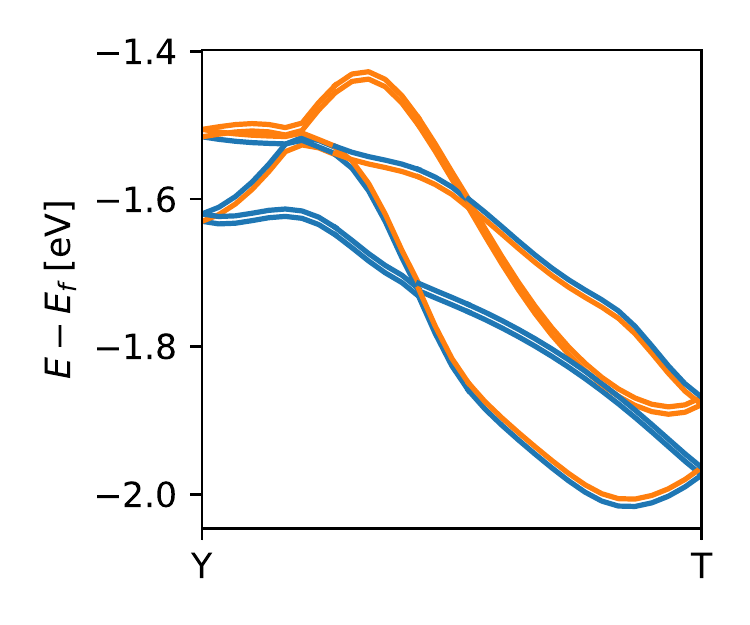}
\hspace{0.3cm}
\label{cuirb_disp}
}
\caption{
(a) Nodal chain in CuIrB formed by the 10th and 11th valence band counted from the band gap at
the Fermi energy.
The green line is within the $k_y=0$ plane and encircles Y, the
red line is restricted to the other mirror plane $k_x=2\pi$ and
winds around T.
(b) Dispersion on the intersection of both planes, shown in
turquoise in (a).
}
\label{fig_nodal_chain}
\end{figure}

Our search has identified CuIrB as an example of a material with
nodal chains.  \ce{CuIrB}, obtained from the elements either by
heating in quartz ampules or by arc-melting and annealing,
crystallizes in SG 43 and exhibits metallic
resistivity~\cite{Kluenter1994}.  The nodal chain made from lines
in the $k_y=0$ and $k_x=2\pi$ planes are shown in
Fig.~\ref{fig_nodal_chain}\subref{cuirb_chain} and the dispersion
along the intersection of the mirror planes is presented in
Fig.~\ref{fig_nodal_chain}\subref{cuirb_disp}.  The dispersion
along the full high-symmetry path as defined in
Fig.~\ref{CuIrB_path} is shown in Fig.~\ref{band_structures}.

\subsection{Almost movable lines}
\label{Sec_almost_movable_lines}

There is one case of symmetry-enforced nodal lines not covered by
the above discussion. The Kramers pairs at the TRIMs S and R in SGs
44 and S in SG 46 have only a mirror symmetry in their little
group and are not part of a pinned nodal line. The mirror
eigenvalues are $\pm\mathrm{i}$ and $\mathcal{T}$ creates pairs
$(+,-)$. In the vicinity of such a TRIM $\vb{K}$, a state at
$\vb{K}+\vb{\delta}$ in the mirror planes is mapped by
$\mathcal{T}$ to a state with opposite eigenvalue at
$\vb{K}-\vb{\delta}$ with the same energy. Consequently, the two
bands with opposite mirror eigenvalues have to exchange somewhere
on any path within the mirror plane connecting these two points.
The exchange of bands is protected by the different symmetry
eigenvalues and the resulting nodal line is only pinned to the
TRIM $\vb{K}$. Otherwise it can be moved freely through the mirror
plane and is therefore called an almost movable
line~\cite{hirschmann2021symmetryenforced}. Similar to movable
nodal lines, we indicate these nodal lines also with a bracket
containing the point of opposite eigenvalue pairing, but instead
of a counterpart with equal eigenvalues we just leave a dash, e.g.
(S;$-$). The corresponding mirror symmetry and its invariant plane
with the nodal line is always the one leaving the high-symmetry
point invariant.

\subsection{Fourfold degeneracies in mirror planes}
\label{mm2_fourfold_points}

In Sec.~\ref{Sec_pinned_lines} we have discussed two different types
of pinned nodal lines. If they intersect a fourfold degeneracy
follows.
These fourfold degeneracies occur in two distinct types. The first
type appears as a point-like degeneracy in the gap between the
nodal lines, whereas the second is crossed by an additional
hourglass nodal line and  a sphere in the BZ enclosing the
fourfold degeneracy will be gapless for every number of filled
bands. 

\begin{figure}
\sidesubfloat[]{
\includegraphics[width=.85\linewidth,valign=t]{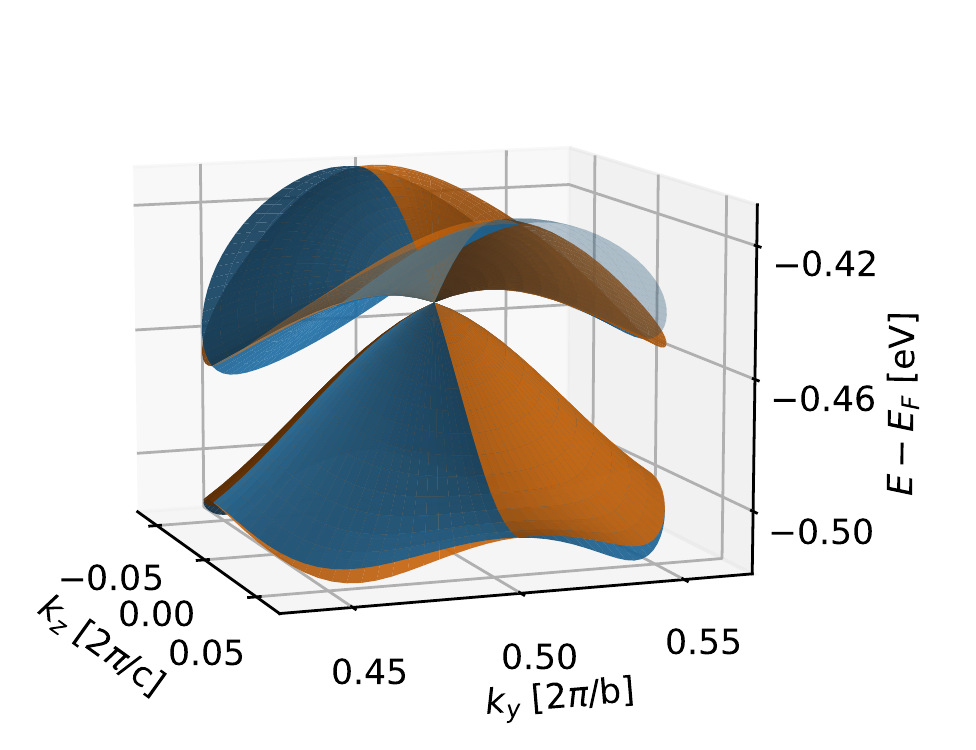}
\label{autlsb_S}
}
 
\sidesubfloat[]{
\includegraphics[width=.85\linewidth,valign=t]{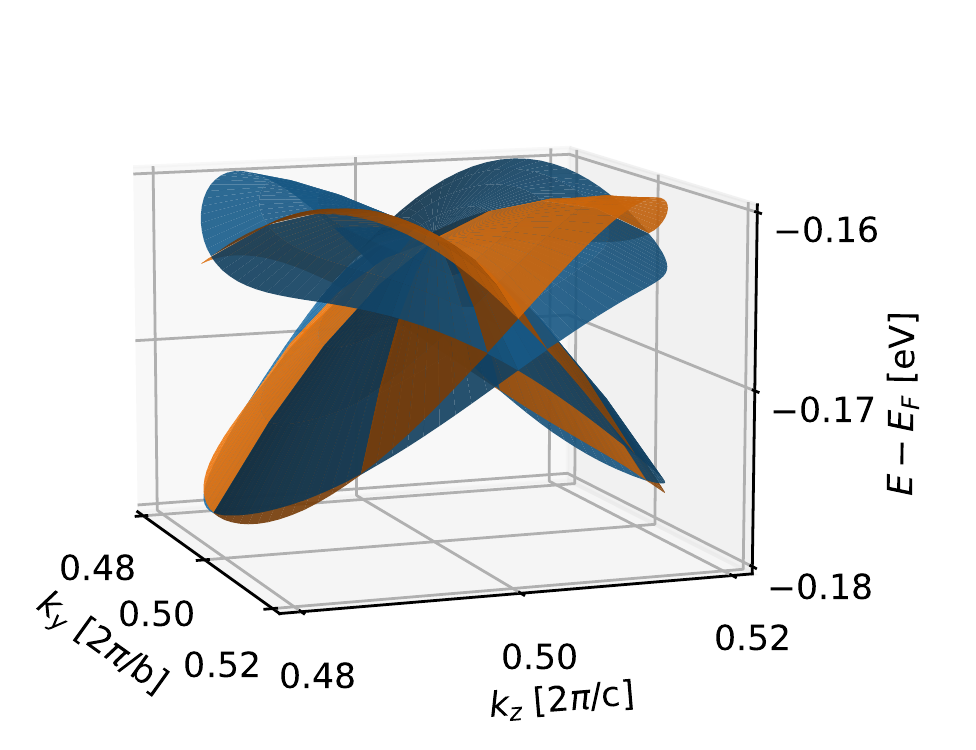}
\label{autlsb_R}
}
\caption{
Dispersion of the two different types of fourfold degenerate
points in AuTlSb in the $k_y$-$k_z$-plane. Colors indicate
the relative sign of $M_{100}$ eigenvalues. 
(a) Around S all lines are made of pairs $(+,-)$ and a point-like
degeneracy is formed.
(b) At R, the nodal lines on R-Z$=(\pi,v,\pi)$ are paired $(+,+)$
and $(-,-)$, and an additional hourglass nodal line closes the gap,
and thus the fourfold degenerate point is not a point crossing
but part of several twofold degenerate nodal lines.
}
\label{4fold_points_mm2}
\end{figure}

In the first type, the anticommutation relation of the two mirror
symmetries $M_1$ and $M_2$ creates always eigenvalue pairs
$(+,-)$.
For the second type, at least one of the mirror symmetries, say
$M_1$, needs to be nonsymmorphic with real eigenvalues at a TRIM.
Time-reversal symmetry then pairs identical eigenvalues, requiring
the fourfold degeneracy. The combined symmetry $M_1\mathcal{T}$
creates a twofold degeneracy along its invariant axis.
Both pinned nodal lines are within the invariant plane of $M_2$.
If its eigenvalues are also real at the fourfold degeneracy, then
the combined antiunitary symmetry $M_1\mathcal{T}$ pairs opposite
eigenvalues because of the anticommutation relation of mirror
symmetries. In that case, the fourfold degeneracy is point-like,
but with vanishing Chern number. We list such cases in the column
``points'' in Tab.~\ref{SGs_noinv_soc} with the degeneracy given
explicitly in brackets. An example is shown in
Fig.~\ref{4fold_points_mm2}\subref{autlsb_S} and the linearized low-energy Hamiltonian
around such a point is given by Eq.~\eqref{mm2_4fold_lowE_gapped} in
Appendix~\ref{App_4fold_mm2}.

The second possibility is complex eigenvalues for $M_2$. In that
case, identical eigenvalues are paired by $M_1\mathcal{T}$ and
there will be an hourglass nodal line in between the two pinned
nodal lines~\cite{ryo_murakami_PRB_17}. Consequently, there is no
gap, at which the fourfold degeneracy appears point like and we
list the fourfold degeneracy in the columns ``lines'' as an
hourglass nodal line with the fourfold degenerate point on both
sides of the semicolon and shown in italic, e.g. (X-$U$;$U$-R).
In this scenario, the two mirror symmetries must differ in their
in-plane translation and their product is always a screw rotation.
See Fig.~\ref{4fold_points_mm2}\subref{autlsb_R} for an example and
Eq.~\eqref{mm2_4fold_lowE_hourglass} for a linearized low-energy Hamiltonian.

\subsubsection{Material example: AuTlSb}

As an example, we present \ce{AuTlSb} crystallizing in an
anion-ordered pyrite-derivative structure in SG 33 and exhibiting
metallic resistivity~\cite{McGuire2006}.
Since the above considerations depend on pairing of $k$-dependent
mirror eigenvalues, both types of fourfold degeneracies can occur in the same SG at
different TRIMs.
In SG 33, there are nodal lines at S-R with
$M_{100}(\tfrac{1}{2},\tfrac{1}{2},\tfrac{1}{2})$-eigenvalues
paired as $(+,-)$, because of the anticommutation relation of the
two glide mirror symmetries.  
On the lines X-S and U-R, the combined antiunitary symmetry
$M_{010}(\tfrac{1}{2},\tfrac{1}{2},0)\mathcal{T}$ creates pairs
$(+,-)$ and $(+,+)$/$(-,-)$, respectively.
A fourfold point can therefore be found in the band structure of
AuTlSb at S, see Fig.~\ref{4fold_points_mm2}\subref{autlsb_S}, whereas \subref{autlsb_R}
shows the fourfold degeneracy with an hourglass nodal line at R.

\section{Rhombic Dipyramidal: SG 47 - SG 74}
\label{sec_VI}

The centrosymmetric SGs in the orthorhombic crystal system, SGs 47 to
74, have crystallographic point group $D_{2h}$ ($mmm$). The
combination of inversion and time-reversal symmetry requires all states to
be spin degenerate throughout the whole BZ via Kramers theorem.
Any additional symmetry-enforced degeneracy in spinful
band structures is therefore a Dirac point or line. This overall twofold
degeneracy already includes many of the degeneracies described above.
For example, there are no nodal planes in centrosymmetric spinful
band structures.
Additionally, the Berry curvature must vanish identically at each
point in the BZ and consequently no degeneracy might carry a
nonzero Chern number.

\rowcolors{1}{spinless_green1}{spinless_green2}
\setlength{\arrayrulewidth}{1.0pt}
\begin{longtable*}{ | r@{ }l || >{\raggedright}p{3.0cm} | >{\raggedright}p{3.5cm} | >{\centering}p{1.7cm} | p{6.0cm} | }
\hline
\multicolumn{2}{|c||}{SG} &  \multicolumn{1}{c|}{points} &  \multicolumn{1}{c|}{lines} & nodal planes & \multicolumn{1}{c|}{notable features} 
\\
\hline
\endhead

%%% SPINLESS %%%%%%%%%%%%%%%%%%%%%%%%%%%%%%%%%%%%%%%%%%%%%%%%%%%%%
47 & ( P$mmm$ )         &   &   &   & \\
48 & ( P$nnn$ )         &   & Z-U-X-S-Y-T-Z &   & \\
49 & ( P$ccm$ )         &   & Z-U-R-T-Z  &   & \\
50 & ( P$ban$ )         &   & X-U-R-T-Y-S-X  &   & \\
51 & ( P$mma$ )         &   &   &  $k_x = \pi$ & \\ 
52 & ( P$nna$ )         &  & U-Z-T, X-S, \mbox{(U-X-$S$;$S$-R)(4),} \mbox{(X-$S$,Y-$S$)(4)}  & $k_y = \pi$ &  precedes unstable $\mathbb{Z}_2$ invariant (with SOC)   \\
53 & ( P$mna$ )         &   & U-X-S-R  & $k_z = \pi$ &  \\
54 & ( P$cca$ )         &  & R-T-Z-U, \mbox{(Z-$U$;$U$-X)(4),} \mbox{(T-$R$;$R$-S)(4),} \mbox{($U$-$R$;$R$-T-Z-$U$)} & $k_x = \pi$  & \\
55 & ( P$bam$ )         &   &   & $k_x,k_y = \pi$  & \\
56 & ( P$ccn$ )         &  & T-Z-U, \mbox{($T$-Z;Y-$T$)(4),} \mbox{($U$-Z-$T$;$U$-R-$T$)(4),} \mbox{(Z-$U$;$U$-X)(4)}  &  $k_x,k_y = \pi$ &  precedes $\mathbb{Z}_2$ invariant (with SOC) \\
57 & ( P$bcm$ )         &   & R-T(4), (Y-$T$;$T$-Z)(4), \mbox{(S-$R$;$R$-U)(4)}  &  $k_y,k_z = \pi$ & \\
58 & ( P$nnm$ )         &   & T-Z-U & $k_x,k_y = \pi$ &     \\
59 & ( P$mmn$ )         &   &   & $k_x,k_y = \pi$ &     \\
60 & ( P$bcn$ )         &  & S-Y-T, U-R(4), \mbox{(Y-$T$;$T$-Z)(4),} \mbox{(Z-$U$;$U$-X)(4),} \mbox{($R$-S-Y-$T$;$T$-$R$)(4)} &  $k_x,k_z = \pi$  & precedes unstable $\mathbb{Z}_2$ invariant (with SOC)  \\
61 & ( P$bca$ )         &  & S-R(4), U-R-T(4), \mbox{(Y-$T$;$T$-Z)(4),} \mbox{(Z-$U$;$U$-X)(4)} \mbox{(X-$S$;$S$-Y)(4)} & $k_{x/y/z} = \pi$ &   \\
62 & ( P$nma$ )         &  & S-R(4), (X-$S$;$S$-Y) & $k_{x/y/z} = \pi$  &  precedes $\mathbb{Z}_2$ invariant (with SOC)  \\
63 & ( C$mcm$ )         &   &   &  $k_z = \pi$ &    \\
64 & ( C$mca$ )         &   & S-R, (S;$-$)(2)   & $k_z = \pi$  & \\
65 & ( C$mmm$ )         &   &   &   & \\
66 & ( C$ccm$ )         &   & A-Z-T &   & \\
67 & ( C$mme$ )         &   & S-R, (S;$-$)(2), \mbox{(R;$-$)(2)}  &   & \\
68 & ( C$cce$ )         &   & A-Z-T,S-R, \mbox{(S;$-$)(2)}, \mbox{(R;$-$)(2)}  &   & \\
69 & ( F$mmm$ )         &   &   &   & \\
70 & ( F$ddd$ )         &   & A-Z-T-Y  &   & \\
71 & ( I$mmm$ )         &   &   &   & \\
72 & ( I$bam$ )         &   & S-W-R, \mbox{(S;$-$)(2),} \mbox{(R;$-$)(2)}  &   & \\
73 & ( I$bca$ ) & W(4) & S-W-R, W-T, \mbox{(S;$-$)(2),} \mbox{(R;$-$)(2),} \mbox{(T;$-$)(2)}  &   & fourfold point ($\mathcal{C}{=}0$) at half filling \\
74 & ( I$mma$ )         &   & W-T, (T;$-$)(2)  &  & \\
\hline
\caption{
\cellcolor{white}
Symmetry-enforced band crossings in \emph{spinless} band
structures of \emph{centrosymmetric} orthorhombic SGs.
The notation is identical to the one introduced in
Tables~\ref{SGs_noinv_nosoc} and~\ref{SGs_noinv_soc}.
\vspace{1cm}
}
\label{SGs_inv_nosoc}
\end{longtable*}

\rowcolors{1}{spinfull_blue1}{spinfull_blue2}
\begin{longtable*}{ | r@{ }l || >{\raggedright}p{3.0cm} | >{\raggedright}p{3.5cm} | >{\centering}p{1.7cm} | p{6.0cm} | }
\hline
\multicolumn{2}{|c||}{SG} &  \multicolumn{1}{c|}{points} &  \multicolumn{1}{c|}{lines} & nontrivial planes & \multicolumn{1}{c|}{notable features} 
\\
\hline
\endhead

%%% SPINFULL %%%%%%%%%%%%%%%%%%%%%%%%%%%%%%%%%%%%%%%%%%%%%%%%%%%%%
47 & ( P$mmm$ )         &   &   &   & \\
48 & ( P$nnn$ )         & S,T,U,X,Y,Z  &   &   & \\
49 & ( P$ccm$ )         & R,T,U,Z &   &   & \\
50 & ( P$ban$ )         & R,S,T,U,X,Y &   &   & \\
51 & ( P$mma$ )         &   & U-X, S-R  &   & \\
52 & ( P$nna$ )         & U,X,Z, X-S(4) & R-S-Y  & ZUR\textbf{T}, XUY\textbf{T}  &  movable Dirac  \\
53 & ( P$mna$ )         & X,S  & T-Z  &   & \\
54 & ( P$cca$ )         & T,Z, U-Z(4), R-T(4)  & S-R-U-X  &   & \\
55 & ( P$bam$ )         &   & X-S-R-U, S-Y, R-T  &   & \\
56 & ( P$ccn$ )         & Z, Z-U(4), Z-T(4)  & Y-T-R-U-X, S-R  & XUS\textbf{R}, YTS\textbf{R}  &  movable Dirac \\
57 & ( P$bcm$ )         &   & U-R,S-Y,T-Z, \mbox{(S-Y;R,T)(4)}  &   & only Dirac nodal line at half filling \\
58 & ( P$nnm$ )         & Z &  X-S-R,S-Y &   & \\
59 & ( P$mmn$ )         &   & X-U,S-R,Y-T  &   &    \\
60 & ( P$bcn$ )         & Y, T-Y(4)  & X-U, R-T-Z, \mbox{(X-U;R)(4),}  \mbox{(R-T-Z;U)(4)}  & X\textbf{S}TZ  & Dirac nodal chain  \\
61 & ( P$bca$ )         &   & T-Z, \mbox{(U-X;S)(4),} \mbox{(S-Y;T)(4),}  \mbox{(T-Z;U)(4)} & XY\textbf{R}Z  &  three Dirac nodal lines at half filling  \\
62 & ( P$nma$ )         &   & Z-U-X, U-R-T, S-Y, (R-U-X;S)  & X\textbf{U}TY, Z\textbf{U}RT   &  \\
63 & ( C$mcm$ )         &                               & \mbox{(R;$-$)}, Z-T &   & \\
64 & ( C$mca$ )         & S                                             & Z-A &   & \\
65 & ( C$mmm$ )         &                                               &       &   & \\
66 & ( C$ccm$ )         & T,Z                                           &       &   & \\
67 & ( C$mme$ )         & R,S                                           &       &   & \\
68 & ( C$cce$ )         & T,Z,R,S                                       &       &   & \\
69 & ( F$mmm$ )         &                                               &       &   & \\
70 & ( F$ddd$ )         & T,Y,Z                                         &       &   & \\
71 & ( I$mmm$ )         &                                               &       &   & \\
72 & ( I$bam$ )         & R,S,W                                         &       &   & \\
73 & ( I$bca$ )         & R,S,T, \mbox{W-[R$\veebar$S$\veebar$T]}(4)&       &   & movable Dirac \\
74 & ( I$mma$ )         & T                                             &       &   & \\
\hline
\caption{
\cellcolor{white}
Symmetry-enforced band crossings in band structures \emph{with
SOC} of \emph{centrosymmetric} orthorhombic SGs.
The notation is identical to the ones introduced in
Tables~\ref{SGs_noinv_nosoc} and~\ref{SGs_noinv_soc}.
All degeneracies are Dirac points and lines and the overall twofold
degeneracy due to $\mathcal{PT}$ is not counted in the number of involved bands.
In the fourth column we list planes with a nontrivial
$\mathbb{Z}_2$ invariant via the TRIMs they contain of which the
TRIM with identical inversion eigenvalue pairing is shown in bold.
}
\label{SGs_inv_soc}
\end{longtable*}
\rowcolors{1}{white}{white}

\subsection{Dirac points at high-symmetry points}
\label{mmm_Diracs}

We call a fourfold point degeneracy a Dirac point, if and only if
it splits into twofold degenerate bands in every direction and its
Chern number vanishes.
Dirac points pinned to TRIMs are also invariant under inversion.
A fourfold degeneracy is enforced, whenever the irreducible
representation for inversion is two-dimensional and has
eigenvalues $\pm1$. Their Kramers partner has to have the same
eigenvalue and $\mathcal{T}$ enforces the fourfold degeneracy.
As shown in Sec.~\ref{Sec_Kramers}, the point W in the BZ of
body-centered lattices hosts a Weyl point in SG 23 and SG 24 in
spinless and spinful band structures, respectively, despite not
being a TRIM.
There are two possibilities of adding inversion to these two SGs.
In spinless (spinful) band structures of SG 72 (SG 73), inversion
combined with time-reversal symmetry doubles this degeneracy.

The other possibility for the inversion center leads to SG 71 (SG
74). There, states at W remain only twofold degenerate and are
part of a nodal line (the overall spin degeneracy). This leads to
a single instance of a pinned Dirac point at a non-TRIM in SG 73.
In the spinless case in SG 72, the fourfold degeneracy is not a
Dirac point in the sense mentioned above, because bands are
non-degenerate away from high-symmetry lines.

\subsection{Movable Dirac points}
\label{Sec_movable_Diracs}

In most cases, the overall Kramers degeneracy of band structures
in centrosymmetric SGs prevents the hourglass dispersion
from occurring on screw rotations, because rotation eigenvalues are paired as
$(+,-)$ on the whole rotation axis.

There are, however, instances where the representation of
$\mathcal{PT}$ anticommutes with the screw rotation. More
concretely, if the screw rotation is additionally off-centered
with respect to the inversion center, exchanging the order of the
two differs by a translation of a lattice vector~\cite{yang_furusaki_PRB_2017}. 
This leads to a
pairing of $(+,+)$ and $(-,-)$ on rotation axes where the
eigenvalue of this translation contributes an additional minus
sign. With Dirac points at the TRIMs, this leads again to the
hourglass dispersion shown in Fig.~\ref{fig_hourglass}\subref{hourglass_sketch}.
The only difference is, that all bands are twofold degenerate
and eigenvalues appear twice. The movable crossing on the axis
forms a Dirac point.
We find these conditions to be met on at least one rotation axis in
SGs 52, 54, 56 and 60. Following our previous notation, movable
Dirac points are given in Tab.~\ref{SGs_inv_soc} in terms of the
two TRIMs on the rotation axis connected by a dash.

We find one additional movable Dirac point in SG 73 from an
extended compatibility relation on the three rotation axis
connecting W with R, S and T. In SU(2)-symmetric band structures
this corresponds to an eightfold degeneracy at W. With nonzero SOC, this
degeneracy is lifted and there are two symmetry related Dirac
points on one of the rotation axes.
Each TRIM has one rotation in its site symmetry group and
eigenvalues are paired as $(+,+,-,-)$ and $(+,+)$ and $(-,-)$ on
the axis. At W, all three rotation eigenvalues are simultaneously
present, but their overall sign is restricted to $+$. This can only
be arranged with a minimum of one band crossing on one of the axes
W-R, W-S and W-T, equivalent to the requirement shown in
Fig.~\ref{Fig_comprel_SG19_SG24_noSOC}\subref{comprel_19}.

\subsection{Movable and almost movable Weyl lines}

In spinless band structures of centrosymmetric SGs, glide
mirror symmetries enforce movable nodal lines from an hourglass
dispersion between high-symmetry lines and planes in the mirror
plane~\cite{ryo_murakami_PRB_17}.
The mechanism is the same as described in
Sec.~\ref{Sec_movable_lines} and we list them using the same
notation in the column ``lines'' in Tab.~\ref{SGs_inv_nosoc}. In
all instances, the pinned and movable nodal lines cross in a
common point, leading to a fourfold degeneracy as described in
Sec.~\ref{mm2_fourfold_points} and shown in
Fig.~\ref{4fold_points_mm2}\subref{autlsb_R}.
Additionally, glide mirror symmetries might introduce almost
movable lines  whenever their eigenvalues at a TRIM are
$\pm\mathrm{i}$, as described in
Sec.~\ref{Sec_almost_movable_lines}.

\subsection{Movable Dirac lines}
\label{Sec_movable_Dirac_lines}

In the spinful case, nodal lines can only be formed as a fourfold
degeneracy, splitting into twofold degenerate bands, called Dirac
lines. Since the Berry phase is a $\mathbb{Z}_2$ invariant, its
topological invariant needs to be evaluated in a spin subspace.
% TODO: Check, be more speficic

A movable Dirac line can only be enforced, if the Kramers pairs
formed by $\mathcal{PT}$ share the same mirror eigenvalue. This is
the case, if and only if the glide mirror symmetry is additionally
off-centered with respect to the inversion center. As an example
we discuss SG 57, which has the glide mirror symmetries
$M_{010}(0,\tfrac{1}{2},\tfrac{1}{2})$ and
$M_{100}(0,\tfrac{1}{2},0)$, of which only the former is also
off-centered.
For a state $\ket{\pm}$ with $M_{010}$-eigenvalue
$\pm\mathrm{i}\exp(\mathrm{i}\tfrac{k_z}{2})$, we can calculate
the eigenvalue for its Kramers partner $\mathcal{PT}\ket{\pm}$ via
the commutation relation of symmetries~\cite{beta_ReO2_2017},
\begin{IEEEeqnarray}{rCl}
M_{010}(0,\tfrac{1}{2},\tfrac{1}{2})\mathcal{PT} 
    &=& 
    t(0,1,1)\mathcal{PT}M_{010}(0,\tfrac{1}{2},\tfrac{1}{2}),
\label{63_PTM}
\end{IEEEeqnarray}
and therefore
\begin{IEEEeqnarray}{rCl}
U_{M_{010}} U_\mathcal{PT}\mathcal{K}\ket{\pm} 
    &=& \mathrm{e}^{\mathrm{i}(k_y+k_z)} U_\mathcal{PT}\mathcal{K} 
        \left( \pm\mathrm{i}\mathrm{e}^{\mathrm{i}\tfrac{k_z}{2}}\right)
        \ket{\pm} \nonumber
\\
    &=& \mp\mathrm{e}^{\mathrm{i}k_y} \mathrm{i}
        \mathrm{e}^{\mathrm{i}\tfrac{k_z}{2}}
        U_\mathcal{PT}\mathcal{K}\ket{\pm}.
\label{57_mirror_eigvals}
\end{IEEEeqnarray}
From this it follows, that equal eigenvalues are paired in the
plane $k_y=\pi$, where $\mathrm{e}^{\mathrm{i}k_y}=-1$.
Additionally, there are further degeneracies at high-symmetry points in
this plane even before taking time-reversal symmetry into account.
On the axis Y-S, the representations of 
$M_{010}(0,\tfrac{1}{2},\tfrac{1}{2})$ and
$M_{001}(0,\tfrac{1}{2},\tfrac{1}{2})$ anticommute.
This readily creates eigenvalue pairs $(+,-)$, and together with
the pairing from $\mathcal{PT}$ a fourfold degeneracy
$(+,+,-,-)$.
At R and T however, the representation of
$M_{010}(0,\tfrac{1}{2},\tfrac{1}{2})$ commutes with all other
symmetries, but the representations of the other two mirror
symmetries anticommute. This implies, that there are two
orthogonal states for each $M_{010}(0,\tfrac{1}{2},\tfrac{1}{2})$
eigenvalue. They can be distinguished by their
$M_{001}(0,0,\tfrac{1}{2})$ eigenvalue, which is left invariant by
$\mathcal{PT}$, therefore the Kramers partners have identical
eigenvalues for both mirror symmetries.  This leads to a total of
four degenerate states and the pairing in terms of
$M_{010}(0,\tfrac{1}{2},\tfrac{1}{2})$ eigenvalues as defined in
Eq.~\eqref{57_mirror_eigvals} is therefore $(+,+,+,+)$ or
$(-,-,-,-)$.
The movable nodal line formed by the exchange of bands $(+,+)$ and
$(-,-)$ on any path connecting the differently paired states is
fourfold degenerate and called a movable Dirac nodal line.  In
spinful centrosymmetric band structures they are found in SGs 57,
60, 61 and 62. Following our previous notation for movable nodal
lines, we indicate them in Tab.~\ref{SGs_inv_soc} by a bracket
containing the degeneracies with pairwise opposite eigenvalues on
the left of the semicolon and the ones with four identical
eigenvalues on the right. The example discussed above reads
(S-Y;R,T).

In analogy to the nodal chains mentioned in
Sec.~\ref{Sec_nodal_chain}, the two movable Dirac nodal lines in
SG 60 both cross through the shared axis U-R and form a Dirac
nodal chain with loops in the $k_x=\pi$ and $k_z=\pi$ planes.
These Dirac nodal chains occur, for example, in the band structure
of the $\beta$-phase of \ce{ReO2}~\cite{beta_ReO2_2017}.

\subsection{Nodal line arrangement in SGs 61 and 73}

Space group 61 is a supergroup of SG 19 with inversion as an
additional generator. This adds further restrictions to the
compatibility relations for bands along the rotation axes discussed in
Sec.~\ref{Sec_movable_Weyl}. 

\begin{figure}
\sidesubfloat[]{
\includegraphics[width=.40\linewidth,valign=t]{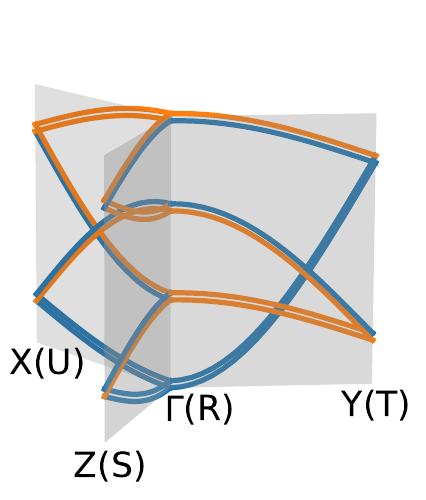}
\label{SG61_GXYZ}
}
\sidesubfloat[]{
\includegraphics[width=.40\linewidth,valign=t]{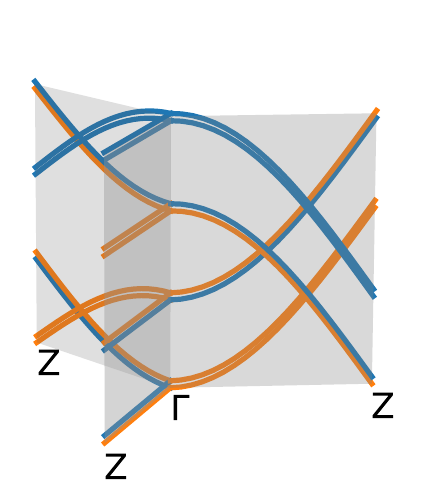}
\label{SG73_GZZZ}
}
\caption{
Possible arrangement of bands along three rotation axes in 
(a) SG 61 in the spinless case for the TRIMs without brackets 
    and the spinful case for the TRIMs in brackets,
(b) SG 73 in the spinless case.
Blue and orange mark the sign of both mirror eigenvalues on each axis
according to Eq.~\eqref{reflection_eigvals}.
}
\label{Fig_comprel_SG61_SG73}
\end{figure}

In the spinless case, SG 61 also has a nodal plane trio enforced
by the three screw rotations. Within the other mirror planes
$k_i=0$, $i=x,y$ or $z$, there is an hourglass nodal line in
between the twofold degeneracies on the intersections with the
nodal planes.
For example, in the plane $k_x=0$ bands form an hourglass
dispersion for any path connecting the lines Z-$T$ and $T$-Y,
where T is fourfold degenerate. 
This includes the path from Z to Y via $\Gamma$ along the rotation
axes. The segments of this path are always part of
two mirror planes and the dispersion along these axes has to
accommodate for the eigenvalue relations of both symmetries simultaneously.
No two bands paired at one of the TRIMs X, Y or Z can be paired
again at some other point. This enforces three crossings on two of the
three lines.  A possible arrangement is shown in
Fig.~\ref{Fig_comprel_SG61_SG73}\subref{SG61_GXYZ}. Each irreducible
representation along the axis is completely determined by the two
mirror eigenvalues of the planes meeting along this line. This is
shown by the double lines, where the color indicates the positive
(blue) or negative (orange) eigenvalue. Note that the order of
colors is important. A crossing can only be removed, when both
eigenvalues are identical. Each crossing is part of an hourglass
nodal line, extending into planes where the mirror eigenvalues differ.
Exchanging the order of bands at $\Gamma$ moves the
crossings to other symmetry axis, but cannot change the overall
picture. The rotation eigenvalue is given by the product of both
mirror eigenvalues. At $\Gamma$ the product of all rotation eigenvalues
equals +1 as in SG 19, c.f.
Fig.~\ref{Fig_comprel_SG19_SG24_noSOC}\subref{comprel_19}.

% with spin:
In the spinful case a similar situation unfolds on the BZ edges.
There are three hourglass nodal lines, (U-X;S) in the $k_x=\pi$
plane, (S-Y;T) in the $k_y=\pi$ plane and (T-Z;U) in the $k_z=\pi$
plane. When connecting two TRIMs out of U,T and S via R, the
eigenvalues of the mirror symmetry whose invariant plane contains
the three points need to exchange for every possible combination.
For example, the path from S via R to T has to provide the
eigenvalue exchange in the $k_y=\pi$ plane, but the segments S-R
and R-T must also be compatible with the hourglass nodal lines in
the $k_x$ and $k_z$ planes, respectively.  This is only possible
with the same connectivity diagram for the three axis R-S, R-U and
R-T as described above for the spinless case and shown in
Fig.~\ref{Fig_comprel_SG19_SG24_noSOC}\subref{comprel_19}.

In SG 73 we also find an extension of the compatibility relations
of its subgroup SG 24. 
All mirror symmetries are non-symmorphic and their
eigenvalues depend on $\vb{k}$. Consequently, an eigenvalue $+$
changes into $-$ when moving along a path which changes the
corresponding coordinate by $2\pi$ modulo reciprocal lattice
translations. For example, the
$M_{010}(0,\tfrac{1}{2},\tfrac{1}{2})$ eigenvalue $+1$ at
$\mathrm{Z}=(0,0,2\pi)$ is labeled with $+$ according to the definition in
Eq.~\eqref{reflection_eigvals}, whereas at the equivalent point
$\mathrm{Z}^\prime=(2\pi,0,0)$ it is the one labeled $-$, i.e. the
bands with $+$ and $-$ have to exchange for any path connecting Z
to itself with a net displacement in $k_y$ of $2\pi$. At the same
time, the TRIM S, R and T are the origin of almost movable nodal
lines in the $k_x=0$, $k_y=0$ and $k_z=0$ planes, respectively,
which facilitate this exchange.
Because $\Gamma$ connects to Z via all
three rotation axes and the above considerations are valid for all
three glide mirrors, there are fewer choices for arranging the
states to avoid crossings. This leads to at least six crossings,
all being part of nodal lines again. A possible arrangement is shown
in Fig.~\ref{Fig_comprel_SG61_SG73}\subref{SG73_GZZZ}.

\subsection{Almost movable Dirac line}

There is only one instance of an almost movable Dirac line in
centrosymmetric orthorhombic SGs with SOC, namely in SG 63.
Because $M_{001}(0,0,\tfrac{1}{2})$ is off-centered, 
the representations $U_{M_{001}}$ and $U_\mathcal{P}$ anticommute
in the $k_z=\pi$ plane and the mirror eigenvalues $\pm\mathrm{i}$
are paired as $(+,+)$ or $(-,-)$ by $\mathcal{PT}$.
At the same time, $\mathcal{T}$ relates $(+,+)$ at R$+(q_x,q_y,0)$
to $(-,-)$ at R$-(q_x,q_y,0)$, requiring the eigenvalues to
exchange on any path in between these two points. This requires a
fourfold degeneracy along a line through R. Following our notation
in Sec.~\ref{Sec_almost_movable_lines} we call this line an almost
movable Dirac line. The same argument applies, in principle, to
every other TRIM in the $k_z=\pi$ plane as well, but there a
pinned nodal line facilitates the exchange.

\subsubsection{Material examples}
The compound LiBH may prove to be a
valuable realization of a material, where a twofold nodal line and
a nodal plane with a pinned fourfold nodal line coexist in the
vicinity of the Fermi energy \cite{LiBH_2020}.  Other known
orthorhombic realizations of movable enforced nodal lines in the
vicinity of the Fermi energy include AgF$_2$  (SG~61)
\cite{AgF2_2019}, Ba$_2$ReO$_5$ (SG~62)
\cite{ExhaustiveHourGlass_2020}, and SrIrO$_3$ (SG~62)
\cite{SrIrO3_2016}.  Accidental nodal lines have been found for
BaLi$_2$Sn (SG~59) \cite{BaLi2Sn_2020}, BaLi$_2$Si (SG~59)
\cite{Li2BaSi_2018}, ZrAs$_2$ (SG~62) \cite{ZrAs2_2020},
Ta$_3$SiTe$_6$ and related compounds (SG~62)
\cite{Ta3SiTe6_and_Nb3SiTe6_2018, Nb3GeTe6_2020}, as well as 3D
$\alpha'$ boron (SG~63) \cite{alphaBoron_2018}.

\subsection{Symmetry-enforced $\mathbb{Z}_2$ topology in planes}
\label{sec_enforced_Z2}

Topologically nontrivial insulators can be understood by an
inversion of bands, which means that the valence bands by
themselves cannot be expressed as a sum of trivial elementary band
representations \cite{CanoBradlynWang_EBR_2018}. Such a
nontrivial topology is not limited to topological insulators, but
can also be found, and even enforced, for gapped subsystems within a
single set of connected bands
\cite{hirschmann2021symmetryenforced,subdimensional_slager}. 

In the following we will discuss the centrosymmetric SGs 56, 61,
and 62 with strong SOC that enforce a nontrivial
weak topological invariant already within a minimal set of
connected bands. To identify whether an effective band inversion
must occur in a subset of the BZ, we calculate the $\mathbb{Z}_2$
index from inversion eigenvalues~\cite{Kane_Fu_InversionInvariant_2007},
which can be obtained for
some SGs directly from the crystalline symmetries. Furthermore,
the argument can be extended by including SGs for which a system
inherits an actual band inversion from the case without spin-orbit
coupling that persists with sufficiently small SOC. 

First, we introduce the $\mathbb{Z}_2$ invariant and apply the
formalism to SG~62, the symmetry group of Ir$_2$Si, for which a
nontrivial weak invariant is enforced. We then discuss SG~52 for
which a nontrivial topology follows from the case of vanishing
SOC, which is confirmed explicitly for
Sr$_2$Bi$_3$.

The weak $\mathbb{Z}_2$ invariant can be calculated on a subset of
the BZ that includes four TRIMs $\Gamma_i$.
The invariant $\nu$ is given by $(-1)^\nu = \prod_{\Gamma_i}
\delta_{\Gamma_i}$ with $\delta_{\Gamma_i} = \prod_{m = 1}^N
\xi_{2m}(\Gamma_i)$, where the inversion eigenvalue
$\xi_{2m}(\Gamma_i)$ is taken for the band $2m$ at the TRIM
$\Gamma_i \in \{ \Gamma, \text{X}, \text{Y}, \text{Z}, \text{U},
\text{S}, \text{T}, \text{R} \}$~\cite{Kane_Fu_InversionInvariant_2007}.
$N$ denotes the number of
occupied bands. Taking only the even-numbered occupied bands is
well-defined because time-reversal symmetry pairs identical inversion eigenvalues
$\xi_{2m}(\Gamma_i) = \xi_{2m +1}(\Gamma_i)$ at each TRIM
$\Gamma_i$.

To illustrate the calculation of the inversion eigenvalues
$\xi_{2m}(\Gamma_i)$ and in extension the invariant $\nu$, we
consider SG~62 as an example. In SG~62 all TRIMs except $\Gamma$
host fourfold degenerate Dirac points and thus we must consider $N
= 2 + 4\mathbb{N}_0$ occupied bands, for which $\delta_{\Gamma_i}$
can be well-defined. Note that for $N$ the bands are counted
without their spin degeneracy, while we consider all the
states in our discussion of inversion eigenvalues. A Dirac point
may consist of either four identical inversion eigenvalues or two
of each possible eigenvalue. 
If one takes into account that the mirror symmetry
$M_{010}(0, \tfrac{1}{2}, 0)$ 
anticommutes with the inversion operation at
TRIMs in the $k_y = \pi$ plane, one finds that these TRIMs contain
both inversion eigenvalues, yielding
$\xi_{2}(\{\text{Y, T, R, S}\}) = \pm 1 = - \xi_{4}(\{\text{Y, T, R, S}\})$.
An analogous argument holds for the mirror symmetry
$M_{001}(\tfrac{1}{2},0,\tfrac{1}{2})$, such that also at the
TRIMs Z and X different eigenvalues are paired. 
Since $\Gamma$ does not host a Dirac point, only the TRIM U
remains to be considered. At U the inversion commutes with the
mirror symmetry $M_{010}(0, \tfrac{1}{2}, 0)$, and thus the bands
can be labeled simultaneously by the eigenvalues of both
symmetries.
While the mirror symmetry $M_{001}(\tfrac{1}{2},0,\tfrac{1}{2})$
commutes with inversion at U, it anticommutes with $M_{010}(0,
\tfrac{1}{2}, 0)$. The composite symmetry $\mathcal{T}
M_{001}(\tfrac{1}{2},0,\tfrac{1}{2})$ squares to $-1$ at U and
relates the mirror eigenvalues as $(+,+)$ and $(-,-)$. 
Since time-reversal symmetry by itself pairs $(+,-)$, there are four distinct states.
All four states carry the same inversion eigenvalue, because it is
invariant under both $M_{001}(\tfrac{1}{2},0,\tfrac{1}{2})$ and
time reversal, while they relate the eigenvalues of $M_{010}(0,
\tfrac{1}{2}, 0)$.
These four orthogonal states form the Dirac point at U. In
summary, for SG~62 we have found the inversion eigenvalues for all
Dirac points at TRIMs with general arguments.

We can thus calculate the weak topological invariant for SG~62
using that $\delta_{\text{X}} = \delta_{\text{Y}} =
\delta_{\text{Z}} = \delta_{\text{T}} = \delta_{\text{S}} =
\delta_{\text{R}} = \xi_{2}(X) \xi_{4}(X) = \pm 1 (\mp 1) = -1$
and $\delta_{\text{U}} = \xi_{2}(U) \xi_{4}(U) = +1$, which is
true for both possible eigenvalues $\xi_{2}(U)= \xi_{4}(U) = \pm
1$. One concludes that the weak invariant of systems in SG~62 is
fixed by symmetry, whereas the strong topology remains material
dependent, because the order of bands at $\Gamma$ is not fixed. 
The latter would become relevant if there is a symmetry breaking
perturbation that gaps the enforced nodal line (R-U-X;S)(4) such
that the system becomes insulating at the filling $N = 2 +
4\mathbb{N}_0$. Thus, the weak topological invariant on any plane
that contains U, but neither $\Gamma$ nor the nodal line, is
always nontrivial.
This argument is independent of the specific order of bands and
thus any material will exhibit the corresponding surface states in
the gap that is also crossed by the nodal line (R-U-X;S)(4). Let
us consider for example the plane ZURT, where the invariant is
given by 
$(-1)^\nu = \delta_{\text{Y}} \delta_{\text{T}} \delta_{\text{R}}
    \delta_{\text{S}} = -1$
and thus $\nu = 1$ indicates a nontrivial subsystem.
Due to the nonzero topological invariant $\nu$, there are surface
states for any termination that truncates the plane ZURT, given
that the projection of the bulk bands into the surface BZ does not
close the gap. For example, for the (100) surface there has to be
a surface state that crosses a bulk gap between $Z$ and
$T$ in the surface BZ, while the bulk bands of the nodal
line only close the gap in the vicinity of $Y$.
We discuss Ir$_2$Si as a material realization for SG~62 in
Sec.~\ref{Sec_Ir2Si}.

The described process for SG~62 also applies analogously to SGs~56
and 61, where SG~61 is a special case because there is no Dirac
point at R. Nevertheless, the weak invariant is independent of
the band order, because without SOC at R there is only one
Dirac point per minimal set of connected bands, for which all
inversion eigenvalues are identical. 
With SOC this degeneracy of identical inversion eigenvalues splits
into four bands with the same representation regarding inversion.
We have listed the TRIMs where identical inversion eigenvalues are
paired and denoted the resulting nontrivial planes in
Table~\ref{SGs_inv_soc}.

In the following we extend the previous approach to consider
cases, where the order of inversion eigenvalues is fixed without
SOC, which leads to a nontrivial weak invariant with SOC as long
as SOC is sufficiently small, such that the bands are not
inverted. We shall consider SG~52, for which by similar arguments
as before the inversion eigenvalues for the occupied bands must
lead to 
$\delta_{\text{X}} = \delta_{\text{Y}} = \delta_{\text{Z}}
    = \delta_{\text{U}} = \delta_{\text{S}} = \delta_{\text{R}}
    = \xi_{2}(X) \xi_{4}(X) = \pm1 (\mp 1) = -1$
and $ \delta_{\text{T}} = 1$ without SOC. 
Once SOC is added to the system, the Dirac point at T splits into
two bands, but as long as no band inversion occurs the value of
$\delta_{\text{T}}$ does not change. Note that this induced
arrangement of inversion eigenvalues is less stable than for
SG~61, where a set of connected bands inherits only identical
inversion eigenvalues from the case without SOC.
In other words, for SG~52 an exchange of energy levels within a set
of connected bands may trivialize the $\mathbb{Z}_2$ invariant,
whereas for SG~61 such an exchange would not affect the topology.
The topological invariant $\nu$ induced by SG~52 is $\nu = 1$ for
any plane that contains T but neither $\Gamma$ nor the movable
Dirac point on X-S(4).
An analogous argument also holds for SG~60.

To show the $\mathbb{Z}_2$ topology inherited from the
case without SOC more clearly, we discuss the material
Sr$_2$Bi$_3$ corresponding to SG~52 in Sec.~\ref{Sec_Sr2Bi3}.
In Sec.~\ref{Sec_Ir2Si} 2 we present a material in SG 62, namely
\ce{Ir2Si}, which has a weak $\mathbb{Z}_2$ invariant enforced by symmetry
alone.

\subsubsection{Material example: \ce{Sr2Bi3}}
\label{Sec_Sr2Bi3}

\ce{Sr2Bi3} crystallizes in SG 52 and was prepared by
melting the elements with a long, subsequent annealing
step~\cite{Merlo1994}. It is very prone to oxidation in air, as
expected based on the Bi$-$Bi bonding. Growth of large single
crystals may be challenging as it appears to be incongruently
melting with a small exposed liquidus~\cite{Smith2017}.

\begin{figure}
\sidesubfloat[]{
\includegraphics[width=.9\linewidth,valign=t]{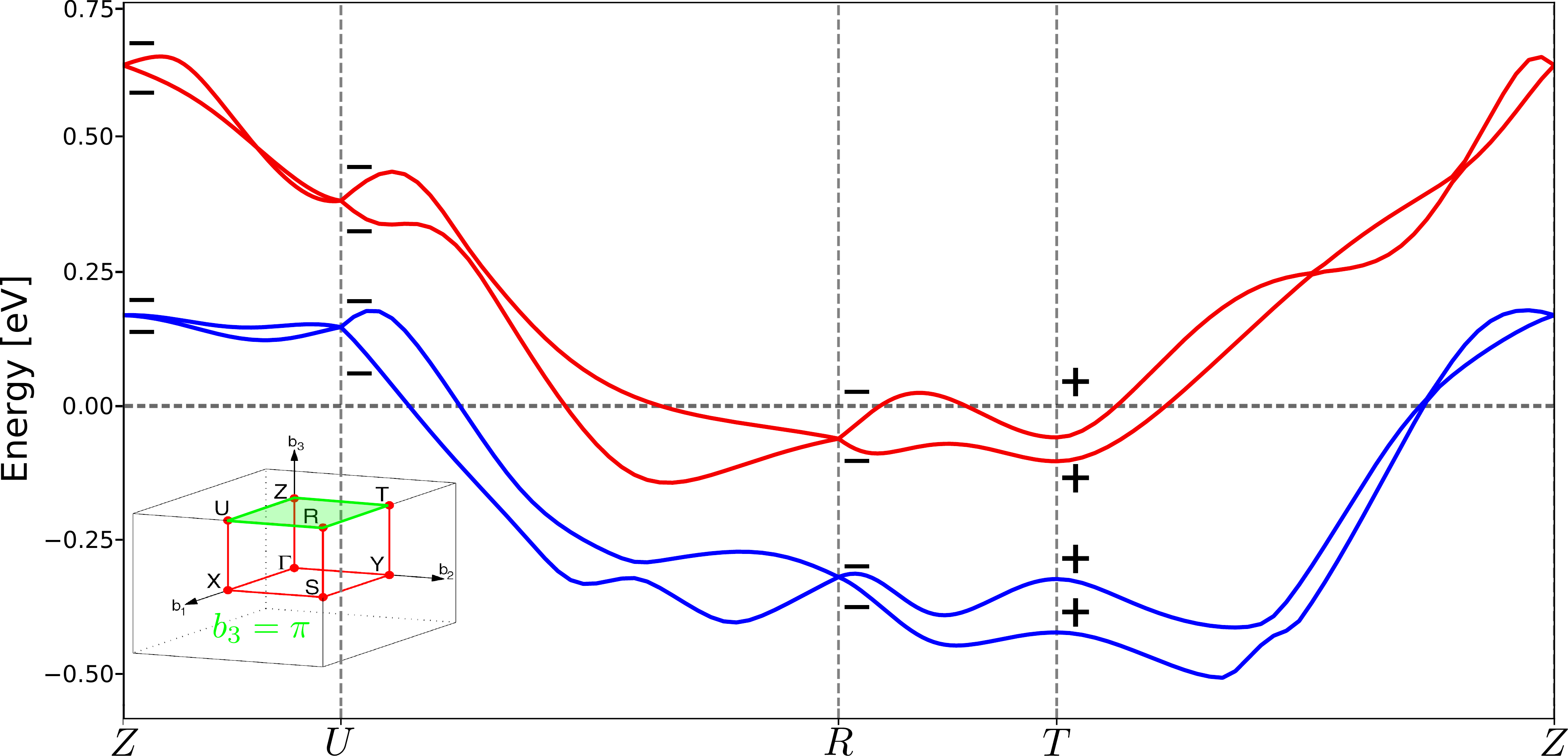}
\label{sr2bi3_irreps}
}\vspace{0.3cm}

\sidesubfloat[]{
\includegraphics[width=.9\linewidth,valign=t]{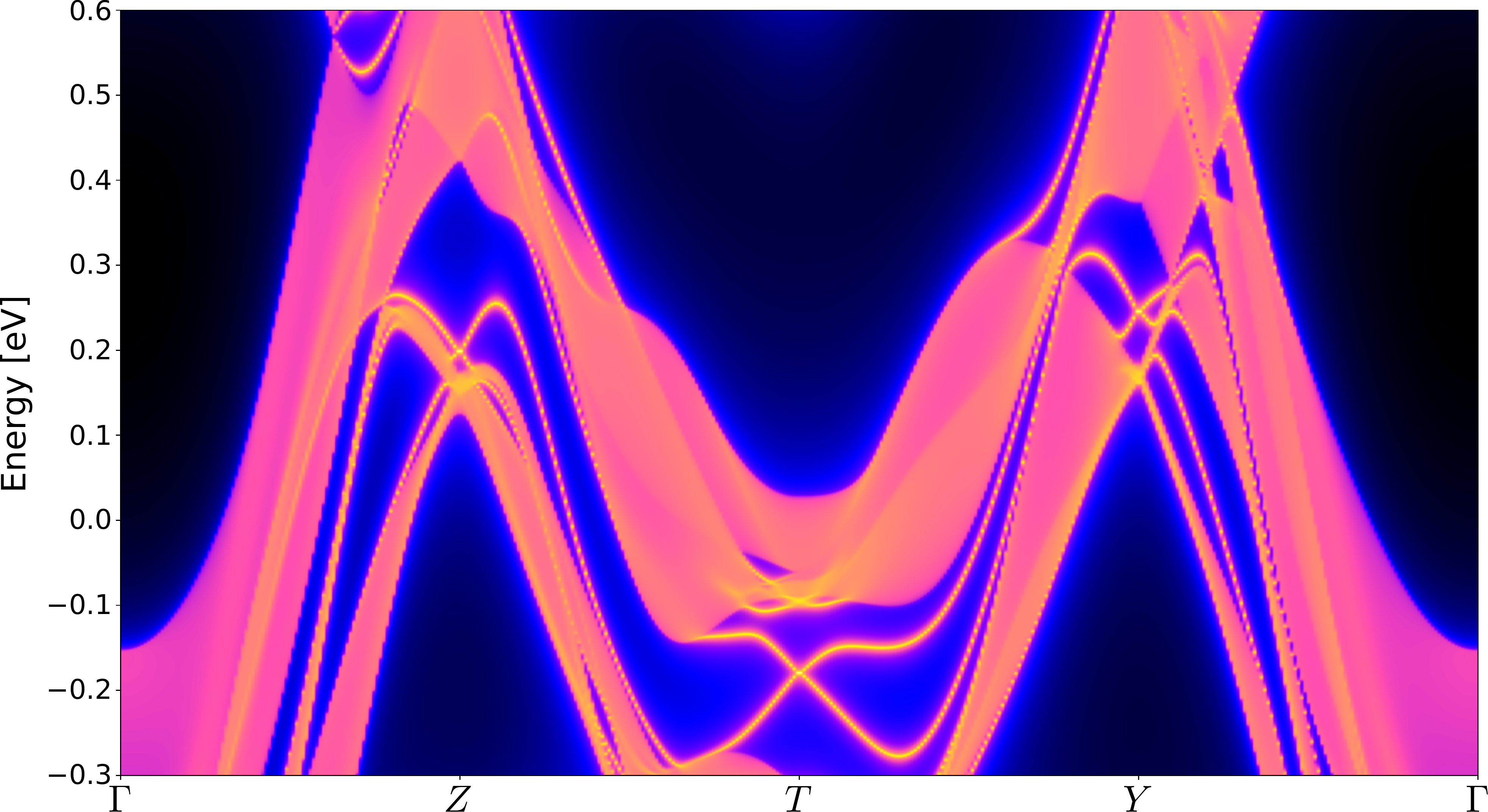}
\label{sr2bi3_SDOS}
}
\caption{
(a) Bulk bands of \ce{Sr2Bi3} in the $k_z=\pi$ plane with the sum
of inversion eigenvalues for each set of band at all TRIMs in the
plane. 
(b) Surface density of states on the $(100)$-surface along a high-symmetry path in the surface BZ.
The nontrivial
plane maps on the line Z-T, see inset in~(a),
with symmetry enforced surface states crossing the projected bulk gap.
\label{Sr2Bi3_Fig}
}
\end{figure}

Projecting the bulk bands in $[100]$ direction preserves the gap
between conduction and valence bands in the ZURT plane.
The nontrivial topology in that plane enforces surface states
crossing that gap in the projection of the plane, see
Fig.~\ref{Sr2Bi3_Fig}. Because time-reversal symmetry is
preserved, the surface states 
form a Dirac point at TRIMs.

\subsubsection{Material example: \ce{Ir2Si}}
\label{Sec_Ir2Si}
\begin{figure}
\sidesubfloat[]{
\includegraphics[width=.90\linewidth,valign=t]{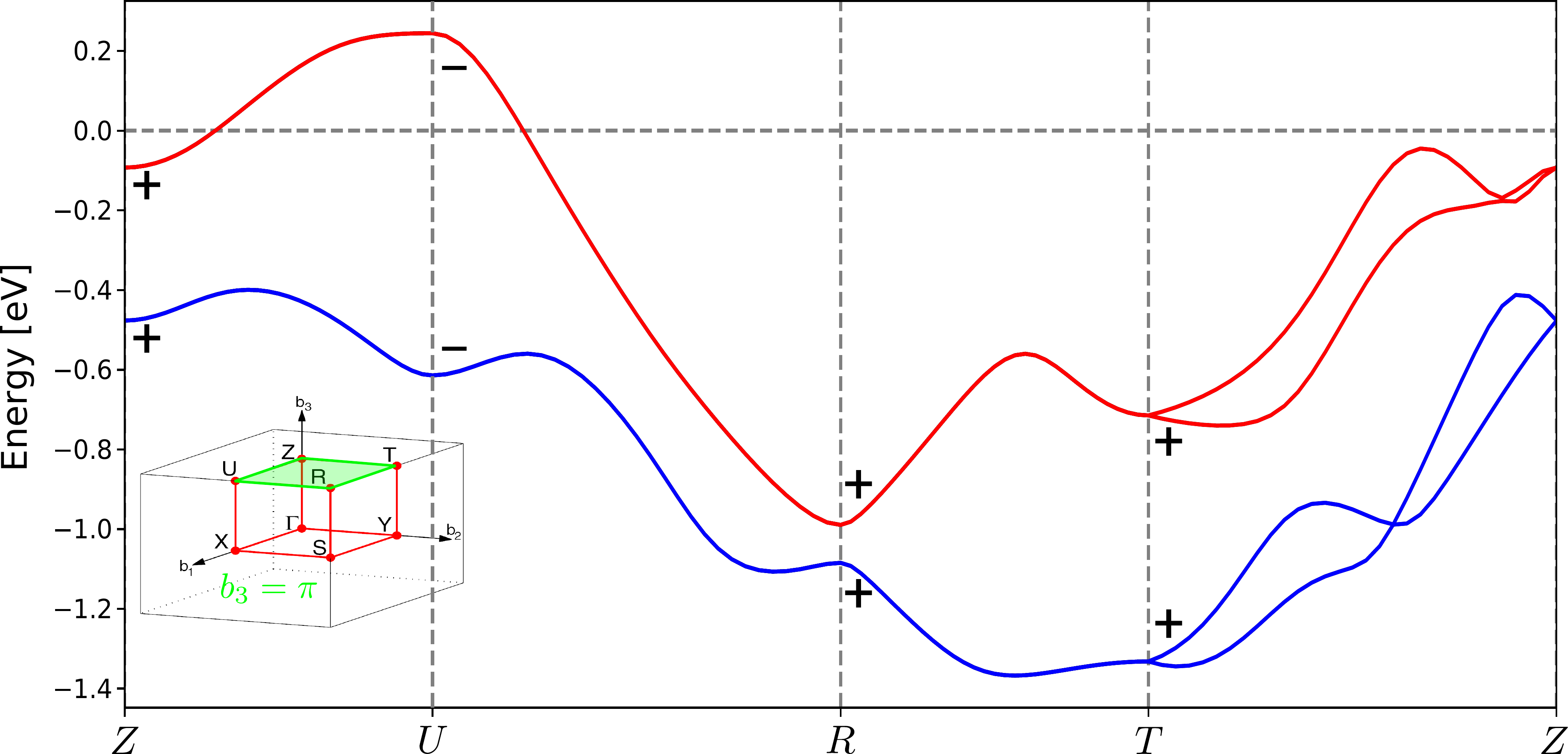}
\label{ir2si_irreps}
}\vspace{0.3cm}

\sidesubfloat[]{
\includegraphics[width=.90\linewidth,valign=t]{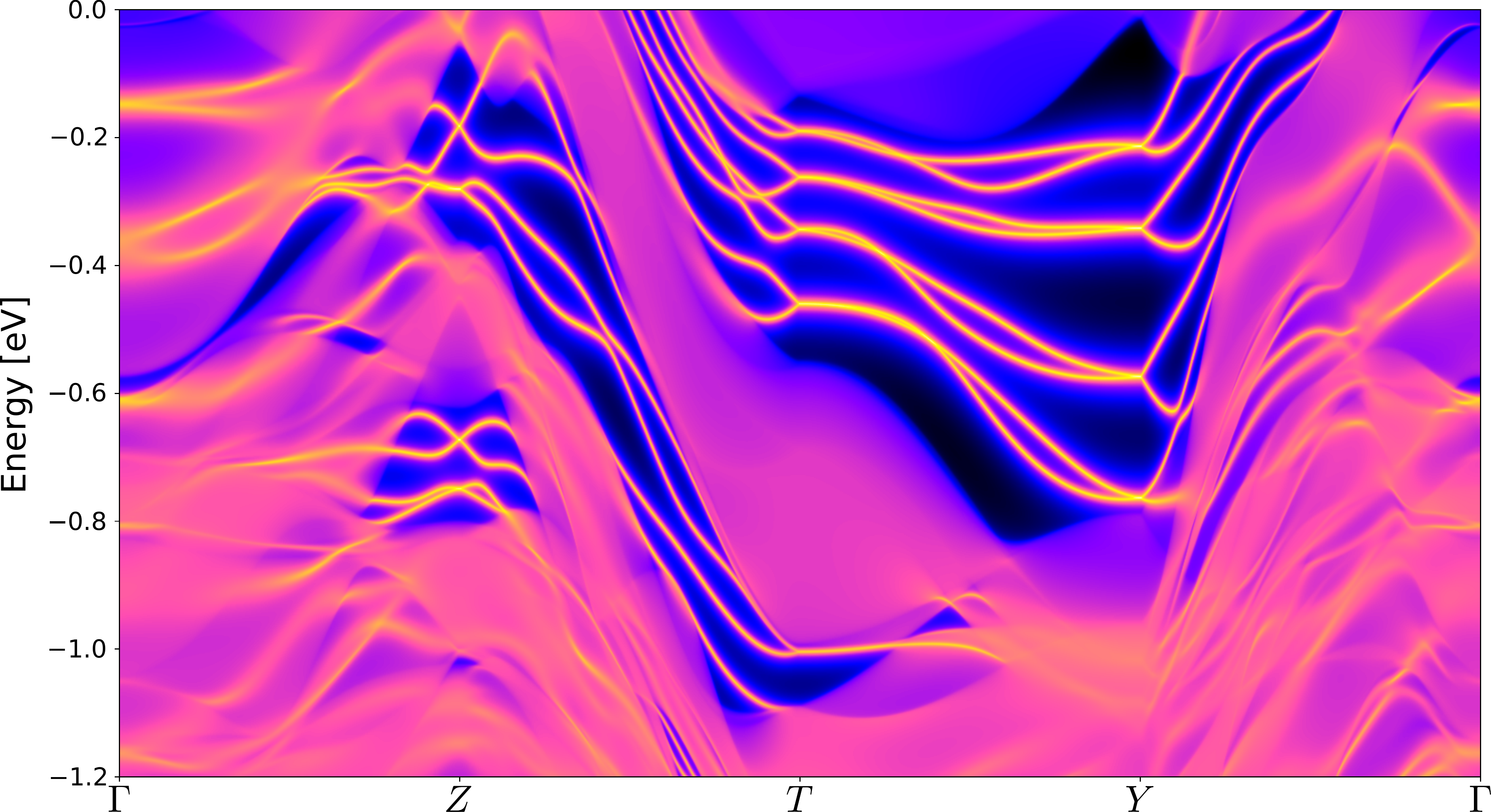}
\label{ir2si_SDOS}
}
\caption{
(a) Bulk bands of \ce{Ir2Si} in the $k_z=\pi$ plane with the sum
of inversion eigenvalues for each set of band at all TRIMs in the
plane. 
(b) Surface density of states  on the $(100)$-surface, where the nontrivial
plane maps on the line Z-T, see inset in~(a).
A pair of time-reversal symmetric surface states connect the bands below the
projected bulk gap to the upper ones.
}
\label{ir2si_Fig}
\end{figure}

Although some ternary iridium silicides have attracted attention
for superconductivity or heavy fermion behavior, very little is
known about the binary \ce{Ir2Si} crystallizing in SG 62.
The initially reported structure~\cite{Schubert1960} has been
corroborated at least once~\cite{Finnie1962}.
In the first four bands below the Fermi energy there is a gap
between pairs of bands in the $k_z=\pi$ plane with symmetry enforced nontrivial
$\mathds{Z}_2$ invariant, shown in
Fig.~\ref{ir2si_Fig}\subref{ir2si_irreps}. The bulk gap of this plane is
preserved for a (100) termination and the nontrivial topology enforces
a surface state traversing the gap on the line Z-T, see
Fig.~\ref{ir2si_Fig}\subref{ir2si_SDOS}. There are several
additional surface states forming Dirac cones in the bulk gaps.
Nevertheless, a clear signature of the nontrivial topology is the
surface band that detaches from the lower bands at the center of
the line Z-T, crosses the gap and connects to higher bands. In the
process the partners of Kramers pairs at TRIMs change when
comparing Z and T, which would not be the case for trivial surface
states.

\section{Conclusion}
\label{sec_VII}

In summary, we have classified all symmetry-enforced band
topologies in orthorhombic crystals with time-reversal symmetry,
both in the presence and absence of spin-orbit coupling. We have
found a vast number of different symmetry-enforced band crossings,
as well as enforced weak $\mathbb{Z}_2$ invariants
(Tables~\ref{SGs_noinv_nosoc}, \ref{SGs_noinv_soc},
\ref{SGs_inv_nosoc}, and \ref{SGs_inv_soc}). Weyl points exist in
the absence of inversion and are enforced to occur by Kramers
theorem (or generalizations thereof), by off-centered rotations
(Sec.~\ref{Sec_Kramers}), or by screw rotations
(Secs.~\ref{Sec_movable_Weyl}  and \ref{mm2_Weyls}). Fourfold
double Weyl points with $|\mathcal{C}|=2$ exist in the presence of
several screw rotations combined with time-reversal symmetry (Sec.
IV D). Dirac points at high-symmetry points are enforced by the
combination of inversion with time-reversal
(Sec.~\ref{mmm_Diracs}), while movable Dirac points arise in the
presence of screw rotations together with space-time inversion
$\mathcal{PT}$ (Sec.~\ref{Sec_movable_Diracs}).

Line degeneracies can be enforced by orthorhombic symmetries as
well. For example, movable Weyl lines with hourglass dispersion
are generated by glide mirror symmetries
(Sec.~\ref{Sec_movable_lines}). These Weyl lines can form nodal
chains if several glide mirror symmetries with orthogonal mirror
planes are present (Sec.~\ref{Sec_nodal_chain}). We find that such
nodal chains exist in the bands of CuIrB near the Fermi level.
Almost movable lines occur in the presence of mirror symmetries
whose mirror symmetries contain TRIMs without any additional
symmetries (Sec.~\ref{Sec_almost_movable_lines}). Movable Dirac
lines are enforced by mirror symmetries that are off-centered with
respect to space-time inversion $\mathcal{PT}$
(Sec.~\ref{Sec_movable_Dirac_lines}).

Finally, orthorhombic symmetries can also enforce nodal planes,
i.e., two-dimensional degeneracies of the bands at the boundary of
the BZ. These arise in chiral SGs and are enforced by the
combination of screw rotations with time-reversal symmetry
(Sec.~\ref{sec_nodal_planes}). Interestingly, in SG 19 with SOC
there are three such nodal planes (i.e., a nodal plane trio) which
are sources (or sinks) of quantized Berry flux, which is absorbed
(emitted) by a single Weyl point at $\Gamma$. 

Besides band degeneracies, orthorhombic symmetries can also
enforce nontrivial $\mathbb{Z}_2$ topologies in two-dimensional, gapped
subspaces of the three-dimensional BZ. The corresponding $\mathbb{Z}_2$
indices are computed from the inversion eigenvalues at the TRIMs
within the two-dimensional subspaces, whose values are fixed by
symmetry in SGs 56 and 62, as well as in SG 61 provided that
elementary band representations with different inversion
eigenvalues do not mix.  Similarly, we find that also SGs 52 and
60 enforced weak $\mathbb{Z}_2$ invariants, however, only in the case of weak
SOC that does not invert the bands. We have identified two
materials, namely \ce{Ir2Si} (SG 62) and \ce{Sr2Bi3} (SG 52), with such
symmetry-enforced weak invariants, which lead to surface helical
modes of topological origin.

Interestingly, there are also some orthorhombic SGs, where the
number of Weyl points of a given band pair can be as low as two.
For example, in SGs 36, 45, and 46 with SOC there can be only two
Weyl points at two TRIMs, while the other TRIMs are part of nodal
planes or nodal lines. Remarkably, in SG 24 without SOC a given
band pair can have only four Weyl points, while the TRIMs are
non-degenerate. In SG 19 without SOC, there are only three Weyl points
for an even number of filled bands, one of which is a fourfold
double Weyl point.

Furthermore, we provide an example of how to use our purely
symmetry-based analysis as a guide to identifying interesting new
materials.  The screening criteria mentioned in Sec.~\ref{sec_III}
can be used as
a starting point for identifying candidates in known compounds, which
in turn might act as a starting point for synthesis of related new
compounds of the same SG. By explicitly including spinless band
structures in our analysis, our paper also applies to phonon bands
and can serve as a blueprint for constructing synthetic materials,
e.g., photonic crystals, acoustic metamaterials or electric
circuit networks. 
The methods used have also been applied to the hexagonal,
trigonal, and tetragonal
SGs~\cite{hexagonals_PhysRevMaterials.2.074201,trigonals_PhysRevMaterials.3.124204,hirschmann2021symmetryenforced}
and can be extended to further SGs and, in a similar manner, to
magnetic SGs.

\begin{acknowledgments}
The authors thank K. Alpin and W. Yau for useful discussions. 
D.H.F. gratefully acknowledges financial support from the
Alexander von Humboldt Foundation.
\end{acknowledgments}

\appendix

\section{Additional band structure calculations}
\label{app_A}

For all the example materials we present in this Appendix the band structure
along the full high-symmetry path, aligned with the symmetries.
We use the standard path for all primitive SGs.
Note that the bath for CuIrB extends beyond the first full BZ
to show the full line T$_1$-Y via A$_1$, where the nodal chain
is formed, see Fig.~\ref{CuIrB_path}.

\begin{figure*}
\subfloat[\ce{Pd7Se4}]{
\includegraphics[width=.45\textwidth]{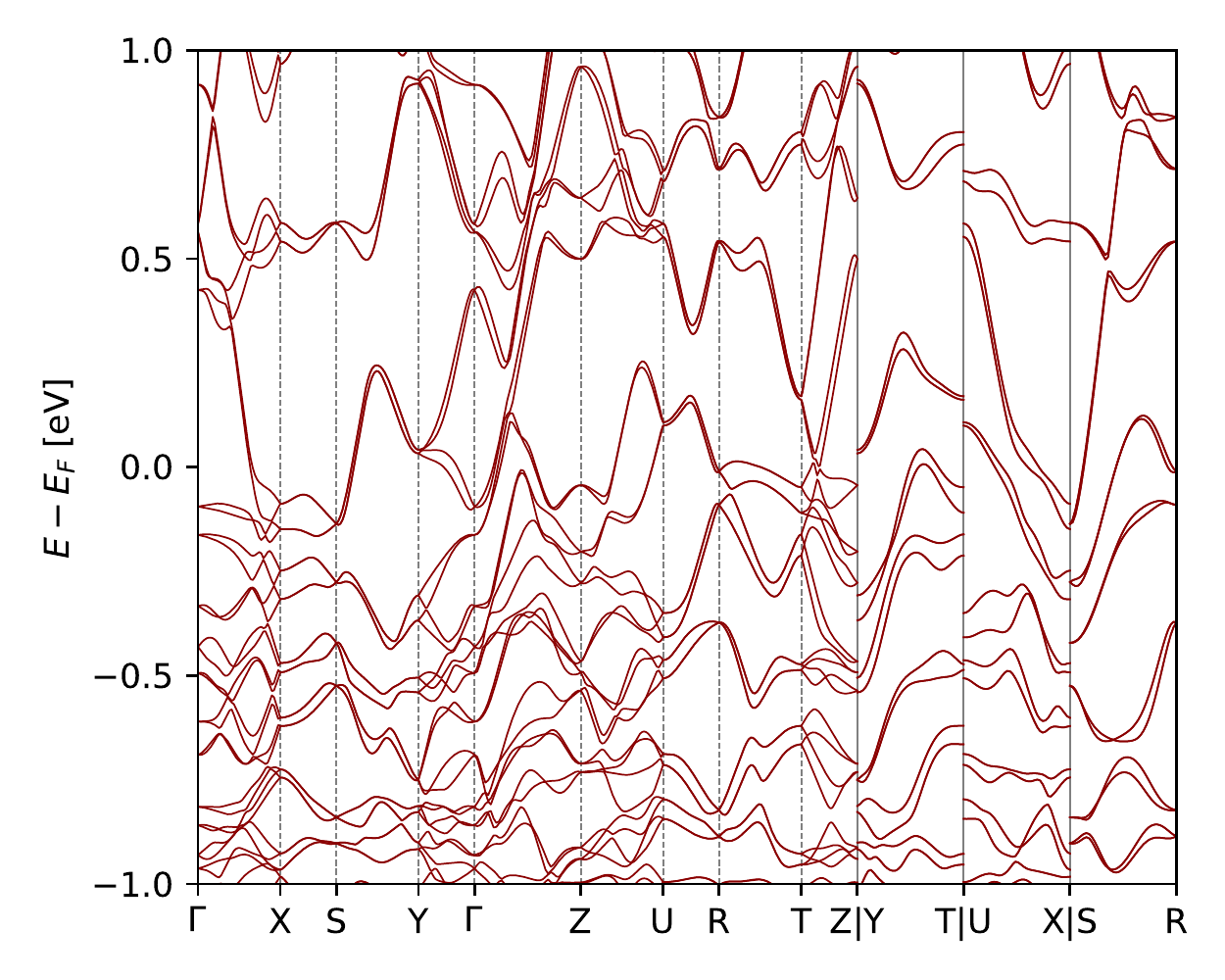}
}
\subfloat[\ce{Ag2Se}]{
\includegraphics[width=.45\textwidth]{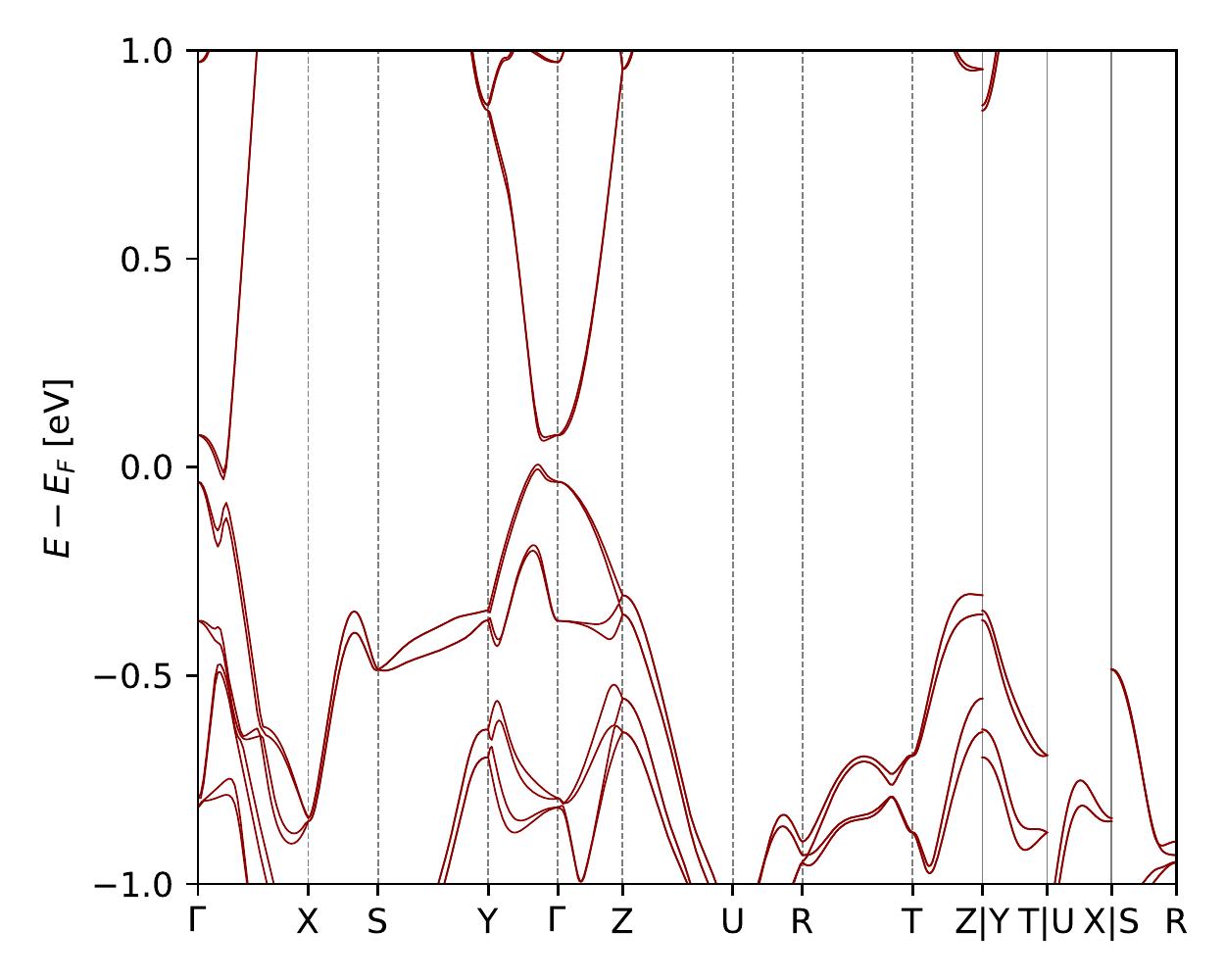}
}
 
\subfloat[\ce{AuTlSb}]{
\includegraphics[width=.45\textwidth]{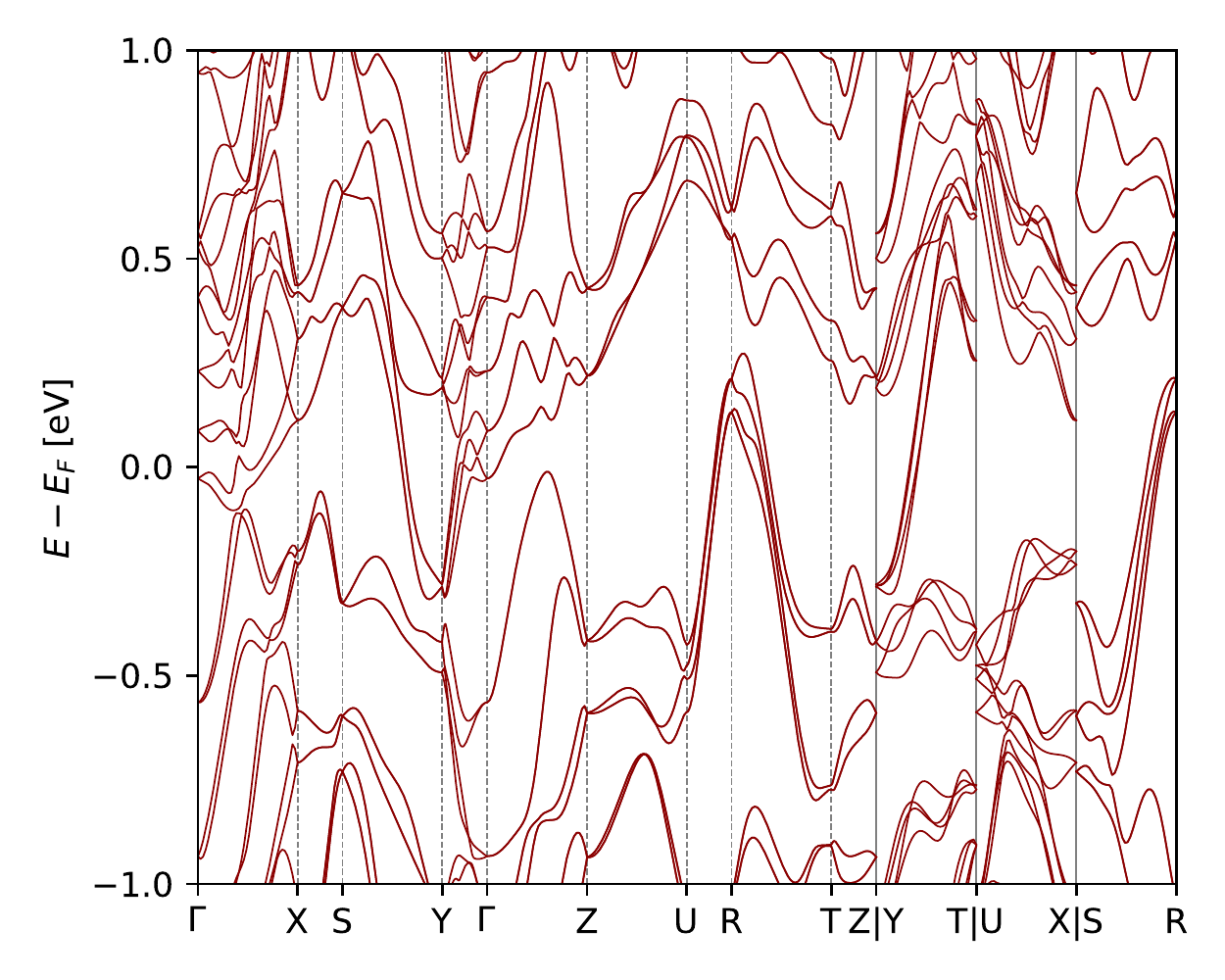}
}
\subfloat[\ce{CuIrB}]{
\includegraphics[width=.45\textwidth]{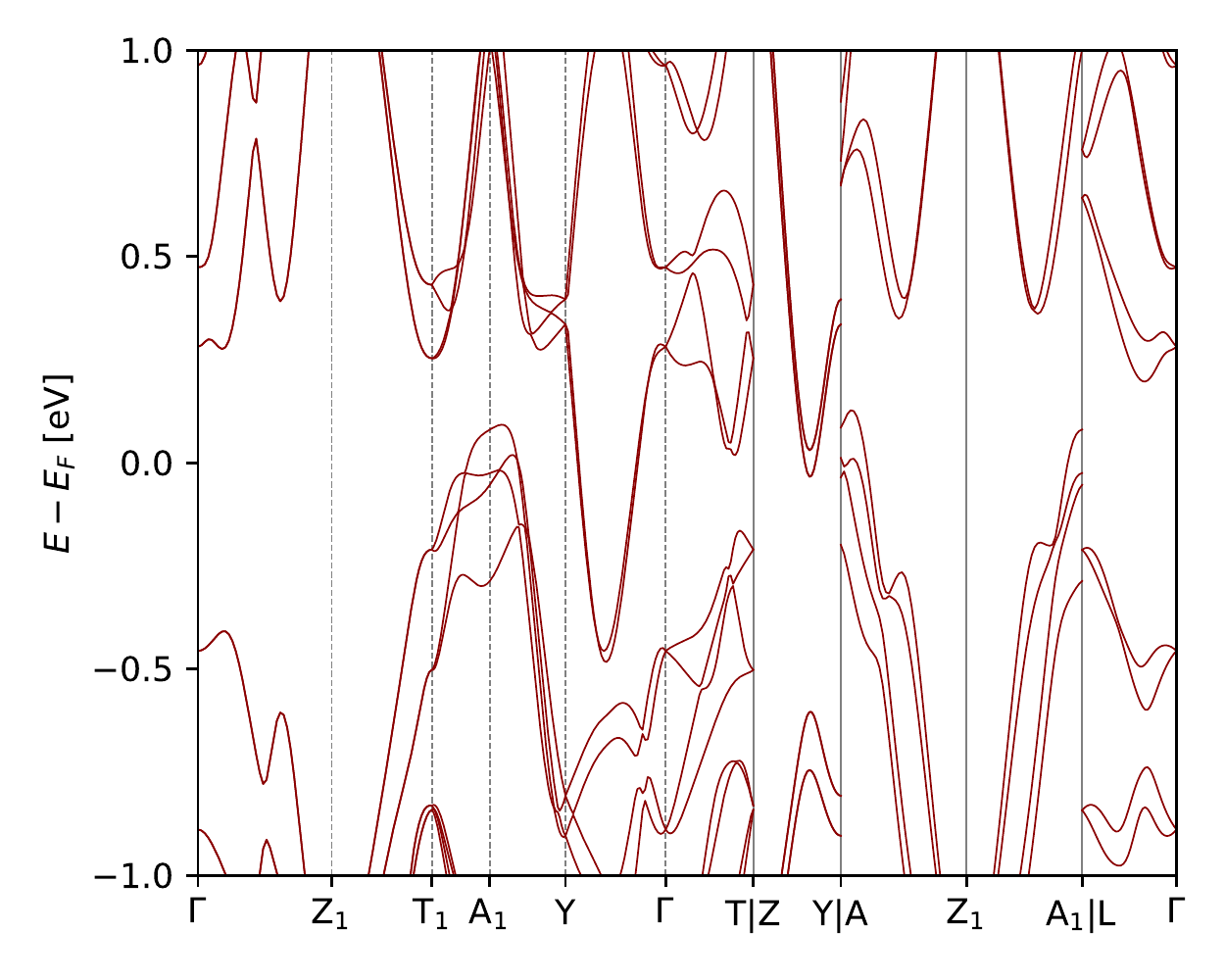}
\label{cuirb_bands}
}

\subfloat[\ce{Sr2Bi3}]{
\includegraphics[width=.45\textwidth]{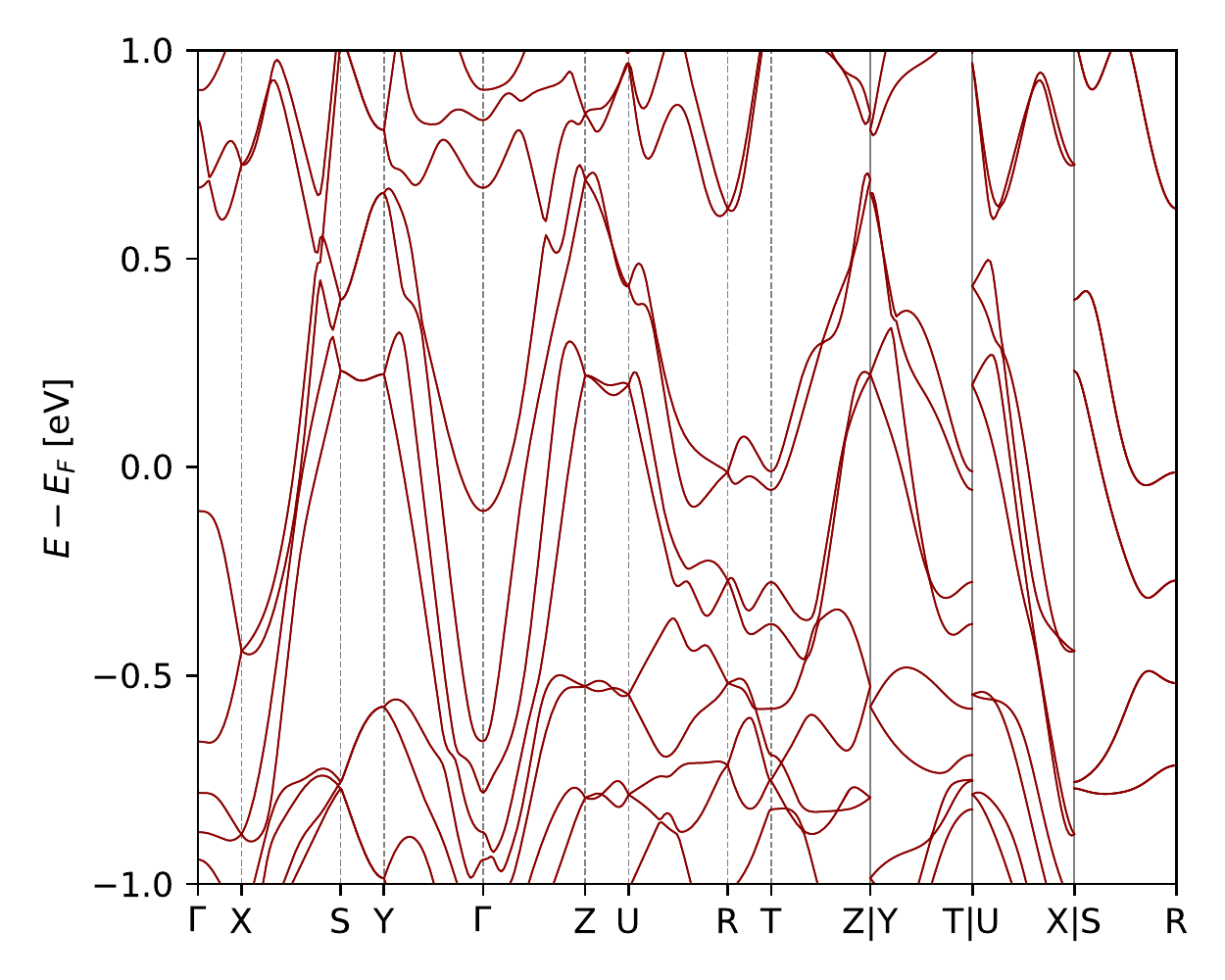}
}
\subfloat[\ce{Ir2Si}]{
\includegraphics[width=.45\textwidth]{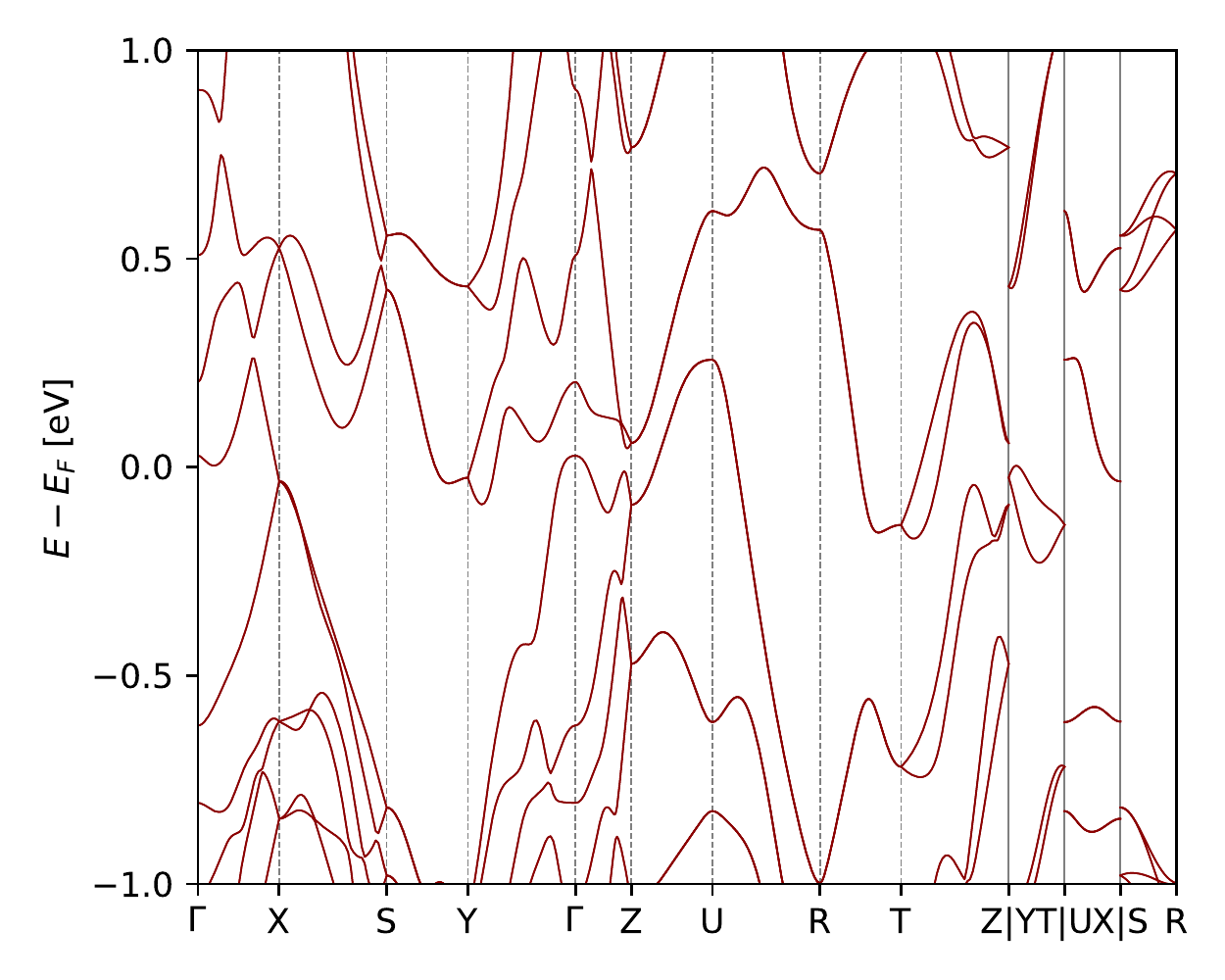}
}
\caption{
DFT band structures including SOC along the high-symmetry paths
(see Fig.~\ref{BZ_body} and Fig-~\ref{CuIrB_path}) for all
examples discussed in the main text.
}
\label{band_structures}
\end{figure*}

\begin{figure}
\includegraphics[width=.5\linewidth]{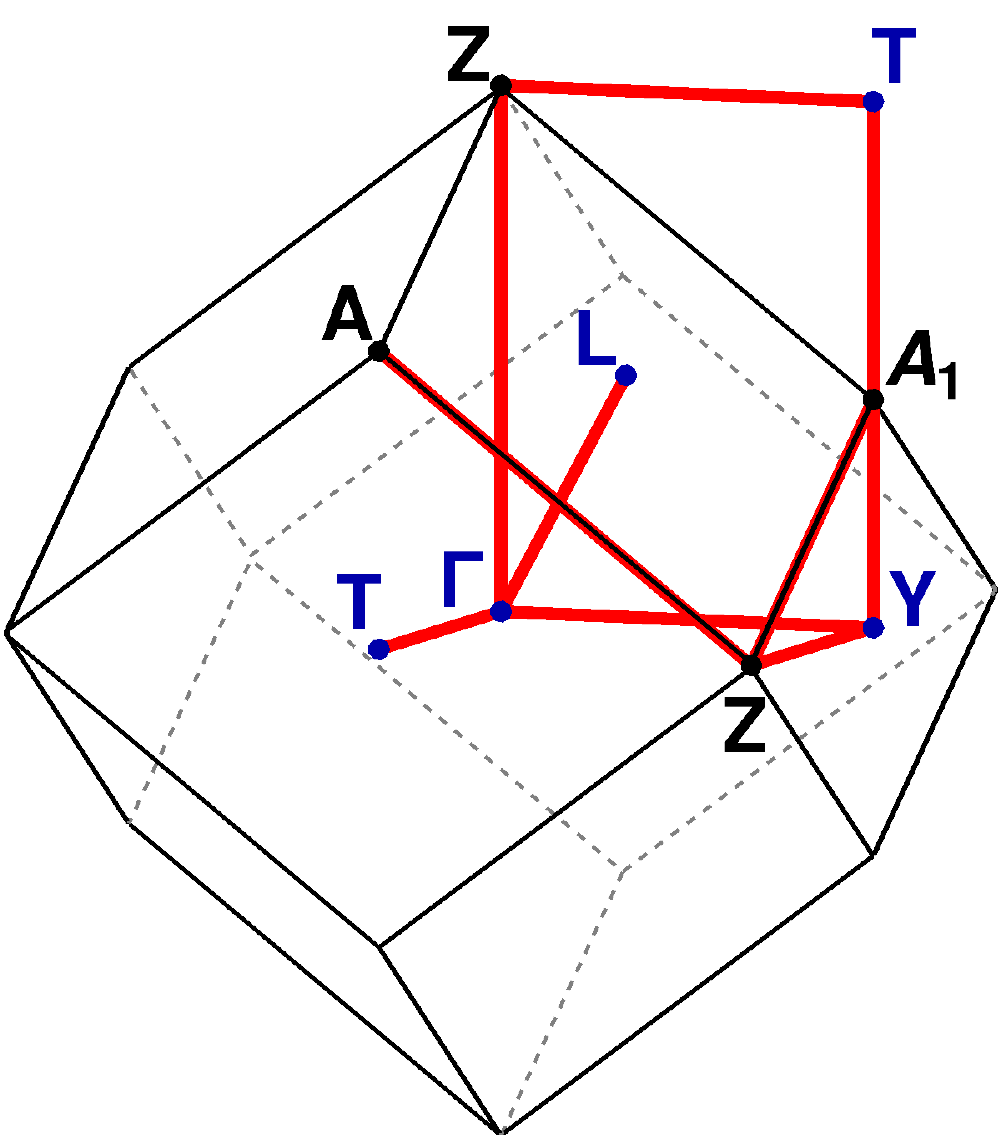}
\caption{
High-symmetry path for CuIrB in SG 43 used in
Fig.~\ref{band_structures}(d).  
}
\label{CuIrB_path}
\end{figure}

\section{Low-energy Hamiltonians for fourfold degenerate points}
\label{app_B}

Here we give low-energy Hamiltonians for the fourfold double Weyl
points with Chern number $\pm2$, and the fourfold degenerate
points formed by intersecting nodal lines, either with an
additional hourglass nodal line in between or as true point
defects.

\subsection{Fourfold double Weyl point}

To define the Hamiltonians describing fourfold double Weyl points, we use the Pauli matrices
$\sigma_i$ and $\tau_i$ ($i=0,x,y,z$), acting in spin and orbital
space, respectively.

The little group of a fourfold double Weyl point
consists of the three twofold rotations of the chiral orthorhombic SGs
and time-reversal symmetry. Their representations can be given as
\begin{IEEEeqnarray}{rCl}
U_{2_{001}} &=& \mathrm{i}\tau_0\tau_z, \\
U_{2_{010}} &=& \tau_0\tau_y, \\
U_{2_{100}} &=& \tau_0\tau_x, \\
U_\mathcal{T}\mathcal{K} &=& \mathrm{i}\tau_y\sigma_x \mathcal{K}.
\end{IEEEeqnarray}
Since two rotations are screw rotations, their eigenvalues are
$\pm1$, consistent with the requirement of squaring to the
identity.

The Hamiltonian up to linear order in $\vb{q}$ around the TRIM
$\vb{K}$, i.e., $\vb{q}=\vb{k}-\vb{K}$, invariant under the above
symmetries is restricted to the form
\begin{IEEEeqnarray}{rCl}
H(\vb{q}) &=&
     v_x q_x \tau_z\sigma_x
    +v_y q_y \tau_z\sigma_y
    +v_z q_z \tau_0\sigma_z
\nonumber \\ &&
    +\lambda_x q_x\tau_x\sigma_x
    +\lambda_y q_y\tau_x\sigma_y.
\end{IEEEeqnarray}
The real parameters $v_i$ define the velocities
or slope of the individual Weyl points. The splitting
from SOC is introduced from the real parameters $\lambda_i$,
which renormalize the velocities $v_i$.
The Chern number of this Hamiltonian is $\pm2$ and it reproduces
the twofold degenerate nodal planes for $q_x=0$ or $q_y=0$.

\subsection{Fourfold degenerate points from multiple intersecting nodal lines}
\label{App_4fold_mm2}

As discussed in Sec.~\ref{mm2_fourfold_points}, fourfold
degeneracies in the rhombic pyramidal SGs are formed by
intersecting nodal lines and can either be point-like or have an
hourglass nodal line running through the fourfold degeneracy.

For the fourfold degenerate point, the representations
for the symmetries
at the TRIM can be given by
\begin{IEEEeqnarray}{rCl}
U_{2_{001}} &=& -\mathrm{i}\tau_0\sigma_z, \\
U_{M_{010}} &=& \tau_0\sigma_y, \\
U_\mathcal{T}\mathcal{K} &=& \mathrm{i}\tau_y\sigma_x \mathcal{K},
\end{IEEEeqnarray}
and the linearized Hamiltonian is restricted to the form
\begin{IEEEeqnarray}{rCl}
H(\vb{q}) &=&
     v_z q_z \tau_z\sigma_0
    +v_x q_x \tau_z\sigma_y
    +v_y q_y \tau_z\sigma_x
\nonumber \\ &&
    +\lambda_x q_x\tau_x\sigma_y
    +\lambda_y q_y\tau_x\sigma_x
    +\lambda_z q_z\tau_x\sigma_0,
\label{mm2_4fold_lowE_gapped}
\end{IEEEeqnarray}
with the real parameters $v_i$ and $\lambda_i$.
The gap between the second and third
band closes only at the TRIM, i.e., for $\vb{q}=0$, but the Chern
number vanishes.
In the case of an hourglass nodal line, the representations are 
\begin{IEEEeqnarray}{rCl}
U_{2_{001}} &=& \tau_0\sigma_z, \\
U_{M_{010}} &=& \tau_0\sigma_y, \\
U_\mathcal{T}\mathcal{K} &=& \mathrm{i}\tau_y\sigma_z \mathcal{K},
\end{IEEEeqnarray}
and the linearized Hamiltonian reads then
\begin{IEEEeqnarray}{rCl}
H(\vb{q}) &=&
     v_z q_z \tau_z\sigma_0
    +v_x q_x \tau_z\sigma_y
    +v_y q_y \tau_0\sigma_x
\nonumber \\ &&
    +\lambda_x q_x\tau_x\sigma_y
    +\lambda_z q_z\tau_x\sigma_0,
\label{mm2_4fold_lowE_hourglass}
\end{IEEEeqnarray}
again with purely real parameters $v_i$ and $\lambda_i$.
There are twofold degeneracies on the axes
$(q_x,0,0)$ and $(0,0,q_z)$ and an additional hourglass nodal line
in between the second and third band, running through $\vb{q}=0$..

\section{Minimal tight-binding model for SG 19}
\label{app_SG19}

SG 19 enforces all notable features of chiral SGs, namely
Kramers and movable Weyl points, fourfold double Weyl points and
topological nodal planes. Therefore, we present a minimal
tight-binding model to give a simple example and to demonstrate the
connection between spinless and spinful band structures.

The minimal number of sites is four per unit cell and we include
nearest neighbor hopping only. This implicitly excludes terms in
the Hamiltonian proportional to $\mathds{1}$, i.e., a uniform
dispersion for the bands and lets us focus on the splitting of 
bands.

\begin{widetext}
\begin{IEEEeqnarray}{rCl}
H(\vb{k}) &=& \mqty( H_0(\vb{k}) & \Lambda(\vb{k}) \\ \Lambda^\dagger(\vb{k}) & H_0^*(-\vb{k}) )
 \label{SG19_TB} \\
H_0(\vb{k}) &=&
\mqty(
0   & (t_0{+}t_1 \mathrm{e}^{+\mathrm{i}k_x})(1{+}\mathrm{e}^{\mathrm{i}k_z}) & (t_2{+}t_3\mathrm{e}^{\mathrm{i}k_y})(1{+}\mathrm{e}^{+\mathrm{i}k_x}) & (t_4{+}t_5\mathrm{e}^{+\mathrm{i}k_z})(1{+}\mathrm{e}^{\mathrm{i}k_y})   \\
    & 0                                                                   & (t_4\mathrm{e}^{-\mathrm{i}k_z}{+}t_5)(1{+}\mathrm{e}^{\mathrm{i}k_y}) & (t_2\mathrm{e}^{\mathrm{i}k_y}{+}t_3)(1{+}\mathrm{e}^{-\mathrm{i}k_x})   \\
    &                                                                     & 0                                                                   & (t_0\mathrm{e}^{-\mathrm{i}k_x}{+}t_1)(1{+}\mathrm{e}^{\mathrm{i}k_z})   \\
    & h.c.                                                                &                                                                     & 0                                                                     )
\\
\Lambda(\vb{k}) &=& \hphantom{+\mathrm{i}}
\mqty(
0                                                                        & (l_0{+}l_1\mathrm{e}^{+\mathrm{i}k_x})(1{+}\mathrm{e}^{+\mathrm{i}k_z}) & (l_2{+}l_3\mathrm{e}^{+\mathrm{i}k_y})(1{+}\mathrm{e}^{+\mathrm{i}k_x}) & (l_4{+}l_5\mathrm{e}^{+\mathrm{i}k_z})(1{-}\mathrm{e}^{+\mathrm{i}k_y})  \\
-(l_0{+}l_1\mathrm{e}^{-\mathrm{i}k_x})(1{+}\mathrm{e}^{-\mathrm{i}k_z}) & 0                                                                       & (l_4\mathrm{e}^{-\mathrm{i}k_z}{+}l_5)(1{-}\mathrm{e}^{+\mathrm{i}k_y}) &-(l_2\mathrm{e}^{+\mathrm{i}k_y}{+}l_3)(1{+}\mathrm{e}^{-\mathrm{i}k_x})  \\
-(l_2{+}l_3\mathrm{e}^{-\mathrm{i}k_y})(1{+}\mathrm{e}^{-\mathrm{i}k_x}) &-(l_4\mathrm{e}^{+\mathrm{i}k_z}{+}l_5)(1{-}\mathrm{e}^{-\mathrm{i}k_y}) & 0                                                                       & -(l_0\mathrm{e}^{-\mathrm{i}k_x}{+}l_1)(1{+}\mathrm{e}^{+\mathrm{i}k_z}) \\
-(l_4{+}l_5\mathrm{e}^{-\mathrm{i}k_z})(1{-}\mathrm{e}^{-\mathrm{i}k_y}) & (l_2\mathrm{e}^{-\mathrm{i}k_y}{+}l_3)(1{+}\mathrm{e}^{+\mathrm{i}k_x}) & (l_0\mathrm{e}^{+\mathrm{i}k_x}{+}l_1)(1{+}\mathrm{e}^{-\mathrm{i}k_z}) & 0 )
\\ 
    & & 
+\mathrm{i}  \mqty(
0                                                                        & (l_6{+}l_7\mathrm{e}^{+\mathrm{i}k_x})(1{+}\mathrm{e}^{+\mathrm{i}k_z}) & (l_8{+}l_9\mathrm{e}^{+\mathrm{i}k_y})(1{-}\mathrm{e}^{+\mathrm{i}k_x}) &-(l_{10}{+}l_{11}\mathrm{e}^{+\mathrm{i}k_z})(1{+}\mathrm{e}^{+\mathrm{i}k_y})  \\
-(l_6{+}l_7\mathrm{e}^{-\mathrm{i}k_x})(1{+}\mathrm{e}^{-\mathrm{i}k_z}) & 0                                                                       & (l_{10}\mathrm{e}^{-\mathrm{i}k_z}{+}l_{11})(1{+}\mathrm{e}^{+\mathrm{i}k_y}) &-(l_8\mathrm{e}^{+\mathrm{i}k_y}{+}l_9)(1{-}\mathrm{e}^{-\mathrm{i}k_x})  \\
-(l_8{+}l_9\mathrm{e}^{-\mathrm{i}k_y})(1{-}\mathrm{e}^{-\mathrm{i}k_x}) &-(l_{10}\mathrm{e}^{+\mathrm{i}k_z}{+}l_{11})(1{+}\mathrm{e}^{-\mathrm{i}k_y}) & 0                                                                       & (l_6\mathrm{e}^{-\mathrm{i}k_x}{+}l_7)(1{+}\mathrm{e}^{+\mathrm{i}k_z}) \\
(l_{10}{+}l_{11}\mathrm{e}^{-\mathrm{i}k_z})(1{+}\mathrm{e}^{-\mathrm{i}k_y}) & (l_8\mathrm{e}^{-\mathrm{i}k_y}{+}l_3)(1{-}\mathrm{e}^{+\mathrm{i}k_x}) &-(l_6\mathrm{e}^{+\mathrm{i}k_x}{+}l_7)(1{+}\mathrm{e}^{-\mathrm{i}k_z}) & 0 )
\nonumber 
\end{IEEEeqnarray}
\end{widetext}
Within each spin subspace, there are six independent real hopping
parameters $t_i$ in $H_0(\vb{k})$ and its time-reversal symmetric
copy $H_0^*(-\vb{k})$. Including SOC adds additional 12 real
parameters $l_i$ in the off-diagonal block $\Lambda(\vb{k})$.

\begin{figure}
\includegraphics[width=1.0\linewidth]{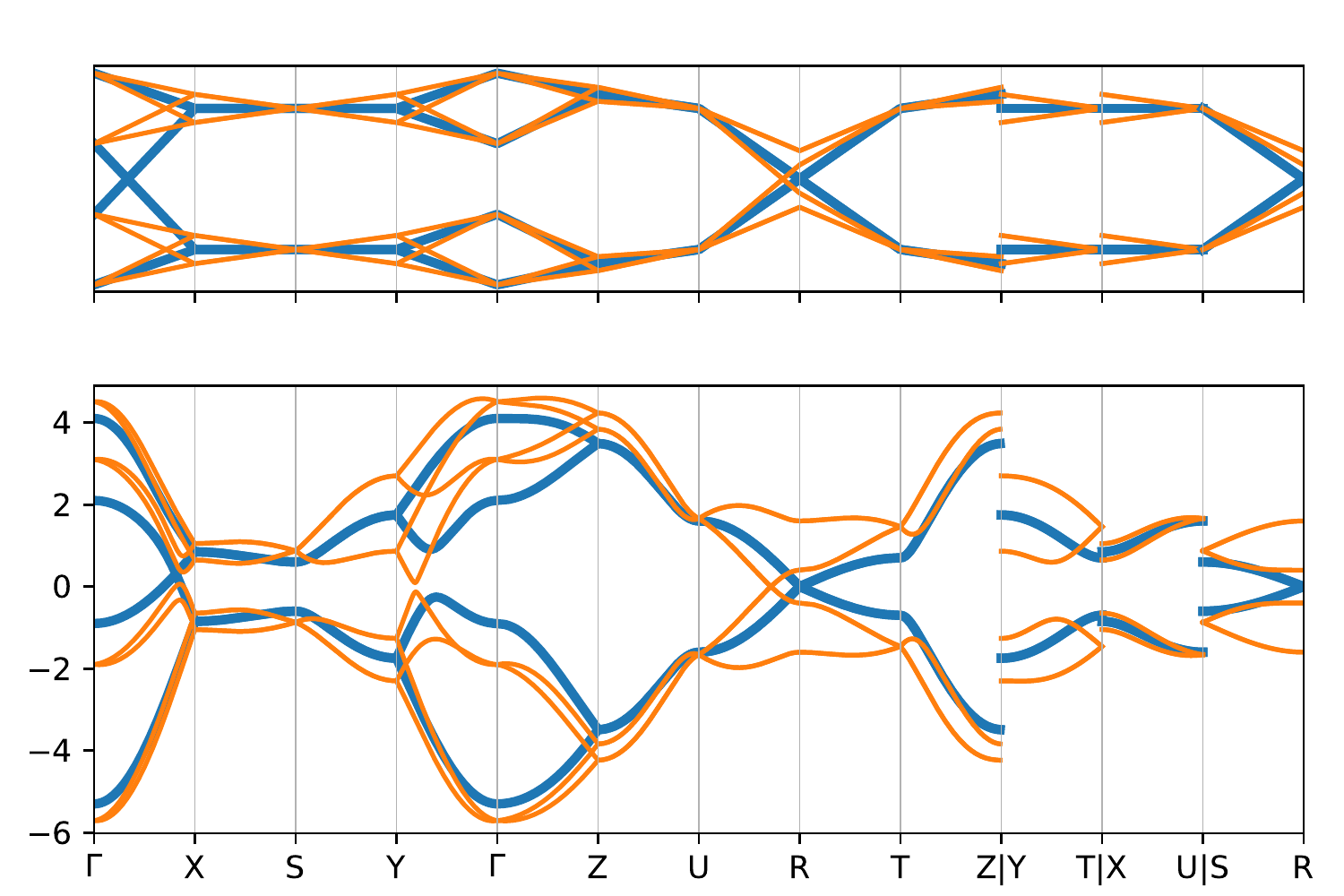}
\caption{
Elementary bands representations in SG 19 without SOC, i.e.,
$l_i=0$ (blue) and with strong SOC (orange). The upper bands are a
schematic band structure with minimal number of crossings, based
on compatibility relations alone. The lower set of bands are from
the minimal tight-binding model defined in Eq.~\eqref{SG19_TB}
with hopping parameters 
$t_0=-t_5=0.25$, $t_1=-t_4=0.55$, $t_2=-0.6$, $t_3=-0.95$,
and for the case with SOC
$l_0=l_1=l_7=0$, $l_2=-l_5=0$, $l_3=0.7$, $l_4=-l_8=0.2$,
$l_5=-0.3$, $l_9=l_{10}=-l_{11}=0.1$. Parameters were chosen
for maximal visibility of the features mentioned in the text.
}
\label{SG19_bands}
\end{figure}

In Fig.~\ref{SG19_bands} we present a schematic
and exemplary band structure along the standard high-symmetry path. 
The nodal planes can be seen in the twofold degenerate bands on
all paths except for the ones in the interior of the BZ, i.e., the
ones connecting to $\Gamma$. On one of these paths, a band
crossing needs to exist in the absence of SOC. In the example, it
can be observed on $\Gamma$-X. The TRIM R is fourfold degenerate.

Switching on SOC lifts spin degeneracy almost everywhere. The
nodal planes and TRIMs remain at least twofold degenerate, while
the degeneracy at R is now only twofold, the TRIMs S, U and T and
a movable point on the U-R axis remain fourfold degenerate.
Further Weyl points are introduced in the hourglass dispersion
along the rotation axes $\Gamma$-X, $\Gamma$-Y and $\Gamma$-Z.
Because of unbalanced Weyl points at $\Gamma$, the nodal planes
are topological, i.e., act as source or sink of Berry flux.

\FloatBarrier

%%%% REFS %%%%%%%%%%%%%%%%%%%%%%%%%%%%%%%%%%%%%%%%%%%%%%%%%%%%%%%%
\newpage
\bibliographystyle{unsrt}
\bibliography{orthorhombics_literature}
%%%%%%%%%%%%%%%%%%%%%%%%%%%%%%%%%%%%%%%%%%%%%%%%%%%%%%%%%%%%%%%%%%

\end{document}

\begin{center}\textbf{$-$ Supporting Information $-$ \\ Symmetry-enforced topological band crossings in orthorhombic crystals: \\
Classification and materials discovery} \\ \vspace{3mm}
Andreas Leonhardt, Moritz M. Hirschmann, Niclas Heinsdorf, \\ Xianxin Wu, Douglas H. Fabini, Andreas P. Schnyder\end{center}

\noindent The following table lists phases in both the Materials Project database and the ICSD (as of February 2021) from space groups 18, 19, 24, 30, 33, 34, 43, 52, 56, 57, 60, 61, 62, and 73 which have 50 sites or fewer in the primitive cell, which lie less than 50 meV~atom$^{-1}$ off the convex hull, and which have a GGA$-$PBE bandgap  less than 0.1~eV. Phases which have a spin-polarized density of states (DOS) stored in the Materials Project database are omitted, but other magnetic phases may be present. $N_{sites}$ = number of sites in the primitive cell, $\Delta E_{hull}$ = energy above the convex hull, $E_g$ = band gap, $Z_{max}$ = maximum atomic number. Heavy atom and light atom projected DOS (PDOS) fractions near $E_F$ are defined in the main text. ``no data'' in the PDOS columns indicates that no DOS calculation is stored in the Materials Project database. For space group 62, only the 15 phases with highest heavy atom PDOS fraction and 15 of the phases with the lowest heavy atom PDOS fraction are included.

\tiny
\begin{longtable}{| l l l l c c c c c c |}
	
	\hline
	& \begin{tabular}{@{}c@{}}Space \\ Group\end{tabular}
	& \begin{tabular}{@{}c@{}}Materials \\ Project ID\end{tabular}
	& Formula & $N_{sites}$
	& \begin{tabular}{@{}c@{}}$\Delta E_{\rm{hull}}$ \\ (meV atom$^{-1}$)\end{tabular}
	& \begin{tabular}{@{}c@{}}$E_g$ \\ (meV)\end{tabular}
	& $Z_{max}$
	& \begin{tabular}{@{}c@{}}heavy atom PDOS \\ on [$-$1 eV, 1 eV] (\%)\end{tabular}
	& \begin{tabular}{@{}c@{}}light atom PDOS \\ on [$-$1 eV, 1 eV] (\%)\end{tabular}
    \\
	\hline
	\endhead

	\hline
	\endfoot

	1. & \#18 & \href{https://materialsproject.org/materials/mp-2503}{mp-2503} & Pd$_{\rm{7}}$Se$_{\rm{4}}$ & 22 & 0.0 & 0 & 46 & 0.0 & 84.8 \\ 
\hline
1. & \#19 & \href{https://materialsproject.org/materials/mp-557299}{mp-557299} & Nd$_{\rm{3}}$MoO$_{\rm{7}}$ & 44 & 12.9 & 66 & 60 & no data & no data \\ 
2. & \#19 & \href{https://materialsproject.org/materials/mp-607479}{mp-607479} & RbEuSiS$_{\rm{4}}$ & 28 & 0.0 & 0 & 63 & no data & no data \\ 
3. & \#19 & \href{https://materialsproject.org/materials/mp-30215}{mp-30215} & Pr$_{\rm{3}}$MoO$_{\rm{7}}$ & 44 & 4.3 & 47 & 59 & no data & no data \\ 
4. & \#19 & \href{https://materialsproject.org/materials/mp-28410}{mp-28410} & ThP$_{\rm{7}}$ & 32 & 39.1 & 0 & 90 & 20.8 & 26.8 \\ 
5. & \#19 & \href{https://materialsproject.org/materials/mp-568936}{mp-568936} & Ag$_{\rm{2}}$Se & 12 & 0.0 & 0 & 47 & 0.0 & 71.8 \\ 
6. & \#19 & \href{https://materialsproject.org/materials/mp-581303}{mp-581303} & NaCuPO$_{\rm{4}}$ & 28 & 40.0 & 4 & 29 & 0.0 & 89.3 \\ 
7. & \#19 & \href{https://materialsproject.org/materials/mp-29966}{mp-29966} & NOF & 12 & 0.0 & 14 & 9 & 0.0 & 78.2 \\ 
8. & \#19 & \href{https://materialsproject.org/materials/mp-1095694}{mp-1095694} & Ag$_{\rm{2}}$S & 12 & 48.7 & 13 & 47 & 0.0 & 73.0 \\ 
\hline
1. & \#33 & \href{https://materialsproject.org/materials/mp-1197254}{mp-1197254} & CaCuGe & 36 & 35.8 & 0 & 32 & no data & no data \\ 
2. & \#33 & \href{https://materialsproject.org/materials/mp-651900}{mp-651900} & CeSiP$_{\rm{3}}$ & 40 & 19.9 & 0 & 58 & no data & no data \\ 
3. & \#33 & \href{https://materialsproject.org/materials/mp-1181633}{mp-1181633} & CuOCl & 48 & 38.6 & 94 & 29 & no data & no data \\ 
4. & \#33 & \href{https://materialsproject.org/materials/mp-542839}{mp-542839} & AuTlSb & 24 & 0.0 & 0 & 81 & 50.9 & 0.0 \\ 
5. & \#33 & \href{https://materialsproject.org/materials/mp-541249}{mp-541249} & LaSiAs$_{\rm{3}}$ & 40 & 11.3 & 0 & 57 & 12.7 & 35.5 \\ 
6. & \#33 & \href{https://materialsproject.org/materials/mp-22889}{mp-22889} & ZnCl$_{\rm{2}}$ & 12 & 0.9 & 40 & 30 & 0.0 & 70.4 \\ 
\hline
1. & \#43 & \href{https://materialsproject.org/materials/mp-1203923}{mp-1203923} & Cu$_{\rm{3}}$B$_{\rm{6}}$O$_{\rm{13}}$H$_{\rm{2}}$ & 48 & 43.8 & 0 & 29 & no data & no data \\ 
2. & \#43 & \href{https://materialsproject.org/materials/mp-28896}{mp-28896} & IrCuB & 12 & 0.0 & 0 & 77 & 55.5 & 28.2 \\ 
3. & \#43 & \href{https://materialsproject.org/materials/mp-11441}{mp-11441} & Hf$_{\rm{2}}$Ga$_{\rm{3}}$ & 10 & 0.0 & 0 & 72 & 46.9 & 18.4 \\ 
4. & \#43 & \href{https://materialsproject.org/materials/mp-846}{mp-846} & Hf$_{\rm{2}}$Al$_{\rm{3}}$ & 10 & 2.2 & 0 & 72 & 45.0 & 6.8 \\ 
5. & \#43 & \href{https://materialsproject.org/materials/mp-28897}{mp-28897} & IrPdB & 12 & 0.0 & 0 & 77 & 41.6 & 50.5 \\ 
6. & \#43 & \href{https://materialsproject.org/materials/mp-542748}{mp-542748} & Ho$_{\rm{3}}$Ge$_{\rm{5}}$ & 16 & 0.0 & 0 & 67 & 37.0 & 27.8 \\ 
7. & \#43 & \href{https://materialsproject.org/materials/mp-505577}{mp-505577} & Dy$_{\rm{3}}$Ge$_{\rm{5}}$ & 16 & 0.7 & 0 & 66 & 36.4 & 27.6 \\ 
8. & \#43 & \href{https://materialsproject.org/materials/mp-2321}{mp-2321} & Tb$_{\rm{3}}$Ge$_{\rm{5}}$ & 16 & 1.3 & 0 & 65 & 35.5 & 27.2 \\ 
9. & \#43 & \href{https://materialsproject.org/materials/mp-580126}{mp-580126} & Sm$_{\rm{3}}$Ge$_{\rm{5}}$ & 16 & 3.5 & 0 & 62 & 27.9 & 25.8 \\ 
10. & \#43 & \href{https://materialsproject.org/materials/mp-2015}{mp-2015} & Nd$_{\rm{3}}$Ge$_{\rm{5}}$ & 16 & 0.0 & 0 & 60 & 26.2 & 25.3 \\ 
11. & \#43 & \href{https://materialsproject.org/materials/mp-567378}{mp-567378} & Ba$_{\rm{5}}$P$_{\rm{9}}$ & 28 & 1.0 & 0 & 56 & 12.0 & 26.5 \\ 
12. & \#43 & \href{https://materialsproject.org/materials/mp-647900}{mp-647900} & TlPbAs$_{\rm{3}}$S$_{\rm{6}}$ & 44 & 4.4 & 0 & 82 & 8.3 & 43.8 \\ 
13. & \#43 & \href{https://materialsproject.org/materials/mp-30686}{mp-30686} & Zr$_{\rm{2}}$Ga$_{\rm{3}}$ & 10 & 0.0 & 0 & 40 & 0.0 & 63.0 \\ 
14. & \#43 & \href{https://materialsproject.org/materials/mp-1482}{mp-1482} & Zr$_{\rm{2}}$Al$_{\rm{3}}$ & 10 & 0.0 & 0 & 40 & 0.0 & 50.4 \\ 
15. & \#43 & \href{https://materialsproject.org/materials/mp-17118}{mp-17118} & Y$_{\rm{3}}$Ge$_{\rm{5}}$ & 16 & 0.0 & 0 & 39 & 0.0 & 55.1 \\ 
16. & \#43 & \href{https://materialsproject.org/materials/mp-568726}{mp-568726} & MoSe$_{\rm{2}}$Cl$_{\rm{12}}$ & 30 & 32.0 & 0 & 42 & 0.0 & 77.3 \\ 
17. & \#43 & \href{https://materialsproject.org/materials/mp-1078326}{mp-1078326} & Ca$_{\rm{2}}$Pd$_{\rm{2}}$Ge & 10 & 0.0 & 0 & 46 & 0.0 & 50.6 \\ 
18. & \#43 & \href{https://materialsproject.org/materials/mp-505508}{mp-505508} & V$_{\rm{2}}$Cu$_{\rm{2}}$O$_{\rm{7}}$ & 22 & 15.3 & 0 & 29 & 0.0 & 90.1 \\ 
\hline
1. & \#52 & \href{https://materialsproject.org/materials/mp-31150}{mp-31150} & Sr$_{\rm{2}}$Bi$_{\rm{3}}$ & 20 & 0.0 & 0 & 83 & 49.6 & 7.3 \\ 
2. & \#52 & \href{https://materialsproject.org/materials/mp-558712}{mp-558712} & Ag$_{\rm{2}}$BiO$_{\rm{3}}$ & 24 & 4.4 & 0 & 83 & 8.5 & 75.6 \\ 
\hline
1. & \#57 & \href{https://materialsproject.org/materials/mp-542663}{mp-542663} & Ce$_{\rm{2}}$SiO$_{\rm{4}}$Te & 32 & 32.2 & 0 & 58 & no data & no data \\ 
2. & \#57 & \href{https://materialsproject.org/materials/mp-1200999}{mp-1200999} & PrFeSb$_{\rm{3}}$ & 40 & 21.1 & 0 & 59 & no data & no data \\ 
3. & \#57 & \href{https://materialsproject.org/materials/mp-16788}{mp-16788} & Gd$_{\rm{2}}$SiO$_{\rm{4}}$Te & 32 & 19.6 & 0 & 64 & no data & no data \\ 
4. & \#57 & \href{https://materialsproject.org/materials/mp-568237}{mp-568237} & CeNiSb$_{\rm{3}}$ & 40 & 4.3 & 0 & 58 & no data & no data \\ 
5. & \#57 & \href{https://materialsproject.org/materials/mp-21597}{mp-21597} & Ce$_{\rm{3}}$Rh$_{\rm{2}}$Ge$_{\rm{2}}$ & 28 & 0.0 & 0 & 58 & no data & no data \\ 
6. & \#57 & \href{https://materialsproject.org/materials/mp-21529}{mp-21529} & La$_{\rm{3}}$Ni$_{\rm{2}}$Ga$_{\rm{2}}$ & 28 & 0.0 & 0 & 57 & no data & no data \\ 
7. & \#57 & \href{https://materialsproject.org/materials/mp-568089}{mp-568089} & CeCoSb$_{\rm{3}}$ & 40 & 0.0 & 0 & 58 & no data & no data \\ 
8. & \#57 & \href{https://materialsproject.org/materials/mp-1196164}{mp-1196164} & PrAuAs$_{\rm{2}}$ & 32 & 0.0 & 0 & 79 & no data & no data \\ 
9. & \#57 & \href{https://materialsproject.org/materials/mp-1201728}{mp-1201728} & SmFeSb$_{\rm{3}}$ & 40 & 0.0 & 0 & 62 & no data & no data \\ 
10. & \#57 & \href{https://materialsproject.org/materials/mp-1201926}{mp-1201926} & NdFeSb$_{\rm{3}}$ & 40 & 18.7 & 0 & 60 & no data & no data \\ 
11. & \#57 & \href{https://materialsproject.org/materials/mp-1194691}{mp-1194691} & Sm$_{\rm{3}}$Ru$_{\rm{2}}$Ge$_{\rm{2}}$ & 28 & 0.0 & 0 & 62 & no data & no data \\ 
12. & \#57 & \href{https://materialsproject.org/materials/mp-1193516}{mp-1193516} & Eu$_{\rm{3}}$Pd$_{\rm{2}}$Sn$_{\rm{2}}$ & 28 & 0.0 & 0 & 63 & no data & no data \\ 
13. & \#57 & \href{https://materialsproject.org/materials/mp-1193872}{mp-1193872} & Pr$_{\rm{3}}$Ru$_{\rm{2}}$Ge$_{\rm{2}}$ & 28 & 0.0 & 0 & 59 & no data & no data \\ 
14. & \#57 & \href{https://materialsproject.org/materials/mp-1193804}{mp-1193804} & Pr$_{\rm{3}}$Ni$_{\rm{2}}$Ga$_{\rm{2}}$ & 28 & 4.0 & 0 & 59 & no data & no data \\ 
15. & \#57 & \href{https://materialsproject.org/materials/mp-1198391}{mp-1198391} & LaAgAs$_{\rm{2}}$ & 32 & 0.0 & 0 & 57 & no data & no data \\ 
16. & \#57 & \href{https://materialsproject.org/materials/mp-1196110}{mp-1196110} & SrCuO$_{\rm{7}}$Te$_{\rm{2}}$ & 44 & 24.7 & 0 & 52 & no data & no data \\ 
17. & \#57 & \href{https://materialsproject.org/materials/mp-555935}{mp-555935} & Ta$_{\rm{2}}$S & 36 & 26.1 & 0 & 73 & no data & no data \\ 
18. & \#57 & \href{https://materialsproject.org/materials/mp-1199267}{mp-1199267} & TbFeSb$_{\rm{3}}$ & 40 & 7.6 & 0 & 65 & no data & no data \\ 
19. & \#57 & \href{https://materialsproject.org/materials/mp-1203135}{mp-1203135} & CuPbO$_{\rm{7}}$Te$_{\rm{2}}$ & 44 & 0.0 & 0 & 82 & no data & no data \\ 
20. & \#57 & \href{https://materialsproject.org/materials/mp-619100}{mp-619100} & LiEuTiO$_{\rm{4}}$ & 28 & 34.9 & 0 & 63 & no data & no data \\ 
21. & \#57 & \href{https://materialsproject.org/materials/mp-16721}{mp-16721} & TmAl & 16 & 0.0 & 0 & 69 & 60.3 & 5.8 \\ 
22. & \#57 & \href{https://materialsproject.org/materials/mp-1188739}{mp-1188739} & ErAl & 16 & 0.0 & 0 & 68 & 59.4 & 5.7 \\ 
23. & \#57 & \href{https://materialsproject.org/materials/mp-1188420}{mp-1188420} & HoAl & 16 & 0.0 & 0 & 67 & 57.5 & 5.7 \\ 
24. & \#57 & \href{https://materialsproject.org/materials/mp-542382}{mp-542382} & ThTl & 24 & 0.0 & 0 & 90 & 56.0 & 0.0 \\ 
25. & \#57 & \href{https://materialsproject.org/materials/mp-433}{mp-433} & DyAl & 16 & 0.0 & 0 & 66 & 56.0 & 5.6 \\ 
26. & \#57 & \href{https://materialsproject.org/materials/mp-11225}{mp-11225} & TbAl & 16 & 0.0 & 0 & 65 & 54.1 & 5.6 \\ 
27. & \#57 & \href{https://materialsproject.org/materials/mp-510188}{mp-510188} & HfGa & 24 & 0.0 & 0 & 72 & 52.3 & 14.5 \\ 
28. & \#57 & \href{https://materialsproject.org/materials/mp-16507}{mp-16507} & LuAl & 16 & 0.0 & 0 & 71 & 51.8 & 5.9 \\ 
29. & \#57 & \href{https://materialsproject.org/materials/mp-978951}{mp-978951} & SmAl & 16 & 0.0 & 0 & 62 & 39.9 & 5.4 \\ 
30. & \#57 & \href{https://materialsproject.org/materials/mp-864637}{mp-864637} & NdAl & 16 & 0.0 & 0 & 60 & 38.4 & 5.3 \\ 
31. & \#57 & \href{https://materialsproject.org/materials/mp-569538}{mp-569538} & LaNiSb$_{\rm{3}}$ & 40 & 0.0 & 0 & 57 & 35.6 & 26.5 \\ 
32. & \#57 & \href{https://materialsproject.org/materials/mp-638141}{mp-638141} & LiRh$_{\rm{3}}$Sn$_{\rm{5}}$ & 36 & 0.0 & 0 & 50 & 25.4 & 47.1 \\ 
33. & \#57 & \href{https://materialsproject.org/materials/mp-569163}{mp-569163} & SmCoSb$_{\rm{3}}$ & 40 & 13.4 & 0 & 62 & 23.7 & 48.9 \\ 
34. & \#57 & \href{https://materialsproject.org/materials/mp-570885}{mp-570885} & LaCoSb$_{\rm{3}}$ & 40 & 6.4 & 0 & 57 & 23.2 & 49.7 \\ 
35. & \#57 & \href{https://materialsproject.org/materials/mp-30718}{mp-30718} & K$_{\rm{5}}$Hg$_{\rm{7}}$ & 48 & 3.2 & 0 & 80 & 22.8 & 4.8 \\ 
36. & \#57 & \href{https://materialsproject.org/materials/mp-630836}{mp-630836} & Nd$_{\rm{3}}$Ru$_{\rm{2}}$Ge$_{\rm{2}}$ & 28 & 0.0 & 0 & 60 & 22.3 & 43.5 \\ 
37. & \#57 & \href{https://materialsproject.org/materials/mp-569153}{mp-569153} & PrCoSb$_{\rm{3}}$ & 40 & 9.9 & 0 & 59 & 22.2 & 50.5 \\ 
38. & \#57 & \href{https://materialsproject.org/materials/mp-570956}{mp-570956} & NdCoSb$_{\rm{3}}$ & 40 & 10.1 & 0 & 60 & 21.8 & 51.7 \\ 
39. & \#57 & \href{https://materialsproject.org/materials/mp-31479}{mp-31479} & Ca$_{\rm{3}}$Pt$_{\rm{2}}$Ga$_{\rm{2}}$ & 28 & 0.0 & 0 & 78 & 20.8 & 27.6 \\ 
40. & \#57 & \href{https://materialsproject.org/materials/mp-1012110}{mp-1012110} & Cs & 4 & 46.3 & 0 & 55 & 8.1 & 0.0 \\ 
41. & \#57 & \href{https://materialsproject.org/materials/mp-553964}{mp-553964} & WO$_{\rm{3}}$ & 16 & 1.6 & 1 & 74 & 1.2 & 75.4 \\ 
42. & \#57 & \href{https://materialsproject.org/materials/mp-640340}{mp-640340} & MgCa$_{\rm{4}}$Al$_{\rm{3}}$ & 32 & 0.0 & 0 & 20 & 0.0 & 25.0 \\ 
43. & \#57 & \href{https://materialsproject.org/materials/mp-504772}{mp-504772} & Y$_{\rm{3}}$Rh$_{\rm{2}}$Si$_{\rm{2}}$ & 28 & 0.0 & 0 & 45 & 0.0 & 58.7 \\ 
44. & \#57 & \href{https://materialsproject.org/materials/mp-31480}{mp-31480} & Ca$_{\rm{3}}$Pd$_{\rm{2}}$Ga$_{\rm{2}}$ & 28 & 0.0 & 0 & 46 & 0.0 & 44.5 \\ 
45. & \#57 & \href{https://materialsproject.org/materials/mp-571039}{mp-571039} & CaPdAl & 12 & 0.0 & 0 & 46 & 0.0 & 40.1 \\ 
\hline
1. & \#60 & \href{https://materialsproject.org/materials/mp-602551}{mp-602551} & EuMo$_{\rm{2}}$TlO$_{\rm{8}}$ & 48 & 0.0 & 0 & 81 & no data & no data \\ 
2. & \#60 & \href{https://materialsproject.org/materials/mp-7228}{mp-7228} & ReO$_{\rm{2}}$ & 12 & 1.7 & 0 & 75 & 67.2 & 20.6 \\ 
3. & \#60 & \href{https://materialsproject.org/materials/mp-1190478}{mp-1190478} & Au$_{\rm{2}}$Pb & 24 & 3.8 & 0 & 82 & 64.9 & 0.0 \\ 
4. & \#60 & \href{https://materialsproject.org/materials/mp-607990}{mp-607990} & Hf$_{\rm{2}}$Co$_{\rm{3}}$Si$_{\rm{4}}$ & 36 & 0.0 & 0 & 72 & 16.2 & 65.7 \\ 
5. & \#60 & \href{https://materialsproject.org/materials/mp-20633}{mp-20633} & PbO$_{\rm{2}}$ & 12 & 5.8 & 0 & 82 & 4.2 & 69.8 \\ 
6. & \#60 & \href{https://materialsproject.org/materials/mp-505229}{mp-505229} & FeAl$_{\rm{3}}$Si$_{\rm{2}}$ & 24 & 0.0 & 0 & 26 & 0.0 & 58.4 \\ 
7. & \#60 & \href{https://materialsproject.org/materials/mp-20648}{mp-20648} & V$_{\rm{2}}$C & 12 & 0.0 & 0 & 23 & 0.0 & 82.4 \\ 
8. & \#60 & \href{https://materialsproject.org/materials/mp-1552}{mp-1552} & Mo$_{\rm{2}}$C & 12 & 0.0 & 0 & 42 & 0.0 & 87.2 \\ 
9. & \#60 & \href{https://materialsproject.org/materials/mp-1105666}{mp-1105666} & Rh$_{\rm{2}}$Se$_{\rm{3}}$ & 20 & 0.0 & 0 & 45 & 0.0 & 79.5 \\ 
\hline
1. & \#61 & \href{https://materialsproject.org/materials/mp-567620}{mp-567620} & CoTlCl$_{\rm{3}}$ & 40 & 7.7 & 0 & 81 & no data & no data \\ 
2. & \#61 & \href{https://materialsproject.org/materials/mp-28301}{mp-28301} & OsBr$_{\rm{4}}$ & 40 & 0.0 & 0 & 76 & no data & no data \\ 
3. & \#61 & \href{https://materialsproject.org/materials/mp-568531}{mp-568531} & PtBi$_{\rm{2}}$ & 24 & 0.0 & 0 & 83 & 75.1 & 0.0 \\ 
4. & \#61 & \href{https://materialsproject.org/materials/mp-1462}{mp-1462} & AuSn$_{\rm{2}}$ & 24 & 0.0 & 0 & 79 & 59.8 & 0.0 \\ 
5. & \#61 & \href{https://materialsproject.org/materials/mp-567945}{mp-567945} & PdSnTe & 24 & 0.0 & 0 & 52 & 42.0 & 21.4 \\ 
6. & \#61 & \href{https://materialsproject.org/materials/mp-1321}{mp-1321} & CdSb & 16 & 0.0 & 53 & 51 & 27.8 & 17.0 \\ 
7. & \#61 & \href{https://materialsproject.org/materials/mp-753}{mp-753} & ZnSb & 16 & 0.0 & 76 & 51 & 27.6 & 20.8 \\ 
8. & \#61 & \href{https://materialsproject.org/materials/mp-648984}{mp-648984} & LaSiAs$_{\rm{3}}$ & 40 & 34.2 & 0 & 57 & 10.8 & 38.5 \\ 
9. & \#61 & \href{https://materialsproject.org/materials/mp-2418}{mp-2418} & PdSe$_{\rm{2}}$ & 12 & 0.0 & 8 & 46 & 0.0 & 72.6 \\ 
10. & \#61 & \href{https://materialsproject.org/materials/mp-505510}{mp-505510} & NiAs$_{\rm{2}}$ & 24 & 0.0 & 0 & 33 & 0.0 & 68.5 \\ 
11. & \#61 & \href{https://materialsproject.org/materials/mp-21598}{mp-21598} & Ni$_{\rm{2}}$SiAs & 32 & 39.9 & 0 & 33 & 0.0 & 79.1 \\ 
12. & \#61 & \href{https://materialsproject.org/materials/mp-27844}{mp-27844} & NiP & 16 & 16.5 & 0 & 28 & 0.0 & 79.5 \\ 
13. & \#61 & \href{https://materialsproject.org/materials/mp-618929}{mp-618929} & Ni$_{\rm{2}}$GeP & 32 & 21.9 & 0 & 32 & 0.0 & 80.8 \\ 
14. & \#61 & \href{https://materialsproject.org/materials/mp-510422}{mp-510422} & Ni$_{\rm{2}}$SiP & 32 & 21.5 & 0 & 28 & 0.0 & 78.2 \\ 
15. & \#61 & \href{https://materialsproject.org/materials/mp-2284}{mp-2284} & AgF$_{\rm{2}}$ & 12 & 8.7 & 0 & 47 & 0.0 & 93.0 \\

\hline

... \\

251. & \#62 & \href{https://materialsproject.org/materials/mp-11534}{mp-11534} & Np & 8 & 0.0 & 0 & 93 & 99.7 & 0.0 \\ 
252. & \#62 & \href{https://materialsproject.org/materials/mp-1101808}{mp-1101808} & CeIrSb & 12 & 0.0 & 8 & 77 & 90.7 & 0.0 \\ 
253. & \#62 & \href{https://materialsproject.org/materials/mp-1006371}{mp-1006371} & CePS & 24 & 0.0 & 0 & 58 & 90.3 & 2.1 \\ 
254. & \#62 & \href{https://materialsproject.org/materials/mp-1095615}{mp-1095615} & CeAsS & 12 & 37.9 & 0 & 58 & 90.1 & 2.7 \\ 
255. & \#62 & \href{https://materialsproject.org/materials/mp-20834}{mp-20834} & CePtSn & 12 & 0.0 & 0 & 78 & 89.3 & 0.0 \\ 
256. & \#62 & \href{https://materialsproject.org/materials/mp-1103038}{mp-1103038} & Ir$_{\rm{2}}$Si & 12 & 7.2 & 0 & 77 & 89.1 & 2.5 \\ 
257. & \#62 & \href{https://materialsproject.org/materials/mp-1105571}{mp-1105571} & UReC$_{\rm{2}}$ & 16 & 0.0 & 0 & 92 & 88.9 & 2.9 \\ 
258. & \#62 & \href{https://materialsproject.org/materials/mp-652608}{mp-652608} & UWC$_{\rm{2}}$ & 16 & 0.0 & 0 & 92 & 88.6 & 2.6 \\ 
259. & \#62 & \href{https://materialsproject.org/materials/mp-20022}{mp-20022} & CeIrGe & 12 & 0.0 & 0 & 77 & 88.3 & 2.6 \\ 
260. & \#62 & \href{https://materialsproject.org/materials/mp-570180}{mp-570180} & YbPt & 8 & 19.9 & 67 & 78 & 87.5 & 0.0 \\ 
261. & \#62 & \href{https://materialsproject.org/materials/mp-1079395}{mp-1079395} & CeGe & 8 & 1.7 & 0 & 58 & 86.2 & 2.0 \\ 
262. & \#62 & \href{https://materialsproject.org/materials/mp-21163}{mp-21163} & Pt$_{\rm{3}}$Si & 16 & 34.7 & 0 & 78 & 86.1 & 3.3 \\ 
263. & \#62 & \href{https://materialsproject.org/materials/mp-1101892}{mp-1101892} & CeAs$_{\rm{2}}$ & 12 & 0.0 & 72 & 58 & 83.3 & 5.8 \\ 
264. & \#62 & \href{https://materialsproject.org/materials/mp-1102133}{mp-1102133} & Tm$_{\rm{2}}$Pt & 12 & 0.0 & 0 & 78 & 82.3 & 0.0 \\ 
265. & \#62 & \href{https://materialsproject.org/materials/mp-22249}{mp-22249} & CeRhSb & 12 & 0.0 & 11 & 58 & 82.2 & 7.9 \\ 

... \\

1155. & \#62 & \href{https://materialsproject.org/materials/mp-1105559}{mp-1105559} & YPd$_{\rm{2}}$Ga & 16 & 0.0 & 0 & 46 & 0.0 & 65.1 \\ 
1156. & \#62 & \href{https://materialsproject.org/materials/mp-1025513}{mp-1025513} & ZrGe & 8 & 0.0 & 0 & 40 & 0.0 & 65.3 \\ 
1157. & \#62 & \href{https://materialsproject.org/materials/mp-605663}{mp-605663} & Pd$_{\rm{5}}$Al & 24 & 0.0 & 0 & 46 & 0.0 & 90.6 \\ 
1158. & \#62 & \href{https://materialsproject.org/materials/mp-21069}{mp-21069} & YRuSi & 12 & 0.0 & 0 & 44 & 0.0 & 70.0 \\ 
1159. & \#62 & \href{https://materialsproject.org/materials/mp-1103148}{mp-1103148} & ZrCoGe & 12 & 0.0 & 0 & 40 & 0.0 & 83.5 \\ 
1160. & \#62 & \href{https://materialsproject.org/materials/mp-1435}{mp-1435} & SrZn$_{\rm{5}}$ & 24 & 0.0 & 0 & 38 & 0.0 & 44.1 \\ 
1161. & \#62 & \href{https://materialsproject.org/materials/mp-680576}{mp-680576} & Ti$_{\rm{2}}$Se & 36 & 4.9 & 0 & 34 & 0.0 & 79.2 \\ 
1162. & \#62 & \href{https://materialsproject.org/materials/mp-1101785}{mp-1101785} & YCoGe & 12 & 0.0 & 0 & 39 & 0.0 & 81.8 \\ 
1163. & \#62 & \href{https://materialsproject.org/materials/mp-1181917}{mp-1181917} & CaRhO$_{\rm{3}}$ & 20 & 13.8 & 0 & 45 & 0.0 & 86.5 \\ 
1164. & \#62 & \href{https://materialsproject.org/materials/mp-1103655}{mp-1103655} & YNiGe & 12 & 0.0 & 0 & 39 & 0.0 & 72.3 \\ 
1165. & \#62 & \href{https://materialsproject.org/materials/mp-1424}{mp-1424} & PdGe & 8 & 0.0 & 0 & 46 & 0.0 & 75.4 \\ 
1166. & \#62 & \href{https://materialsproject.org/materials/mp-2058}{mp-2058} & Ni$_{\rm{3}}$B & 16 & 0.0 & 0 & 28 & 0.0 & 94.5 \\ 
1167. & \#62 & \href{https://materialsproject.org/materials/mp-20327}{mp-20327} & TiP$_{\rm{2}}$ & 12 & 0.0 & 0 & 22 & 0.0 & 70.4 \\ 
1168. & \#62 & \href{https://materialsproject.org/materials/mp-20948}{mp-20948} & NbNiP$_{\rm{2}}$ & 16 & 0.0 & 0 & 41 & 0.0 & 78.4 \\ 
1169. & \#62 & \href{https://materialsproject.org/materials/mp-22262}{mp-22262} & MoCoP & 12 & 0.0 & 0 & 42 & 0.0 & 90.2 \\ 
\hline
1. & \#73 & \href{https://materialsproject.org/materials/mp-568329}{mp-568329} & Na$_{\rm{2}}$K$_{\rm{3}}$SnBi$_{\rm{3}}$ & 36 & 0.0 & 6 & 83 & 48.6 & 3.1 \\

\end{longtable}
\normalsize